\documentclass[9pt]{article}
\usepackage[margin=1.2in]{geometry}
\usepackage[utf8]{inputenc}
\usepackage[english]{babel}

\usepackage{natbib}
\usepackage{authblk}
\usepackage{amsmath}

\usepackage{hyperref}
\usepackage{tikz}
\usepackage{booktabs}
\usepackage{titlesec}

\newcommand{\cip}{\mbox{$\perp\!\!\!\perp$}}

\allowdisplaybreaks

\newcommand*\rot{\rotatebox{90}}

\titleformat*{\section}{\large\bfseries}
\titleformat*{\subsection}{\normalsize\bfseries}

\usetikzlibrary{calc,positioning,arrows,shapes.multipart}

\title{\Large \bf Handling time-dependent exposures and confounders when estimating attributable fractions --- bridging the gap between multistate and counterfactual modeling}
\author[1,2,3]{Johan Steen}
\author[4]{Pawe\l{} Morzywo\l{}ek}
\author[1,2]{Wim Van Biesen}
\author[1,3]{\\Johan Decruyenaere}
\author[4,5]{Stijn Vansteelandt}
\affil[1]{\scriptsize Department of Internal Medicine and Pediatrics, Ghent University}
\affil[2]{\scriptsize Renal Division, Ghent University Hospital}
\affil[3]{\scriptsize Department of Intensive Care Medicine, Ghent University Hospital}
\affil[4]{\scriptsize Department of Applied Mathematics, Computer Science and Statistics, Ghent University}
\affil[5]{\scriptsize Department of Medical Statistics, London School of Hygiene and Tropical Medicine}
\date{January 19, 2023}

\begin{document}

\maketitle

\begin{figure}[b]
    \fbox{%
      \footnotesize
        \parbox{\textwidth}{
                \textbf{This is the pre-peer reviewed version of the following article:\\
                \textit{Steen J, Morzywołek P, Van Biesen W, Decruyenaere J, Vansteelandt S. Dealing with time-dependent exposures and confounding when defining and estimating attributable fractions—Revisiting estimands and estimators. Statistics in Medicine. 2023;1-23. doi: 10.1002/sim.9988},\\
                which has been published in final form at \url{https://onlinelibrary.wiley.com/doi/full/10.1002/sim.9988}.\\
                This article may be used for non-commercial purposes in accordance with Wiley Terms and Conditions for Use of Self-Archived Versions.}
        }%
    }
\end{figure}

\begin{abstract}
  The population-attributable fraction (PAF) expresses the proportion of events that can be ascribed to a certain exposure in a certain population. It can be strongly time-dependent because either exposure incidence or excess risk may change over time. Competing events may moreover hinder the outcome of interest from being observed. Occurrence of either of these events may, in turn, prevent the exposure of interest. Estimation approaches therefore need to carefully account for the timing of events in such highly dynamic settings. The use of multistate models has been widely encouraged to eliminate preventable yet common types of time-dependent bias. Even so, it has been pointed out that proposed multistate modeling approaches for PAF estimation fail to fully eliminate such bias. In addition, assessing whether patients die \emph{from} rather than \emph{with} a certain exposure not only requires adequate modeling of the timing of events but also of their confounding factors. While proposed multistate modeling approaches for confounding adjustment may adequately accommodate baseline imbalances, unlike \emph{g-methods}, these proposals are not generally equipped to handle time-dependent confounding. However, the connection between multistate modeling and g-methods (e.g. inverse probability of censoring weighting) for PAF estimation is not readily apparent. In this paper, we provide a weighting-based characterization of both approaches to illustrate this connection, to pinpoint current shortcomings of multistate modeling, and to enhance intuition into simple modifications to overcome these. R code is made available to foster the uptake of g-methods for PAF estimation.
\end{abstract}

\section{Introduction}

The population-attributable fraction (PAF) expresses the proportion of events that can be ascribed to a certain exposure or risk factor in a certain population, thereby quantifying the contribution of this risk factor to the burden of mortality or morbidity.
This quantity involves a comparison of the observable probability of an event (in this world) and the unobservable probability of an event in a hypothetical world where exposure could be eradicated.

A major focus in the field of hospital epidemiology is accurate quantification of the burden of hospital-acquired infections (HAIs) on hospital mortality. Since HAIs are acquired in the course of hospitalization and because the burden may depend on the actual onset and duration, there is special interest in quantifying this burden as a function of time since hospital admission.
Accordingly, the time-dependent PAF of hospital mortality due to HAIs can be expressed in terms of cumulative incidences or incidence proportions
\begin{eqnarray*}
  \text{PAF}(t) &= \dfrac{\Pr(T\le t, \epsilon = 1)-\Pr(T^0\le t, \epsilon^0 = 1)}{\Pr(T\le t, \epsilon = 1)},
\end{eqnarray*}
where $T$ denotes the observed time from hospital admission (i.e. the time origin) to hospital death or discharge, $\epsilon$ the type of the observed event ($\epsilon = 1$ indicating hospital death and $\epsilon = 2$ indicating hospital discharge), and $T^0$ and $\epsilon^0$ the hypothetical or \emph{counterfactual} event time and event type indicator that would (hypothetically) have been observed if --- possibly \emph{counter to the fact} --- no infection were acquired.

$\Pr(T \le t, \epsilon = 1)$ and $\Pr(T^0\le t, \epsilon^0 = 1)$ respectively denote the observed (factual) and the hypothetical (counterfactual) cumulative incidence of hospital mortality at time $t$. The observable quantity $\Pr(T\le t, \epsilon = 1)$ expresses the cumulative incidence under observed circumstances (i.e. under standard preventive and therapeutic care). Given complete follow-up (until hospital death or discharge), it can be consistently estimated from the observed data without any causal assumptions. Estimation of the unobservable quantity $\Pr(T^0\le t, \epsilon^0 = 1)$ is arguably more challenging, due to its hypothetical nature, and necessitates strong and untestable causal assumptions. Notwithstanding this, consistent estimation of this quantity is required to warrant the intended interpretation of the PAF as a measure of \emph{causal} attribution (although to some readers this may sound like a tautology).

In this manuscript, we illustrate that proposed approaches for estimating $\Pr(T^0\le t, \epsilon^0 = 1)$ each aim to answer the counterfactual question of how the future of infected patients had unfolded had they not developed an HAI; they do so by letting infected patients transfer their weight in the analysis to uninfected patients.
Intuitively, transferring weight from infected to uninfected patients compensates for the depletion of infected patients from the `counterfactual risk set', so as to reconstruct the original study population in a hypothetical world where HAI is eradicated.
To avoid bias, these weight transfers should ideally account for subtle aspects related to the timing of observable events.
Proposed estimation approaches differ in the way these aspects are taken into account.
A weighting based characterization of these approaches therefore sheds light on the biases they may produce and provides a conceptually appealing bridge between multistate and counterfactual modeling approaches.

In the next section, we further formalize and extend notation.
In Section~\ref{sec:main-estimands}, we more formally define the PAF as an estimand expressed in terms of factual and counterfactual risks. Estimation of the factual risks is discussed and introduced as basis for estimation of the counterfactual risks.
In Section~\ref{sec:main-ident}, we present conditions that permit identifying the counterfactual risk from the observed data.
In Sections~\ref{sec:main-timedepexp} and~\ref{sec:main-timedepconf}, we discuss and compare different proposed approaches for estimating the counterfactual risk $\Pr(T^0\le t, \epsilon^0 = 1)$ using a weighting based characterization of each of these approaches. By means of a toy example, we first illustrate in Section~\ref{sec:main-timedepexp} how estimation approaches differ in the way their corresponding weights account for the time-dependent nature of exposure and its competing events. We illustrate that multistate modeling approaches can be organized hierarchically in terms of how well they succeed to eliminate different forms of time-dependent bias, and that the preferred multistate modeling approach corresponds to artificially censoring the counterfactual event time at the time of exposure onset.
In Section~\ref{sec:main-timedepconf}, we discuss how further accounting for the time-dependent nature of prognostic factors related to the exposure may reduce selection bias due to artificial censoring using inverse probability of censoring weighting (IPCW).
In Section~\ref{sec:main-empex}, we present a direct comparison of the discussed estimation approaches in an empirical example using data from the Ghent University Hospital ICUs. Finally, in Section~\ref{sec:main-discussion}, we end with a discussion in which we briefly touch upon challenges related to causal interpretation of the PAF in terms of well-defined interventions.

\section{Setting and notation}\label{sec:main-setnot}

\begin{figure}[t!]
\centerline{\includegraphics[width=250pt]{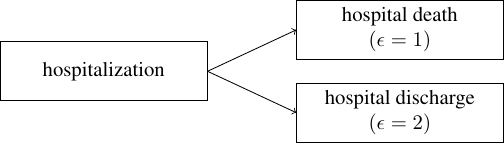}}
\caption{A competing risk model in which hospital discharge is treated as a competing event.\label{fig:competingrisk1}}
\end{figure}

\begin{figure}[t!]
\centerline{\includegraphics[width=250pt]{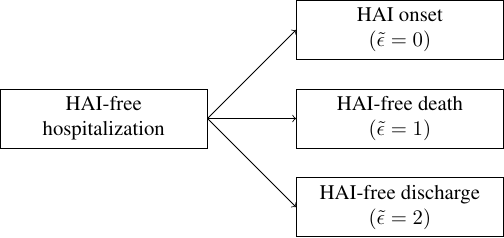}}
\caption{A competing risk model in which HAI onset is additionally treated as a competing event.\label{fig:competingrisk2}}
\end{figure}

For each patient $i = 1,...,n$, we observe the time $T_i$ from hospital admission to either hospital death or discharge, whichever occurs first, and the event type $\epsilon_i=j$, $j\in \{1,2\}$, with 1 indicating hospital death, the event of interest, and 2 indicating hospital discharge, the competing event (as in the simple competing risk model depicted in Figure~\ref{fig:competingrisk1}). In patients who acquire infection, we also observe the time $C_i$ from hospital admission to infection onset. In patients who get discharged or die at the hospital without having acquired infection, $C_i$ is set to $+\infty$. To quantify the probability of experiencing an exposure- or infection-free event, we also consider $\tilde T_i \equiv \text{min}(T_i, C_i)$, the time from hospital admission to either HAI, hospital death or discharge, whichever occurs first, and the event type $\tilde \epsilon_i \equiv \delta_i \epsilon_i = j, j \in \{0,1,2\}$, with $\delta_i \equiv I(T_i < C_i)$ and $I(\cdot)$ the indicator function (as in the competing risk model depicted in Figure~\ref{fig:competingrisk2}, with 0 indicating infection onset, 1 infection-free hospital death and 2 infection-free hospital discharge). Throughout, the observed data are assumed to be independent and identically distributed across $i$, so we will generally suppress the index $i$ to simplify notation, unless otherwise noted.

We consider equally spaced discrete time follow-up intervals indexed by $k = 0,1,...,K$, with interval $k$ defined as $(t_{k-1},t_k]$, $t_0 \equiv 0$ indicating the time origin (hospital admission) and $K$ the follow-up interval of interest, at the end of or before the maximum possible follow-up interval $\tau$.
We define the discrete-time counting processes
\begin{eqnarray*}
  A_k &\equiv& I\left(\tilde T \le t_k, \tilde \epsilon = 0\right)\\
  D_k &\equiv& I\left(T \le t_k, \epsilon = 2\right)\\
  Y_k &\equiv& I\left(T \le t_k, \epsilon = 1\right),
\end{eqnarray*}
which denote monotone indicators for infection, hospital discharge and hospital death by the end of interval $k$, respectively \citep[see][for similar notation]{Bekaert2010,Young2020}. Because all patients are admitted to the hospital without HAIs, we have $A_0 \equiv D_0 \equiv Y_0 \equiv 0$. Furthermore, note that $A_{\tau} \equiv I(\tilde T \le t_{\tau}, \tilde \epsilon = 0) = I(\tilde T \le T, \tilde \epsilon = 0) = I(\tilde \epsilon = 0)$. At the end of each interval $k$, we observe a vector $L_k$ of time-varying patient-specific covariates (i.e. patient characteristics and prognostic factors). For $k \ge 1$, this may, for instance, include indicators for intercurrent events and time-updated values of either baseline measurements (which form a subset of $L_0$) or measurements for which registration only starts at some point after baseline (e.g. measurements that may be contingent on an intercurrent event).

In many applications, including ours, follow-up may not be periodic (e.g. once daily) for HAI onset and hospital death and discharge, but discretization may nonetheless be desirable to some level to warrant appropriate alignment with periodic follow-up measures included in $L_k$. To account for potential ties that may arise from this, within each follow-up interval $k \ge 1$, we assume the temporal ordering $(A_k, D_k, Y_k, L_k)$. As the length of chosen follow-up intervals approximates zero, so does the probability of ties. By consequence, this ordering assumption then becomes practically irrelevant \citep{Young2020,Stensrud2021}.
This allows us to define time-dependent indicators for infection-free hospital discharge and infection-free hospital death as follows
\begin{eqnarray*}
  \tilde D_k &\equiv& I\left(\tilde T \le t_k, \tilde \epsilon = 2\right) = \left(1-A_k\right) D_k\\
  \tilde Y_k &\equiv& I\left(\tilde T \le t_k, \tilde \epsilon = 1\right) = \left(1-A_k\right) Y_k.
\end{eqnarray*}
For any variable $Z_k$, we define the history of measures from start of follow-up as $\overline{Z}_k \equiv \{Z_0, ..., Z_k\}$ and the future of measures until the end of follow-up of interest as $\underline Z_k \equiv \{Z_k, ..., Z_K\}$. As soon as a patient has experienced one of the competing events in any interval $k$, s/he can no longer experience the other event(s). For instance, whenever a patient is discharged from the hospital in interval $k$ ($D_{k-1} = 0, D_k = 1$), all future event indicators $Y_k$ are deterministically zero.

Let $\overline a \equiv \overline a_{\tau} = 0$ denote a HAI-free counterfactual or hypothetical exposure path, and $T_i^{\overline a = 0}$ the counterfactual event time of patient $i$ under a hypothetical regime where --- possibly counter to the fact --- patient $i$ remained uninfected while hospitalized.
Furthermore, let $\epsilon_i^{\overline a = 0} = j, j \in \{1,2\},$ denote the counterfactual event type of patient $i$,
where, as before, $j=1$ in case of hospital death, or $j=2$ in case of hospital discharge. In the previous section, we have used the shorthand notation $T^0$ and $\epsilon^0$ to indicate $T^{\overline a = 0}$ and $\epsilon^{\overline a = 0}$, respectively \citep{Bekaert2010}. Equivalently, we define the counterfactual discrete-time counting processes
\begin{eqnarray*}
  D^{\overline a = 0}_k &\equiv& I\left(T^{\overline a = 0} \le t_k, \epsilon^{\overline a = 0} = 2\right)\\
  Y^{\overline a = 0}_k &\equiv& I\left(T^{\overline a = 0} \le t_k, \epsilon^{\overline a = 0} = 1\right).
\end{eqnarray*}
Finally, for notational convenience, we define
\begin{eqnarray*}
  \phi_K &\equiv& \Pr\left(T\le t_K, \epsilon = 1\right) = \Pr\left(Y_K = 1\right)\\
  \tilde \phi_K &\equiv& \Pr\left(\tilde T\le t_K, \tilde \epsilon = 1\right) = \Pr\left(\tilde Y_K = 1\right)\\
  \alpha_K &\equiv& \Pr\left(\tilde T\le t_K, \tilde \epsilon = 0\right) = \Pr\left(A_K = 1\right)\\
  \varphi_K &\equiv& \Pr\left(T^{\overline a = 0}\le t_K, \epsilon^{\overline a = 0} = 1\right) = \Pr\left(Y^{\overline a = 0}_K = 1\right).
\end{eqnarray*}

\section{Factual and counterfactual estimands}\label{sec:main-estimands}

The fraction of deceased in-patients that have died \emph{with} infection by the end of follow-up interval of interest $K$ is defined as
\begin{eqnarray*}
  \dfrac{\phi_K-\tilde \phi_K}{\phi_K},
\end{eqnarray*}
which involves a contrast of two observable or factual quantities: $\phi_K$, the cumulative incidence of hospital death, and $\tilde \phi_K$, the (event-specific) cumulative incidence or \emph{crude risk} of infection-free hospital death.

The fraction of deceased in-patients that have died \emph{from} infection or the PAF by the end of interval $K$ is defined as
\begin{eqnarray*}
  \dfrac{\phi_K-\varphi_K}{\phi_K},
\end{eqnarray*}
which involves a contrast of a factual and counterfactual quantity: $\phi_K$, the cumulative incidence of hospital death,
and $\varphi_K$, the cumulative incidence or \emph{net risk} of hospital death \emph{under (hypothetical) elimination of infection}. Causal contrasts between observed and hypothetical scenarios, as expressed in the numerator, have been coined \emph{population intervention effects} \citep{Hubbard2008a,Westreich2017}.

\subsection{Factual risks}
The factual risks $\phi_K$ and $\tilde \phi_K$ are identifiable from the observed data without causal assumptions, except, perhaps, in the presence of loss to follow-up or administrative censoring. In the absence of such censoring, $\phi_K$ is identified by the following functional of the observed data
\begin{eqnarray*}
  \sum_{k = 1}^K E\left[Y_k\left(1-D_k\right)\left(1-Y_{k-1}\right)\right],
\end{eqnarray*}
which can be estimated by the empirical cumulative distribution function of hospital death
\begin{eqnarray*}
    \widehat\phi_K &\equiv& n^{-1} \sum_{k = 1}^K d_{1k},
\end{eqnarray*}
with $d_{1k} = \sum_{i=1}^n Y_{ik} (1-D_{ik}) (1-Y_{i,k-1})$ denoting the number of hospital deaths in interval $k$.
In the absence of censoring, $\widehat\phi_K$ is algebraically equivalent to the Aalen-Johansen estimator (see Appendix~\ref{sec:app-estobsdata1} in the Supplementary Material for a formal proof)
\begin{eqnarray*}
  \widehat\phi_K^{\text{\tiny AJ}} &\equiv& \sum_{k = 1}^K \widehat{h}^{(1)}_k \left(1-\widehat{h}^{(2)}_k\right) \prod_{s = 1}^{k-1} \left(1-\widehat{h}^{(1)}_s\right)\left(1-\widehat{h}^{(2)}_s\right)
\end{eqnarray*}
with discrete-time event-specific hazards
\begin{eqnarray*}
  h^{(1)}_{k} &\equiv& \Pr\left(Y_k = 1 \middle\vert D_k = Y_{k-1} = 0\right)\\
  h^{(2)}_{k} &\equiv& \Pr\left(D_k = 1 \middle\vert Y_{k-1} = D_{k-1} = 0\right),
\end{eqnarray*}
and $\widehat{h}^{(1)}_k$ and $\widehat{h}^{(2)}_k$ their corresponding sample proportions.
Note that the set of patients at risk of hospital death or discharge at the start of interval $k$ are those for whom $T>t_{k-1}$ or, similarly, $\{D_{k-1} = Y_{k-1} = 0\}$. Given the assumed temporal ordering of event types within each interval $k$ (see previous section), this risk set is only clearly reflected in the
conditioning set of $h^{(2)}_{k}$.

Similarly, in the absence of censoring, $\tilde\phi_K$ is identified by the following functional from the observed data
\begin{eqnarray*}
  \sum_{k = 1}^K E\left[Y_k\left(1-D_k\right)\left(1-A_k\right)\left(1-Y_{k-1}\right)\right],
\end{eqnarray*}
which can be estimated by the empirical cumulative distribution of HAI-free hospital death
\begin{eqnarray*}
\widehat{\tilde\phi}_K \equiv \hspace{0.1cm} n^{-1} \sum_{k = 1}^K \tilde d_{1k},
\end{eqnarray*}
with $\tilde d_{1k} = \sum_{i=1}^n Y_{ik} (1-D_{ik}) (1-A_{ik}) (1-Y_{i,k-1})$ denoting the number of HAI-free hospital deaths in interval $k$.
Again, $\widehat{\tilde\phi}_K$ is algebraically equivalent to the following Aalen-Johansen estimator (see Appendix~\ref{sec:app-estobsdata2} in the Supplementary Material for a formal proof)
\begin{eqnarray*}
  \widehat{\tilde\phi}_K^{\text{\tiny AJ}} \equiv \sum_{k = 1}^K \widehat{\tilde h}^{(1)}_k \left(1-\widehat{\tilde h}^{(2)}_k\right) \left(1-\widehat{\tilde h}^{(0)}_k\right) \prod_{s = 1}^{k-1} \left(1-\widehat{\tilde h}^{(0)}_s\right)\left(1-\widehat{\tilde h}^{(1)}_s\right)\left(1-\widehat{\tilde h}^{(2)}_s\right),
\end{eqnarray*}
with discrete-time event-specific hazards
\begin{eqnarray*}
  \tilde h^{(0)}_k &\equiv& \Pr\left(A_k = 1 \middle\vert Y_{k-1} = D_{k-1} = A_{k-1} = 0\right)\nonumber\\
  \tilde h^{(1)}_k &\equiv& \Pr\left(Y_k = 1 \middle\vert D_k = A_k = Y_{k-1} = 0\right)\nonumber\\
  \tilde h^{(2)}_k &\equiv& \Pr\left(D_k = 1 \middle\vert A_k = Y_{k-1} = D_{k-1} = 0\right).
\end{eqnarray*}
and $\widehat{\tilde h}^{(0)}_k$, $\widehat{\tilde h}^{(1)}_k$ and $\widehat{\tilde h}^{(2)}_k$ their corresponding sample proportions.
The set of patients at risk of HAI-free hospital death or discharge at the start of interval $k$ are those for whom $\tilde T>t_{k-1}$ or, similarly, $\{A_{k-1} = D_{k-1} = Y_{k-1} = 0\}$. Given the assumed temporal ordering of event types within each interval $k$, this risk set is only clearly reflected in the conditioning set of $\tilde h^{(0)}_{k}$.

\subsection{Counterfactual risks}
Estimation of the counterfactual risk $\varphi_K$ is more challenging due to its hypothetical nature.
Ideally, we would wish to observe every patient's event time and event status if HAIs could somehow have been eradicated.
Suppose that, counter to the fact, we could. The counterfactual risk $\varphi_K$ could then be estimated from this counterfactual data by the following
Aalen-Johansen estimator
\begin{eqnarray*}
  \sum_{k = 1}^K \widehat{h}^{(1),\overline a = 0}_k \left(1-\widehat{h}^{(2),\overline a = 0}_k\right) \prod_{s = 1}^{k-1} \left(1-\widehat{h}^{(1),\overline a = 0}_s\right)\left(1-\widehat{h}^{(2),\overline a = 0}_s\right)
\end{eqnarray*}
with discrete-time counterfactual event-specific hazards
\begin{eqnarray*}
  h^{(1), \overline a = 0}_k &\equiv& \Pr\left(Y^{\overline a = 0}_k = 1 \middle\vert D^{\overline a = 0}_k = Y^{\overline a = 0}_{k-1} = 0\right)\\
  h^{(2), \overline a = 0}_k &\equiv& \Pr\left(D^{\overline a = 0}_k = 1 \middle\vert Y^{\overline a = 0}_{k-1} = D^{\overline a = 0}_{k-1} = 0\right),
\end{eqnarray*}
or by the complement of the following Kaplan-Meier like estimator
\begin{eqnarray*}
  \prod_{k = 1}^{K} \left(1-\widehat{h}^{\text{sd},\overline a = 0}_k\right)
\end{eqnarray*}
with discrete-time counterfactual subdistribution hazards
\begin{eqnarray*}
  h^{\text{sd},\overline a = 0}_{k} &\equiv& \Pr\left(Y^{\overline a = 0}_k = 1 \middle\vert Y^{\overline a = 0}_{k-1} = 0\right).
\end{eqnarray*}
Unfortunately, we cannot observe this counterfactual data. However, under the consistency assumption that, among patients who did not acquire infection during their hospitalization, $T^{\overline a = 0} = T$ and $\epsilon^{\overline a = 0} = \epsilon$ (or, similarly, $D_k = \tilde D_k = D^{\overline a = 0}_k$ and $Y_k = \tilde Y_k = Y^{\overline a = 0}_k$ for each $k \le K$), and the monotonicity assumption that no exposed patient benefits from being exposed (i.e. no exposed patient would die sooner during hospitalization had s/he not been exposed), a natural lower bound for $\varphi_K$ is the crude risk of HAI-free hospital mortality $\tilde\phi_K$.
Equating these two risks is only justified if every patient who dies \emph{with} infection also dies \emph{from} infection.

We next illustrate that proposed approaches for estimating $\varphi_K$ each aim to adjust for the fact that not all patients who die \emph{with} infection also die \emph{from} infection by upweighing the increments of $\tilde\phi_K$ (or its estimators) but differ in the way their corresponding weights account for subtle but important aspects related to the time-dependent nature of exposure, its competing events and their confounding factors. In the next section, we first review the conditions required for identifying $\varphi_K$ from the observed data and illustrate that an inverse probability weighted representation of the identification result provides the basis for understanding different sources of potential bias, inherent to estimation approaches that will be discussed in Section~\ref{sec:main-timedepexp}.

\section{Identification of the counterfactual risk}\label{sec:main-ident}

Following the work of Robins and colleagues \citep{Robins1986,Robins2008a,Young2020}, the counterfactual risk $\varphi_K$ is identified by Robins' non-parametric \emph{g-formula} \citep{Robins1986} under the following three causal assumptions, for each $k = 0,...,K$ (for a formal proof, see Appendix~\ref{sec:app-ident1} in the Supplementary Material):
\begin{enumerate}
  \item Sequential exchangeability
  \begin{eqnarray}\label{main-seqexch}
  \underline{Y}_k^{\overline{a} = 0} \cip  A_k \vert \overline{L}_{k-1}, \overline{D}_{k-1}, Y_{k-1} = A_{k-1} = 0
  \end{eqnarray}
  This assumption states that conditional on the measured past, at each interval $k = 0, ..., K$, future counterfactual hospital death is independent of the current infection status among previously uninfected patients (that are either still hospitalized or discharged alive).
  \item Positivity
  \begin{eqnarray}\label{main-positivity}
  &\Pr\left(A_k = 0 \middle\vert \overline{L}_{k-1}, A_{k-1} = D_{k-1} = Y_{k-1} = 0\right) > 0 \text{ w.p.1}
  \end{eqnarray}
  This assumption states that, at each interval $k = 0, ..., K$, each patient in the risk set for HAI-free hospital death ($A_{k-1} = D_{k-1} = Y_{k-1} = 0$), has a strictly positive probability of remaining uninfected until the end of interval $k$, given his or her measured past.
  \item Consistency
  \begin{eqnarray}\label{main-consistency}
  \text{If } A_k = 0 \text{ then } \overline{L}_k = \overline{L}_k^{\overline{a} = 0} \text{, } \overline{D}_k = \overline{D}_k^{\overline{a} = 0} \text{ and } \overline{Y}_k = \overline{Y}_k^{\overline{a} = 0}.
  \end{eqnarray}
  This assumption provides the necessary link between the observed and counterfactual outcomes. For this assumption to be met, uninfected patients’ observed outcomes need to be the same as their (counterfactual) outcomes that would have been observed had they received a preventive intervention that eradicates infection. This assumption may be problematic when such intervention also affects outcomes through factors other than infection.
\end{enumerate}

In Appendices~\ref{sec:app-ident2} and~\ref{sec:app-ident3} in the Supplementary Material, respectively, we demonstrate that, following \cite{Young2020}, the discrete-time counterfactual subdistribution hazards $h^{\text{sd},\overline a = 0}_{k}$ (defined in Section~\ref{sec:main-estimands}) are identified from the observed data under the same set of causal assumptions, and the discrete-time counterfactual event-specific hazards $h^{(1), \overline a = 0}_k$ and $h^{(2), \overline a = 0}_k$ (defined in Section~\ref{sec:main-estimands}) are identified from the observed data under an additional sequential exchangeability assumption. In Appendix~\ref{sec:app-swigs} in the Supplementary Material, we give a general template of a causal diagram under which these sequential exchangeability assumptions hold.

The g-formula, which is a non-parametric functional of the observed data distribution (given by expression~\eqref{gformula} in Appendix~\ref{sec:app-ident1} in the Supplementary Material), can be re-expressed as a weighted cumulative distribution function of HAI-free hospital death
\begin{eqnarray}\label{IPCWcdf}
  \sum_{k=1}^K E\left[Y_k(1-D_k)(1-A_k)(1-Y_{k-1})W_k\right],
\end{eqnarray}
with weights
\begin{eqnarray}\label{IPCweights}
  W_{k} &\equiv& \prod_{s=1}^k \dfrac{1}{\Pr\left(A_s=0 \middle\vert \overline L_{s-1}, A_{s-1} = D_{s-1} = Y_{s-1} = 0\right)},
\end{eqnarray}
and $\overline{L}_{k}$ denoting the history up to $t_k$ of baseline and time-dependent prognostic factors of both future HAI onset and hospital death or discharge, assumed to be sufficient for confounding adjustment.

It is important to note here that each HAI-free death in interval $k$ is weighed by a factor that only takes into account the history of events and covariate levels up to interval $k$, and does so in a way that respects the temporal ordering of those events up to interval $k$. The weights thereby avoid potential forms of time-dependent bias. Moreover, under the sequential exchangeability assumption, the weights $W_k$ also adequately account for all (time-varying) confounding factors and thereby eliminate confounding bias. If $\overline{L}_{k}$ comprises only a subset of relevant confounders, such that sequential exchangeability only holds approximately, the weights may, at best, reduce confounding bias.

In the next section, we demonstrate the different sources of bias that plague naive, but also established estimation approaches based on multistate modelling when the goal (either explicit or implicit) is estimation of the causal target parameter $\varphi_K$. We illustrate that different types of bias may arise by conflating this counterfactual estimand with a purely \emph{statistical} estimand that may not carry a clear causal interpretation. Weighting based representations of these statistical estimands (or their corresponding estimators) permit a comparison with the `reference' weights from this section and may give more insight into the incremental removal of bias related to (i) inappropriate accounting for the timing of exposure, (ii) inappropriate accounting for the temporal order of exposure and its competing events, and (iii) inappropriate adjustment for confounding of the time-varying relation between exposure and its competing events.
We use a toy example to provide intuition into how weight in the analytic sample is being transferred from infected to uninfected patients.

\section{Accounting for time-dependent exposures: different (weighted) means to an (hypothetical) end}\label{sec:main-timedepexp}

\begin{figure}[t!]
\centerline{\includegraphics[width=300pt]{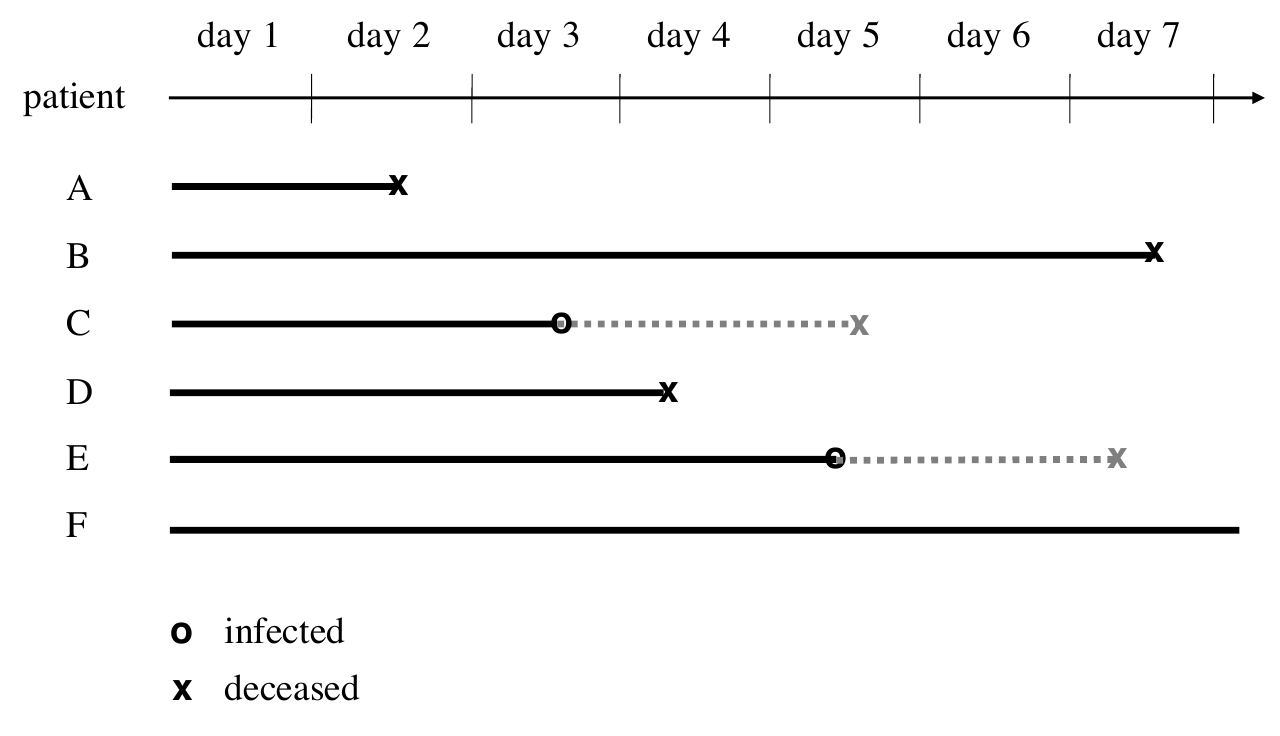}}
\caption{Toy example trajectories for 6 patients with follow-up of 7 days.\label{fig:toyexample}}
\end{figure}

Consider the six patients depicted in the toy example in Figure~\ref{fig:toyexample}, who are followed up for a week, from day 1 (the day of hospital admission) to day 7. Patients A, B and D die while uninfected, whereas patients C and E acquire infection during their hospitalization. For the former three patients, the factual (observed) scenario fully corresponds to the counterfactual scenario (of remaining without infection until death or discharge). This also holds for patient F who gets discharged alive after day 7. The factual scenario of the latter two patients corresponds to their counterfactual scenario only until infection onset. In other words, all patients carry relevant information for estimating $\varphi_K$ because each patient is uninfected for a certain amount of time.

\subsection{Treating exposure as an exclusion criterion}
A naive approach for estimating $\varphi_K$ is to completely ignore that also infected patients carry relevant information, and hence to ignore HAI onset. This is implicitly done when only considering patients who never got infected during hospitalization.\footnote{Although, to our knowledge, the specific approach discussed in this section (which considers a time-to-event outcome subject to a competing event) has never been advocated as such, it is similar in spirit to application of traditional (e.g. regression- or matching-based) estimation methods for the PAF in cross-sectional or time-fixed settings in a setting with a time-dependent exposure, in that it compares outcomes between ever and never exposed patients. For a recent example, see \cite{Suzuki2019} or references in \cite{Cube2020}.}
This approach implicitly conflates the counterfactual risk $\varphi_K$ with $$\Pr\left(\tilde T \le t_K, \tilde \epsilon = 1 \middle\vert \tilde \epsilon \ne 0\right) = \Pr\left(\tilde Y_K = 1 \middle\vert A_{\tau} = 0\right) = \tilde\phi_K/(1-\alpha_{\tau}),$$ the factual risk of HAI-free hospital death by interval $K$ among patients that remain uninfected until the maximal end of follow-up time. In the absence of censoring due to loss to follow-up or administrative end of the study, this corresponds to a naive way of rescaling the cumulative distribution of HAI-free hospital death $\tilde \phi_K$ by a constant (time-fixed) factor (see Appendix~\ref{sec:app-wcharest1} in the Supplementary Material)
\begin{eqnarray*}
  \sum_{k = 1}^K E\left[Y_k(1-D_k)(1-A_k)(1-Y_{k-1})W^{\star}\right],
\end{eqnarray*}
with weights
\begin{eqnarray*}
  W^{\star} &\equiv& \dfrac{1}{1-\Pr(\tilde \epsilon = 0)} = \dfrac{1}{\Pr\left(A_{\tau} = 0\right)} = 1 + \dfrac{\Pr\left(A_{\tau} = 1\right)}{\Pr\left(A_{\tau} = 0\right)}.
\end{eqnarray*}
The corresponding estimator is given by
\begin{eqnarray*}
  \widehat\varphi_K^{\star} \equiv \hspace{0.1cm} n^{-1}\sum_{k = 1}^K \tilde d^{w}_{1k},
\end{eqnarray*}
where $\tilde d^{w}_{1k} \equiv \sum_{i=1}^n Y_{ik}\left(1-D_{ik}\right)\left(1-A_{ik}\right)\left(1-Y_{i,k-1}\right) \widehat{W}^{\star}$ denotes the weighted number of HAI-free hospital deaths in interval $k$, $\widehat{W}^{\star} = P_n^{-1}\left(A_{\tau} = 0\right)$ and $P_n(X=x) \equiv n^{-1} \sum_{i=1}^n I(X_i = x)$ denotes the empirical probability. An expression of the corresponding (and algebraically equivalent) Aalen-Johansen estimator $\widehat\varphi_K^{\star \text{\tiny AJ}}$ is given in Appendix~\ref{sec:app-wcharest1} in the Supplementary Material.

Only considering uninfected patients corresponds to weighing each uninfected patient's contribution by a fixed amount $\widehat{W}^{\star}$ equal to the inverse probability of remaining uninfected until the end of hospitalization. In other words, the weight transferred from infected patients to each uninfected patient equals the odds of acquiring infection at any point during hospitalization and hence does not depend on the time of HAI onset. This weight is transferred uniformly at the start of follow-up. Let's return to our toy example. Two out of 6 patients (patients C and E) acquire an HAI during their hospitalization. Hence, the marginal odds of HAI is 2/4, which is the weight transferred at day 0 from patients who eventually acquire HAI (possibly after day 0) to patients who remain uninfected during hospitalization (Table~\ref{tab:toyexample2}).

This approach systematically fails to account for HAI onset time by treating patients' future infection status (at hospital death or discharge) as if it were already known at hospital admission and using it as an exclusion criterion, thereby violating the key principle of not conditioning on the future.
Failing to appropriately apportion infection-free time to infected patients (by erroneously coding patient time as infected time) is equivalent to censoring infected patients at hospital admission instead of at their HAI onset. This artificially and inadvertently makes infected patients immortal until infection onset and produces so-called \emph{immortal time bias} \citep{Suissa2003,VanWalraven2004}. This bias is also reflected in the corresponding weight transferal scheme, where weights incorporate information that is not yet available at the time they are transferred and therefore fail to adequately account for the relative timing of events.
Because the apportioned weights compensate for the depletion of all future infected patients from the risk set, they tend to overcompensate.
Accordingly, early HAI-free deaths get disproportionally large weights, resulting in systematic overestimation of $\varphi_K$, especially early on during follow-up.
In certain cases, estimates of this counterfactual risk may even exceed the factual risk $\phi_K$, translating into negative PAF estimates that may possibly lead to erroneous interpretations in terms of a seemingly protective effect of early onset HAIs. Certain longevity studies have been criticized on similar grounds because for the same reason, in contrast to reported findings, becoming a jazz musician or winning an Oscar doesn't necessarily make you live longer \citep{Rothman1992,Sylvestre2006}.
Considering HAI-free time of infected patients is therefore imperative to ensure that weights adequately quantify the likelihood of each deceased uninfected patient to have remained without infection until her \emph{individual} time of death (rather than at the maximal follow-up time). In the next section, however, we illustrate that some approaches that account for HAI onset time don't succeed in removing all forms of time-dependent bias.

\begin{table}[t]%
\small
\centering
\begin{tabular}{@{\extracolsep\fill}lccccccc@{\extracolsep\fill}}
\toprule
\textbf{patient} & \textbf{day 1} & \textbf{day 2}  & \textbf{day 3}  & \textbf{day 4}  & \textbf{day 5} & \textbf{day 6} & \textbf{day 7} \\
\midrule
A & 1.5 & \textbf{1.5} & \textbf{1.5} & \textbf{1.5} & \textbf{1.5} & \textbf{1.5} & \textbf{1.5}\\
B & 1.5 & 1.5 & 1.5 & 1.5 & 1.5 & 1.5 & \textbf{1.5}\\
C & 0 & 0 & \textcolor{gray}{0} & \textcolor{gray}{0} & \textcolor{gray}{0} & \textcolor{gray}{0} & \textcolor{gray}{0}\\
D & 1.5 & 1.5 & 1.5 & \textbf{1.5} & \textbf{1.5} & \textbf{1.5} & \textbf{1.5}\\
E & 0 & 0 & 0 & 0 & \textcolor{gray}{0} & \textcolor{gray}{0} & \textcolor{gray}{0}\\
F & 1.5 & 1.5 & 1.5 & 1.5 & 1.5 & 1.5 & 1.5\\ 
\midrule
$\widehat\varphi^{\star}_K$ & 0 & 1.5/6 & 1.5/6 & 3/6 & 3/6 & 3/6 & 4.5/6\\
\bottomrule
\end{tabular}
\caption{Overview of accumulated weight over time for each patient in the toy example when exposure onset is treated as an exclusion criterion. Bold and gray typeface indicate a patient has died without infection or has acquired infection the day before, respectively. \label{tab:toyexample2}}%
\end{table}

\subsection{Treating exposure onset as a time-dependent exclusion criterion}
One of the dominant estimation approaches based on multistate modeling, proposed by \cite{Schumacher2007}, takes HAI-free time of infected patients into account in estimation of the counterfactual risk $\varphi_K$.
This approach, however, implicitly conflates $\varphi_K$ with $$\Pr\left[\tilde T \le t_K, \tilde \epsilon = 1 \middle\vert \tilde T > t_K \cup (\tilde T \le t_K, \tilde \epsilon \ne 0)\right] = \Pr\left(\tilde Y_K = 1 \middle\vert A_K = 0\right) = \tilde\phi_K/(1-\alpha_K),$$ the factual risk of HAI-free hospital death by interval $K$ among (previously) uninfected patients. It thereby excludes all patients who have acquired an infection by landmark interval of interest $K$ from the analytic sample for estimating $\varphi_K$. The estimation procedure of this approach involves combining cumulative incidence estimates from a progressive illness death model (Figure~\ref{fig:extillnessdeath} in Appendix~\ref{sec:app-wcharest2} in the Supplementary Material), where infection is considered an intermediate (rather than an absorbing) state. However, in the absence of censoring due to loss to follow-up or administrative end of the study, their targeted statistical estimand is, again, simply a weighted version of $\tilde \phi_K$, the cumulative distribution of HAI-free hospital death at each landmark interval $K$,
\begin{eqnarray*}
  \sum_{k = 1}^K E\left[Y_k(1-D_k)(1-A_k)(1-Y_{k-1})W^{\dagger}_K\right],
\end{eqnarray*}
with increments scaled by the inverse probability of having remained without infection until (and including) interval $K$ (see Appendix~\ref{sec:app-wcharest2} in the Supplementary Material)
\begin{eqnarray*}
  W^{\dagger}_K &\equiv& \dfrac{1}{1-\Pr\left(\tilde T \le t_K, \tilde \epsilon = 0\right)} = \dfrac{1}{\Pr\left(A_K = 0\right)} = 1 + \dfrac{\Pr\left(A_K = 1\right)}{\Pr\left(A_K = 0\right)}.
\end{eqnarray*}
Their estimator can then be expressed as follows
\begin{eqnarray*}
  \widehat\varphi_K^{\dagger} \equiv \hspace{0.1cm} n^{-1} \sum_{k=1}^K \tilde d^{w_K}_{1k},
\end{eqnarray*}
with $\tilde d^{w_K}_{1k} \equiv \sum_{i=1}^n Y_{ik}\left(1-D_{ik}\right)\left(1-A_{ik}\right)\left(1-Y_{i,k-1}\right) \widehat{W}^{\dagger}_{K}$ the weighted number of HAI-free hospital deaths in interval $k$ and $\widehat{W}^{\dagger}_K = P_n^{-1}\left(A_K = 0\right)$.
An expression of the corresponding (and algebraically equivalent) Aalen-Johansen estimator $\widehat\varphi_K^{\dagger \text{\tiny AJ}}$ is given in Appendix~\ref{sec:app-wcharest2} in the Supplementary Material.

The implicit weighting scheme of this approach accommodates the progressive selection of uninfected patients over time by rescaling these patients' contributions
to compensate for the depletion of infected patients from the risk set of infection-free hospital death.
It does so by assigning time-dependent (instead of time-fixed) weights that are smaller for earlier landmarks and thereby eliminates the immortal time bias that plagues the naive estimator $\widehat\varphi_K^{\star}$ discussed in the previous section.
More specifically, as soon as patients acquire infection, they are no longer considered to be at risk of HAI-free hospital death and transfer their weight to patients who have until that time remained uninfected. The total amount of weight transferred to each uninfected patient by $t_K$ corresponds to the marginal odds of HAI by $t_K$.
However, implicit to this approach is that infected patients also distribute their weight to uninfected patients who have already died or have been discharged, i.e. patients who are no longer at risk of HAI. This can be seen upon noting that HAI-free hospital deaths in interval $k$ are weighed by a function  $\widehat{W}^{\dagger}_{K}$ that does not depend on $k$ but on $K$, the succeeding landmark at which the empirical cumulative distribution function is evaluated. Because of this inappropriate selection of time-dependent weights, this approach fails to entirely eliminate all forms of time-dependent bias.

\begin{table}[t!]%
\small
\centering
\begin{tabular}{@{\extracolsep\fill}lccccccc@{\extracolsep\fill}}
\toprule
\textbf{patient} & \textbf{day 1} & \textbf{day 2}  & \textbf{day 3}  & \textbf{day 4}  & \textbf{day 5} & \textbf{day 6} & \textbf{day 7} \\
\midrule
A & 1 & \textbf{1} & \textbf{1.2} & \textbf{1.2} & \textbf{1.5} & \textbf{1.5} & \textbf{1.5}\\
B & 1 & 1 & 1.2 & 1.2 & 1.5 & 1.5 & \textbf{1.5}\\
C & 1 & 1 & \textcolor{gray}{0} & \textcolor{gray}{0} & \textcolor{gray}{0} & \textcolor{gray}{0} & \textcolor{gray}{0}\\
D & 1 & 1 & 1.2 & \textbf{1.2} & \textbf{1.5} & \textbf{1.5} & \textbf{1.5}\\
E & 1 & 1 & 1.2 & 1.2 & \textcolor{gray}{0} & \textcolor{gray}{0} & \textcolor{gray}{0}\\
F & 1 & 1 & 1.2 & 1.2 & 1.5 & 1.5 & 1.5\\ 
\midrule
marginal odds & 0/6 & 0/6 & 1/5 & 1/5 & 2/4 & 2/4 & 2/4\\
\midrule
$\widehat\varphi_K^{\dagger}$ & 0 & 1/6 & 1.2/6 & 2.4/6 & 3/6 & 3/6 & 4.5/6\\
\bottomrule
\end{tabular}
\caption{Overview of accumulated weight over time for each patient in the toy example when exposure onset is treated as a time-dependent exclusion criterion. Bold and gray typeface indicate a patient has died without infection or has acquired infection the day before, respectively. \label{tab:toyexample3}}%
\end{table}

Let's reconsider our toy example. According to this approach, weights get transferred at each HAI onset time (Table~\ref{tab:toyexample3}).
For instance, when patient C acquires infection on day 3, she distributes her weight among all patients who have so far remained without HAI, irrespective of whether they have already died or been discharged. That is, each of these patients increases his/her weight with the odds of having acquired infection by that day (=1/5). In contrast to the naive approach discussed earlier, weight is also transferred to patient E, who is uninfected on day 3 but develops infection two days later. On day 4, patient E similarly distributes her weight to all patients who have remained without infection (each receives a weight of 1/4), along with the weight she has (in the meantime) received from patient C (each indirectly receives a weight of $(1/4)\times(1/5)$ from patient C via patient E).
As a result, each patient that has remained without infection by day 4, accumulates weight received from both patient C and patient E, corresponding to a total transferred weight of $1/5+1/4+(1/4)\times(1/5) = 1/2$, which equals the odds of having acquired infection by day 4.
Importantly, patient A now receives a smaller weight when she dies than patients B and D at their respective death times.
This is because patients B and D have accumulated more weight (from patient C, and from patients C and E, respectively) by their time of death than patient A by the time she died. However, after their death, patient A and patient D receive additional weight, such that it eventually matches the weight accumulated by patient B.

Even though patients who die early have accumulated less weight by their time of death than patients who die at later time points,
they eventually accumulate the same weight by a fixed landmark interval of interest $K$.
As with the naive approach, their accumulated weight by $t_K$ is identical to that of patients who die later in time, even though it should be smaller (and fixed from the time of their death onward) so as to reflect their shorter time at risk for HAI and, accordingly, the smaller degree of depletion of infected patients by their time of death that needs to be compensated for.
Due to this weighting scheme, it seems as if patients who died (or got discharged) without infection are implicitly (re)considered to still be at risk of infection-free hospital death
because, even after their time of death (or discharge), they continue to accumulate weight via newly infected patients. This finding is not readily apparent when adopting multistate model notation, but can intuitively be appreciated by resorting to a weighting-based characterization of the corresponding multistate model based estimator $\widehat\varphi_K^{\dagger}$.

The notion of an extended risk set that includes patients who are strictly no longer at risk for the event of interest is also characteristic for the subdistribution hazard \citep{Gray1988,Fine1999b,Putter2020}. However, the extended risk set for the subdistribution hazard of HAI-free hospital death includes patients that have experienced either of the \emph{competing} events (HAI onset or HAI-free hospital discharge). In contrast, the weighting scheme of $\widehat\varphi_K^{\dagger}$ implicitly reconsiders patients that have died without HAI to still be at risk for the \emph{same} event.
An event that has already occurred hence gets reweighed based on information (on future infections) that is not available at the time of that event, again violating the key principle of not conditioning on the future.
Estimation of the subdistribution hazard at a certain interval $k$, on the other hand, does not violate this principle as it only considers information up to $k$ (see Appendices~\ref{sec:app-estobsdata2} and~\ref{sec:app-wcharest2} in the Supplementary Material for more details).
This failure to fully respect the temporal order of events leads to a residual form of time-dependent bias that is more subtle than immortal time bias (especially when the incidence of infection is low). In the next section, we illustrate that treating HAI onset as a censoring event can eliminate this residual form of bias.

\subsection{Treating exposure onset as a censoring event}\label{sec:main-timedepexp-cens}

The consistency assumption~\eqref{main-consistency} from Section~\ref{sec:main-ident} states that, when a patient is uninfected up to interval $k$, her counterfactual indicator of hospital death $Y^{\overline a = 0}_k$ equals her observed or factual indicator of hospital death $Y_k$. In other words, under this assumption, we observe the counterfactual time to hospital death in patients who die while uninfected, but we don't observe this counterfactual event time in patients who get infected.
Framing the fundamental problem of causal inference as a missing data problem \citep{Holland1986} then naturally leads to conceiving HAI onset as a censoring event for the counterfactual event of interest.

Artificially censoring patients at their HAI onset time implicitly aims to \emph{emulate} this hypothetical world (in which these patients would have further remained without infection beyond their observed HAI onset time) \citep{Arjas1993,Pepe1993,Pepe1993a,Keiding1999,Keiding2001} and results in the following Aalen-Johansen estimator
\begin{eqnarray*}
\widehat\varphi_K^{\circ \text{\tiny AJ}} \equiv \sum_{k = 1}^K \widehat{\tilde h}^{(1)}_k \left(1-\widehat{\tilde h}^{(2)}_k\right) \prod_{s = 1}^{k-1} \left(1-\widehat{\tilde h}^{(1)}_s\right) \left(1-\widehat{\tilde h}^{(2)}_s\right),
\end{eqnarray*}
with discrete-time event-specific hazards $\tilde h^{(1)}_k$ and $\tilde h^{(2)}_k$ as defined in Section~\ref{sec:main-estimands}. This estimator is equivalent to $\widehat\phi_K^{\text{\tiny AJ}}$ upon excluding patients that have gotten infected by interval $k$ from the risk set of interval $k$ and to $\widehat{\tilde\phi}_K^{\text{\tiny AJ}}$ upon setting the event-specific hazards of infection $\tilde h^{(0)}_k$ to zero for each $k = 1,...,K$.

Artificial censoring ensures that as time progresses and as patients get infected, they transfer their weight in the analysis to hospitalized uninfected patients (Table~\ref{tab:toyexample1}). For instance, on day 1 and day 2, corresponding to intervals $(0,1]$ and $(1,2]$, all patients belong to the risk set. From day 3, patient A, who died on day 2, and patient C, who gets infected on day 3, are excluded from the risk set. Because patient A died without having acquired infection, she does not transfer weight to the five uninfected patients in the risk set on day 3. In contrast, patient C got infected on day 3 and distributes his weight over the remaining four patients in the risk set. By the time patient D dies, on day 4, he has an accumulated weight of 1+1/4 = 1.25.
On day 5, patient E acquires infection and, accordingly gets censored and is excluded from the risk set that day. She also distributes her accumulated weight (which includes the weight she received from patient C) over the remaining two patients in the risk set ($= 1/2 \times (1 + 1/4)$). By now, these remaining patients have accumulated a weight of $1+1/4+1/2\times(1+1/4) = 1.875$.

\begin{table}[t!]%
\small
\centering
\begin{tabular}{@{\extracolsep\fill}lccccccc@{\extracolsep\fill}}
\toprule
\textbf{patient} & \textbf{day 1} & \textbf{day 2}  & \textbf{day 3}  & \textbf{day 4}  & \textbf{day 5} & \textbf{day 6} & \textbf{day 7} \\
\midrule
A & 1 & \textbf{1} & \textbf{1} & \textbf{1} & \textbf{1} & \textbf{1} & \textbf{1}\\
B & 1 & 1 & 1.25 & 1.25 & 1.875 & 1.875 & \textbf{1.875}\\
C & 1 & 1 & \textcolor{gray}{0} & \textcolor{gray}{0} & \textcolor{gray}{0} & \textcolor{gray}{0} & \textcolor{gray}{0}\\
D & 1 & 1 & 1.25 & \textbf{1.25} & \textbf{1.25} & \textbf{1.25} & \textbf{1.25}\\
E & 1 & 1 & 1.25 & 1.25 & \textcolor{gray}{0} & \textcolor{gray}{0} & \textcolor{gray}{0}\\
F & 1 & 1 & 1.25 & 1.25 & 1.875 & 1.875 & 1.875\\ 
\midrule
conditional odds & 0 & 0 & 1/4 & 0 & 1/2 & 0 & 0\\
\midrule
$\widehat{\varphi}^{\circ}_K$ & 0 & 1/6 & 1/6 & 2.25/6 & 2.25/6 & 2.25/6 & 4.125/6\\
\bottomrule
\end{tabular}
\caption{Overview of accumulated weight over time for each patient in the toy example when exposure onset is treated as a censoring event. Bold and gray typeface indicate a patient has died without infection or has acquired infection the day before, respectively. \label{tab:toyexample1}}%
\end{table}

This weight transferal scheme may not be readily apparent from the usual formulation of $\widehat\varphi_K^{\circ \text{\tiny AJ}}$, but follows from an alternative representation of this Aalen-Johansen estimator, in terms of an inverse probability of censoring (IPC) weighted average \citep[see][and Appendix~\ref{sec:app-wcharest3} in the Supplementary Material]{Antolini2006} \citep[for a similar IPC weighted representation of the Kaplan-Meier estimator see][]{Robins2000b,Satten2001}, which, in the absence of remaining forms of censoring, is algebraically equivalent to the following estimator
\begin{eqnarray*}
  \widehat\varphi_K^{\circ} \equiv \hspace{0.1cm} n^{-1}\sum_{k = 1}^K \tilde d^{w_k}_{1k},
\end{eqnarray*}
where $\tilde d^{w_k}_{1k} \equiv \sum_{i=1}^n Y_{ik}\left(1-D_{ik}\right)\left(1-A_{ik}\right)\left(1-Y_{i,k-1}\right) \widehat{W}^{\circ}_k$ denotes the IPC weighted number of HAI-free hospital deaths in interval $k$, and the weights
\begin{eqnarray}
  W^{\circ}_k &\equiv& \prod_{s = 1}^k \dfrac{1}{1-\Pr\left(\tilde T \in (t_{s-1}, t_s], \tilde \epsilon = 0 \middle\vert \tilde T > t_{s-1}\right)} = \prod_{s = 1}^k \dfrac{1}{\Pr\left(A_s=0 \middle\vert Y_{s-1} = D_{s-1} = A_{s-1} = 0\right)}\label{AJk0_keidingW}\\
&=& 1+\sum_{s = 1}^k \dfrac{\Pr\left(A_s =1 \middle\vert Y_{s-1} = D_{s-1} = A_{s-1} = 0\right)}{\Pr\left(A_{s} =0 \middle\vert Y_{s-1} = D_{s-1} = A_{s-1} = 0\right)}
\prod_{s' = 1}^{s-1} \left\{ 1+ \dfrac{\Pr\left(A_{s'} =1 \middle\vert Y_{s'-1} = D_{s'-1} = A_{s'-1} = 0\right)}{\Pr\left(A_{s'} =0 \middle\vert Y_{s'-1} = D_{s'-1} = A_{s'-1} = 0\right)} \right\}\nonumber\\ \label{AJk0_keidingW2}
\end{eqnarray}
are non-parametrically estimated using sample proportions.

Note that $\widehat\varphi_K^{\circ}$ is again a weighted version of the empirical cumulative distribution of HAI-free hospital death $\widehat{\tilde\phi}_K$ where the weight $\widehat W^{\circ}_k$ apportioned to each HAI-free death in interval $k$ equals the inverse probability of having remained uninfected while hospitalized by interval $k$.
The product in the denominator of~\eqref{AJk0_keidingW} corresponds to the Kaplan-Meier estimator of remaining without infection up to interval $k$, treating both hospital death and discharge as (independent) censoring events.
The alternative formulation of $W^{\circ}_k$ in~\eqref{AJk0_keidingW2} illustrates that these apportioned weights can be decomposed in a way that formalizes the above transferal scheme (see Appendix~\ref{sec:app-decompw} in the Supplementary Material for intermediate steps). That is, over the course of time, each patient accumulates weight that consists of her own unit-weight and weight transferred by patients who got censored during previous time waves, both directly (the first factor in the summation) and indirectly through intermediate transferals (the second factor in the summation).
This weight transferal scheme fully respects the timing of events of interest by only prospectively transferring weight from hospitalized infected patients to hospitalized uninfected patients and no sooner than time of infection onset. The weight being directly transferred from one patient to another at $t_{k-1}$ can be expressed as the odds of infection in interval $k$ in the risk set at $t_{k-1}$.

A more formal justification for this estimation approach follows from the fact that assuming (artificial) censoring is \emph{independent} of the counterfactual event times\footnote{Under the aforementioned assumed temporal order of event types within an interval $k$, this implies that throughout, we assume artificial censoring times (infection onset times) to occur right before possibly tied event times (infection-free hospital death or discharge times) within interval $k$.} is equivalent to assuming that the counterfactual event-specific hazards (defined in Section~\ref{sec:main-estimands}) are identical to the observable or factual event-specific hazards among patients who have remained uninfected up until (and including) interval $k$
\begin{eqnarray}\label{eqhaz}
  h^{(1), \overline a = 0}_k &=& \tilde h^{(1)}_k\nonumber\\
  h^{(2), \overline a = 0}_k &=& \tilde h^{(2)}_k.
\end{eqnarray}
Under this assumption, $\widehat\varphi_K^{\circ}$ and $\widehat\varphi_K^{\circ \text{\tiny AJ}}$ are consistent estimators of the counterfactual risk $\varphi_K$. Moreover, in Appendix~\ref{sec:app-ident3} in the Supplementary Material, we demonstrate that~\eqref{eqhaz} is implied by conditions~\eqref{main-seqexch}-\eqref{main-consistency} with $\overline{L}_{k-1} = \emptyset$ for $k = 1,...,K$, and an additional exchangeability assumption. The link between independent censoring and marginal sequential exchangeability or the `no unmeasured confounding' assumption also becomes apparent upon noting that the weights $W_k$ (defined in Section~\ref{sec:main-ident}) correspond to $W^{\circ}_k$ upon choosing $\overline{L}_k = \emptyset$, i.e. under the assumption of no confounding of the association between HAI, on the one hand, and hospital death or discharge, on the other hand.

\subsection{Hypothetical interventions implicitly emulated by different estimation approaches}

As recently pointed out and formalized by \cite{Young2020}, the choice whether or not to treat a competing event as a censoring event depends on whether one targets an estimand \emph{with} or \emph{without} hypothetical elimination of that competing event \citep[also see][]{VonCube2019,VonCube2019a,VonCube2022}.
Censoring an event of interest at a competing event time re-weighs the events of interest in a way that fully respects the temporal ordering of all considered event types.

Artificial censoring is also implicitly achieved by multistate modeling based approaches that impose constraints on the intensities or hazards between health states to emulate a hypothetical world where entry into states corresponding to that competing event is somehow prevented \citep{Arjas1993,Klein1993,Keiding2001}. More specifically, this corresponds to calculating the risk of HAI-free hospital mortality upon setting to zero the intensity for the transition from the state `hospitalized without infection' to `HAI' in the progressive disability model (depicted in Figure~\ref{fig:extillnessdeath} in Appendix~\ref{sec:app-wcharest2} in the Supplementary Material), as proposed by \cite{Schumacher2007} \citep[also see][]{VonCube2019,VonCube2019a,VonCube2022} or, equivalently, the time-dependent event-specific hazards of infection in the competing risk model depicted in Figure~\ref{fig:competingrisk2}. This can be appreciated by the fact that
\begin{eqnarray*}
  \widehat\varphi_K^{\circ \text{\tiny AJ}} &&= \sum_{k=1}^K \widehat{\tilde h}^{(1)}_k \left(1 - 0 \widehat{\tilde h}^{(0)}_k \right) \left(1 - \widehat{\tilde h}^{(2)}_k \right) \prod_{s=1}^{k-1} \left(1 - 0 \widehat{\tilde h}^{(0)}_s \right) \left(1 - \widehat{\tilde h}^{(1)}_s \right) \left(1 - \widehat{\tilde h}^{(2)}_s \right),
\end{eqnarray*}
or, in other words, $\widehat\varphi_K^{\circ \text{\tiny AJ}}$ corresponds to $\widehat{\tilde\phi}_K^{\text{\tiny AJ}}$ upon setting $\tilde h^{(0)}_k = 0$ for each $k = 1,...,K$ (see Appendix~\ref{sec:app-wcharest3} in the Supplementary Material).
Intuitively, artificially setting these hazards to zero corresponds to artificially censoring infected patients, such that, as patients get infected, their weight in the analysis is transferred to patients who may still enter one of the other (infection-free) health states, i.e. those at risk of HAI-free hospital death or discharge, but not to those who have already entered one of these health states (see Appendix~\ref{sec:app-wcharest3} in the Supplementary Material).

This stands in contrast with multistate modeling based approaches that treat HAI onset either as a baseline or time-dependent exclusion criterion rather than a censoring event because these approaches also (tacitly) transfer weight to patients who have already died or been discharged. While $\widehat\varphi_K^{\star}$ and $\widehat\varphi_K^{\dagger}$ also reweigh the contributions of HAI-free hospital deaths to the cumulative incidence curve, weighing is done in a way that does not correspond to treating the competing event HAI onset as a censoring event. As a result, the applied weights deviate from the corresponding IPC weights and fail to fully respect the temporal ordering of all considered event types. When treating HAI onset as a baseline exclusion criterion, as in estimator $\widehat\varphi_K^{\star}$ (and $\widehat\varphi_K^{\star \text{\tiny AJ}}$ in Appendix~\ref{sec:app-wcharest1} in the Supplementary Material), a patient's weight already gets transferred at baseline, when her future exposure status is still unknown. This approach hence fails to appropriately account for the timing of exposure.

When treating HAI onset as a time-dependent exclusion criterion, as in estimator $\widehat\varphi_K^{\dagger}$ (and $\widehat\varphi_K^{\dagger \text{\tiny AJ}}$ in Appendix~\ref{sec:app-wcharest2} in the Supplementary Material), which is implicit in the approach proposed by \cite{Schumacher2007}, a patient's weight only gets transferred at exposure onset. However, it gets distributed not only among patients still at risk, but also among patients who have already died or been discharged without infection. This approach therefore accounts for the timing of exposure, yet fails to appropriately account for the temporal order of exposure and its competing events (HAI-free hospital death or discharge). Nonetheless, the implicit rationale behind this approach is also hypothetical elimination of the competing event, at least when one wishes to endow the estimated PAF a causal interpretation. Upon noting that
\begin{eqnarray*}
  \widehat\varphi_K^{\dagger \text{\tiny AJ}} &&= \sum_{k=1}^K \widehat{c}^{\dagger (1)}_{k,K} \widehat{\tilde h}^{(1)}_k \left(1 - 0 \widehat{\tilde h}^{(0)}_k \right) \left(1 - \widehat{c}^{\dagger (2)}_{k,K} \widehat{\tilde h}^{(2)}_k \right) \prod_{s=1}^{k-1} \left(1 - 0 \widehat{\tilde h}^{(0)}_s \right) \left(1 - \widehat{c}^{\dagger (1)}_{s,K} \widehat{\tilde h}^{(1)}_s \right) \left(1 - \widehat{c}^{\dagger (2)}_{s,K} \widehat{\tilde h}^{(2)}_s \right),
\end{eqnarray*}
with
\begin{eqnarray*}
  c^{\dagger (1)}_{k,K} &\equiv& \dfrac{1}{\Pr\left(A_K = 0 \middle\vert D_k = A_k = Y_{k-1} = 0\right)}\\
  c^{\dagger (2)}_{k,K} &\equiv& \dfrac{1}{\Pr\left(A_K = 0 \middle\vert A_k = Y_{k-1} = D_{k-1} = 0\right)},
\end{eqnarray*}
one could claim that, from a conceptual and counterfactual point of view, this estimation approach indeed achieves hypothetical elimination of exposure, but with certain unintended and unrealistic side-effects that render death and discharge recurrent events (see Appendix~\ref{sec:app-wcharest2} in the Supplementary Material for more details).

In this section we have illustrated that, even in the absence of confounding, bias may occur in the estimation of the counterfactual risk $\varphi_K$ when failing to adequately account for the temporal ordering of events. However, while time-dependent bias can be eliminated by treating HAI onset (the time-dependent exposure of interest) as a censoring event, the resulting estimators $\widehat\varphi_K^{\circ}$ and $\widehat\varphi_K^{\circ \text{\tiny AJ}}$ only enable bias-free estimation in the absence of confounding of the association between HAI onset, on the one hand, and hospital mortality and discharge, on the other hand. In other words, these estimators are consistent only under the assumption of independent (artificial) censoring. Under this assumption, progressive depletion of infected patients from the risk set over time, does not introduce bias due to differential selection of uninfected hospitalized patients.
However, in most (if not all) applications, this assumption is unrealistic. In the next section, we discuss how estimation can be modified to accommodate settings in which censoring at each time can be assumed independent conditional on a set of baseline and time-varying prognostic factors or confounders measured up to that time.

\section{Accounting for time-dependent confounding: tackling dependent censoring of a hypothetical endpoint}\label{sec:main-timedepconf}

The progressive selection of a risk set of uninfected patients poses additional challenges regarding confounding adjustment or, equivalently, modeling the censoring mechanism when treating HAI onset as a censoring event.
Bias-free estimation necessitates confounding adjustment for indicators of disease severity, e.g. by standardization of estimates obtained from stratified multistate models \citep{Walter1976,Whittemore1982,Benichou2001}.
As such, estimation approaches based on multistate modeling enable to reduce selection bias due to imbalances in the distribution of baseline confounders between censored and uncensored patients at hospital admission.
However, because HAIs are rarely acquired upon hospital admission, but are inherently time-dependent, confounding of their effects on hospital death and discharge inevitably is also time-dependent. Equivalently, progressive selection of HAI-free patients in the risk set over time is driven by a censoring or selection mechanism that is likely to also depend on prognostic factors that evolve over time.

For instance, prior to acquiring infection, patients may deteriorate further and may therefore be at increased risk of infection, even if, at admission, their prognosis is similar to that of patients who eventually do not acquire infection.
Consequently, confounding adjustment should not only be made at baseline (e.g. for severity of illness indicators recorded at time of admission), but also for the evolution of such indicators over time.
Even though current approaches for estimating the PAF based on multistate models may account for the time-dependent nature of acquiring HAI, they do not readily permit to account for the time-dependent nature of prognostic factors that drive the progressive selection of uninfected patients over time, except upon adequate reweighting (see Appendix~\ref{sec:app-connest1} in the Supplementary Material for more details).
They are therefore bound to provide biased estimates of the fraction of hospital mortality that can \emph{causally} be attributed to HAIs.

Generalized methods, abbreviated g-methods \citep{Robins2008a,Hernan2020}, enable us to additionally tackle the time-dependent nature of confounding or, equivalently, selection of HAI-free patients. This class of methods, in particular inverse probability (IP) weighting, can be characterized as a natural generalization of the censoring approach discussed in section~\ref{sec:main-timedepexp-cens}. That is, estimators $\widehat\varphi_K^{\circ}$ and $\widehat\varphi_K^{\circ \text{\tiny AJ}}$ can be further refined so as to also consider the impact of relevant time-dependent confounders on the censoring process.
As such, IPCW estimation can be tailored to not only eliminate time-dependent bias, but also to reduce bias due to the differential selection of uninfected patients as characterized by baseline and time-dependent prognostic factors \citep{Robins2000b,Satten2001a}.

As highlighted in Section~\ref{sec:main-ident}, in the absence of censoring due to loss to follow-up or administrative end of the study, the counterfactual risk $\varphi_K$ is identified from the observed data by the non-parametric g-formula under conditions~\eqref{main-seqexch}-\eqref{main-consistency}. As before, this functional can be expressed as a weighted version of the cumulative distribution function of HAI-free hospital death $\tilde\phi_K$, given by expression~\eqref{IPCWcdf} with weights given in~\eqref{IPCweights}. This leads to the following estimator, corresponding to an IPC weighted empirical cumulative distribution function
\begin{eqnarray*}
  \widehat\varphi_K \equiv \hspace{0.1cm} n^{-1}\sum_{k=1}^K \tilde d^{w_{ik}}_{1k},
\end{eqnarray*}
where $\tilde d^{w_{ik}}_{1k} \equiv \sum_{i=1}^n Y_{ik}\left(1-D_{ik}\right)\left(1-A_{ik}\right)\left(1-Y_{i,k-1}\right) W_{ik}(\widehat \theta)$ denotes the IPC weighted number of HAI-free hospital deaths in interval $k$ and the weights
\begin{eqnarray*}
  W_{ik}(\widehat \theta) &\equiv& \displaystyle\prod_{s = 1}^{k} \dfrac{1}{1-p(s, \overline L_{i,s-1}; \widehat \theta)},
\end{eqnarray*}
can be computed using the fitted values of a pre-specified model $p(k, \overline L_{k-1};\theta)$ (e.g. a pooled logistic regression model or an extended Cox regression model with time-varying covariates, and possibly time-varying effects), indexed by (vector-valued) parameter $\theta$, for the discrete-time event-specific hazard of exposure in interval $k$ conditional on the confounder history $\Pr(A_{k}=1 | \overline L_{k-1}, A_{k-1} = D_{k-1} = Y_{k-1} = 0)$.
However, even under correct specification of $p(k, \overline L_{k-1};\theta)$, due to sampling variability, $\widehat\varphi_K$ may produce estimates $> 1$. To accommodate for this, sample bounded IPC weighted estimators \citep{Tan2010b} can be used, such as the one proposed by \cite{Bekaert2010} (see Appendix~\ref{sec:app-connest3} in the Supplementary Material) or the following IPC weighted Aalen-Johansen estimator
\begin{eqnarray*}
  \widehat\varphi_K^{\text{\tiny AJ}} \equiv \sum_{k = 1}^K \widehat{\tilde h}^{(1),w_{ik}}_k(\widehat \theta) \left(1-\widehat{\tilde h}^{(2),w_{ik}}_k(\widehat \theta)\right) \prod_{s = 1}^{k-1} \left(1-\widehat{\tilde h}^{(1),w_{is}}_s(\widehat \theta)\right) \left(1-\widehat{\tilde h}^{(2),w_{is}}_s(\widehat \theta)\right),
\end{eqnarray*}
with $\tilde h^{(1),w_{ik}}_k$ and $\tilde h^{(2),w_{ik}}_k$ (defined in Appendix~\ref{sec:app-wcharest4} in the Supplementary Material) denoting discrete-time event-specific hazards in the IPC weighted risk sets, and their respective estimators
\begin{eqnarray*}
  \widehat{\tilde h}^{(1),w_{ik}}_k(\widehat \theta) &\equiv& \dfrac{\sum_{i=1}^n Y_{ik} (1-D_{ik}) (1-A_{ik}) (1-Y_{i,k-1}) W_{ik}(\widehat \theta)}{\sum_{i=1}^n (1-D_{ik}) (1-A_{ik}) (1-Y_{i,k-1}) W_{ik}(\widehat \theta)}\\
  \widehat{\tilde h}^{(2),w_{ik}}_k(\widehat \theta) &\equiv& \dfrac{\sum_{i=1}^n D_{ik} (1-A_{ik}) (1-Y_{i,k-1}) (1-D_{i,k-1}) W_{ik}(\widehat \theta)}{\sum_{i=1}^n (1-A_{ik}) (1-Y_{i,k-1}) (1-D_{i,k-1}) W_{ik}(\widehat \theta)}.
\end{eqnarray*}
Re-weighting the risk sets aims to recover the counterfactual risk sets where all patients still hospitalized had remained HAI-free.
When the weights $W_{k}$ are non-parametrically estimated, $\widehat\varphi_K$ and $\widehat\varphi_K^{\text{\tiny AJ}}$ can again be shown to be algebraically equivalent (see Appendix~\ref{sec:app-wcharest4} in the Supplementary Material).
As such, these estimators again aim to construct a pseudo-population that matches the original population, but in which HAIs are eradicated, by assigning weights to patient time contributions that do not only depend on time, but also on individual prognostic factors that may evolve over time.

Incorporating information on time-dependent prognostic factors for both HAI onset and hospital death and discharge in the weights relaxes the strong assumption of (marginally) independent censoring by HAI onset and replaces it with the weaker assumption of \emph{conditionally} independent censoring. In Appendix~\ref{sec:app-wcharest4} in the Supplementary Material, we demonstrate that conditionally independent censoring is implied by conditions~\eqref{main-seqexch}-\eqref{main-consistency}, with $\overline{L}_{k-1}$ chosen to be a non-empty set of (time-dependent) covariates for $k = 1,...,K$, and an additional exchangeability assumption. Moreover, in Appendix~\ref{sec:app-wcharest4} in the Supplementary Material, we illustrate that this assumption is equivalent to assuming that, for every interval $k$, within strata defined by observed levels of $\overline L_k$, the observable event-specific hazards of HAI-free hospital death and discharge equal the counterfactual event-specific hazards of HAI-free hospital death and discharge.
More specifically, because the relation between timing of infection and covariate history is explicitly accounted for in the weights, censored (infected) patients prospectively transfer their weight to uncensored (uninfected) patients with the same covariate history.
In doing so, potential imbalances between infected and uninfected patients with respect to relevant prognostic factors can be restored at each single time point and selection bias can be eliminated, at least insofar as the selected set of time-dependent covariates $\overline L_k$ is sufficient to adjust for confounding of the effect of infection on hospital death and hospital discharge or insofar as censoring is rendered independent or \emph{explainable} conditional on this covariate set (see Appendix~\ref{sec:app-connest} in the Supplementary Material for more details on the connection with other proposed estimation approaches \citep{Schumacher2007,Bekaert2010,Pouwels2017a,Young2020}).

\section{Empirical data example}\label{sec:main-empex}

In this section we illustrate the differences between the four estimation approaches using an empirical example. In particular, we highlight that different sources of bias can be attributed to weight transfer schemes that do not adequately account for the temporal ordering of events or time-dependent prognostic factors of those events. Given the unboundedness property of $\widehat\varphi_K$ we chose to compare the (weighted) Aalen-Johansen estimators $\widehat\varphi_K^{\star \text{\tiny AJ}}$, $\widehat\varphi_K^{\dagger \text{\tiny AJ}}$, $\widehat\varphi_K^{\circ \text{\tiny AJ}}$ and $\widehat\varphi_K^{\text{\tiny AJ}}$.
These estimators may accommodate settings with censoring due to loss to follow-up or administrative end of study. Moreover, in the absence of such censoring (as in our motivating example) the former three estimators are algebraically equivalent to $\widehat\varphi_K^{\star}$, $\widehat\varphi_K^{\dagger}$ and $\widehat\varphi_K^{\circ}$, respectively.

The population attributable fraction of ICU death due to HAIs was estimated in a cohort of 1,478 patients hospitalized at the Ghent University Hospital medical and surgical intensive care units for at least 48 hours between 2013 and 2017. Patients were included in the cohort if they had received mechanical ventilation and had remained without (suspected) infection within the first 48 hours following admission. Pseudonymized records were extracted from the Intensive Care Information System (ICIS) database (GE Healthcare Centricity Critical Care). Unique admission identifiers allowed to link these records to corresponding records from the infection surveillance system (\textbf{Co}mputer-based \textbf{S}urveillance and \textbf{A}lerting of nosocomial infections, Antimicrobial \textbf{R}esistance and \textbf{A}ntibiotic consumption in the ICU; COSARA) database \citep{Steurbaut2012,DeBus2014,DeBus2018} that included detailed information on HAI diagnosis, presumed onset and antibiotic therapy. This cohort forms a subgroup of a larger cohort that has been described in more detail in a clinical paper focusing on estimation of the population attributable fraction of ICU death due to ventilator-associated pneumonia \citep{Steen2020b}. The Ghent University Hospital Ethics Committee approved this study (registration number B670201732106) and waived informed consent since all analyses were performed retrospectively on pseudonymized records.

For illustrative purposes and to demonstrate clear differences between estimation approaches, we here focus on estimation of the population attributable fraction of ICU death within the first 30 ICU days due to ICU-acquired bacterial infections (including abdominal, catheter-related, respiratory, and urinary tract infections acquired at least 48 hours following ICU admission) developed during this 30-day time window in this patient cohort. These analyses are merely intended as an illustration of the discussed estimation approaches. Results should therefore not be interpreted as conclusive evidence as insufficient attention was paid to correct HAI classification or careful confounder selection based on expert opinion or available substantive knowledge.

\begin{table}[t!]%
\centering
\begin{tabular}{lcccc}
\hline\\[0.5ex]%
\shortstack[l]{\textit{Required data depending on whether}\\\textit{estimation approach treats HAI onset as...}} & \rot{\shortstack[c]{exclusion\\criterion ($\widehat\varphi_K^{\star \text{\tiny AJ}}$)}} & \rot{\shortstack[c]{time-dependent\\exclusion\\criterion ($\widehat\varphi_K^{\dagger \text{\tiny AJ}}$)}} & \rot{\shortstack[c]{(marginally)\\ independent\\censoring ($\widehat\varphi_K^{\circ \text{\tiny AJ}}$)}} & \rot{\shortstack[c]{conditionally\\ independent\\censoring ($\widehat\varphi_K^{\text{\tiny AJ}}$)}}\\[0.5ex]\hline
time of ICU death/discharge               & $\times$ & $\times$ & $\times$ & $\times$ \\%
HAI status at death/discharge             & $\times$ & $\times$ & $\times$ & $\times$ \\%
time of HAI onset                         &          & $\times$ & $\times$ & $\times$ \\%
baseline and time-dependent confounders   &          &          &          & $\times$ \\[0.5ex]\hline%
\end{tabular}
\caption{Overview of required data for application of each of the four discussed estimation approaches. \label{tab:data}}%
\end{table}

As listed in Table~\ref{tab:data}, due to different levels of complexity, the different estimation approaches also differ in terms of required data, either in terms of granularity or type of data. The naive estimator $\widehat\varphi_K^{\star \text{\tiny AJ}}$ treats HAI onset as an exclusion criterion and therefore does not require any data on HAI onset time. In addition, when accounting for timing of HAI onset but not for differential selection of uninfected patients (the second and third estimation approaches listed in Table~\ref{tab:data}, corresponding to estimators $\widehat\varphi_K^{\dagger \text{\tiny AJ}}$ and $\widehat\varphi_K^{\circ \text{\tiny AJ}}$), estimation can, in principle, be done based on aggregated data (i.e. so-called life tables which list the number of events of each type $\tilde \epsilon$ at each discrete time point) and hence no individual patient data is required. To tackle selection bias due to informative censoring (the fourth approach listed in Table~\ref{tab:data}, corresponding to estimator $\widehat\varphi_K^{\text{\tiny AJ}}$), however, individual patient characteristics and prognostic factors need to be taken into account in the IPC weights $W_{ik}$. To estimate these weights, a Cox model was fitted for the event-specific discrete time hazard of developing an ICU-acquired bacterial infection conditional on gender, admission category (medical verus surgical ICU), admission year, age at admission, updated Charlson comorbidity index (restricted cubic spline with 2 knots) and the Sequential Organ Failure Assessment (SOFA) score, both at admission (restricted cubic spline with 4 knots) and two days prior to potential development of infection (to acknowledge that this severity-of-illness score may be a surrogate marker for an incubating infection) (restricted cubic spline with 4 knots). For each of the four estimators, percentile-based 95\% confidence intervals were calculated from the 2.5 and 97.5 percentiles of the distribution of 1,000 non-parametric bootstrap samples. R code to replicate these analyses is available as supplementary material.

\begin{figure}[t!]
\centerline{\includegraphics[width=330pt]{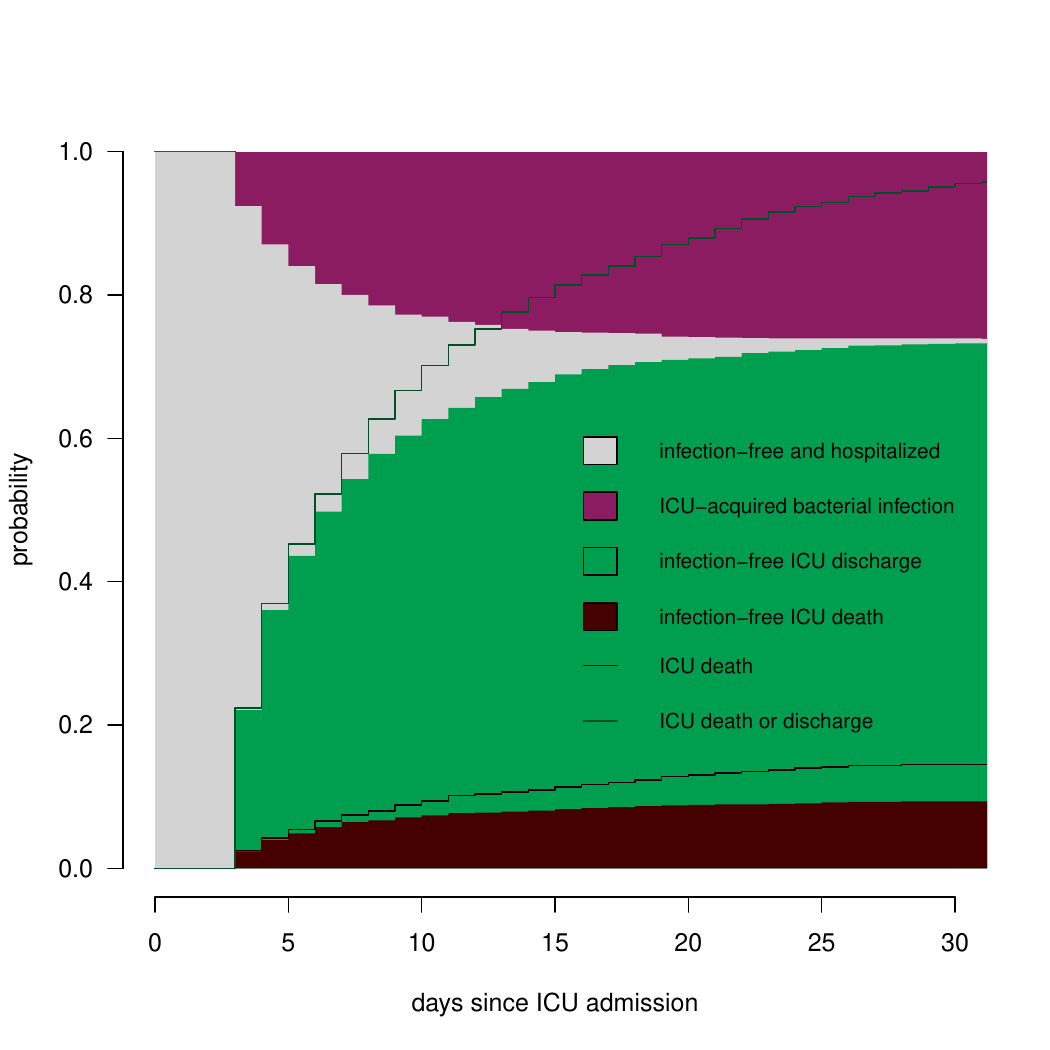}}
\caption{Unadjusted state occupation probabilities as estimated by the Aalen-Johansen estimator corresponding to the (multistate) competing risk models depicted in Figure~\ref{fig:competingrisk1} (solid lines representing stacked probabilities) and Figure~\ref{fig:competingrisk2} (shaded areas also representing stacked probabilities).\label{fig:emp_example1}}
\end{figure}

\begin{figure}[h!]
\centerline{\includegraphics[width=400pt]{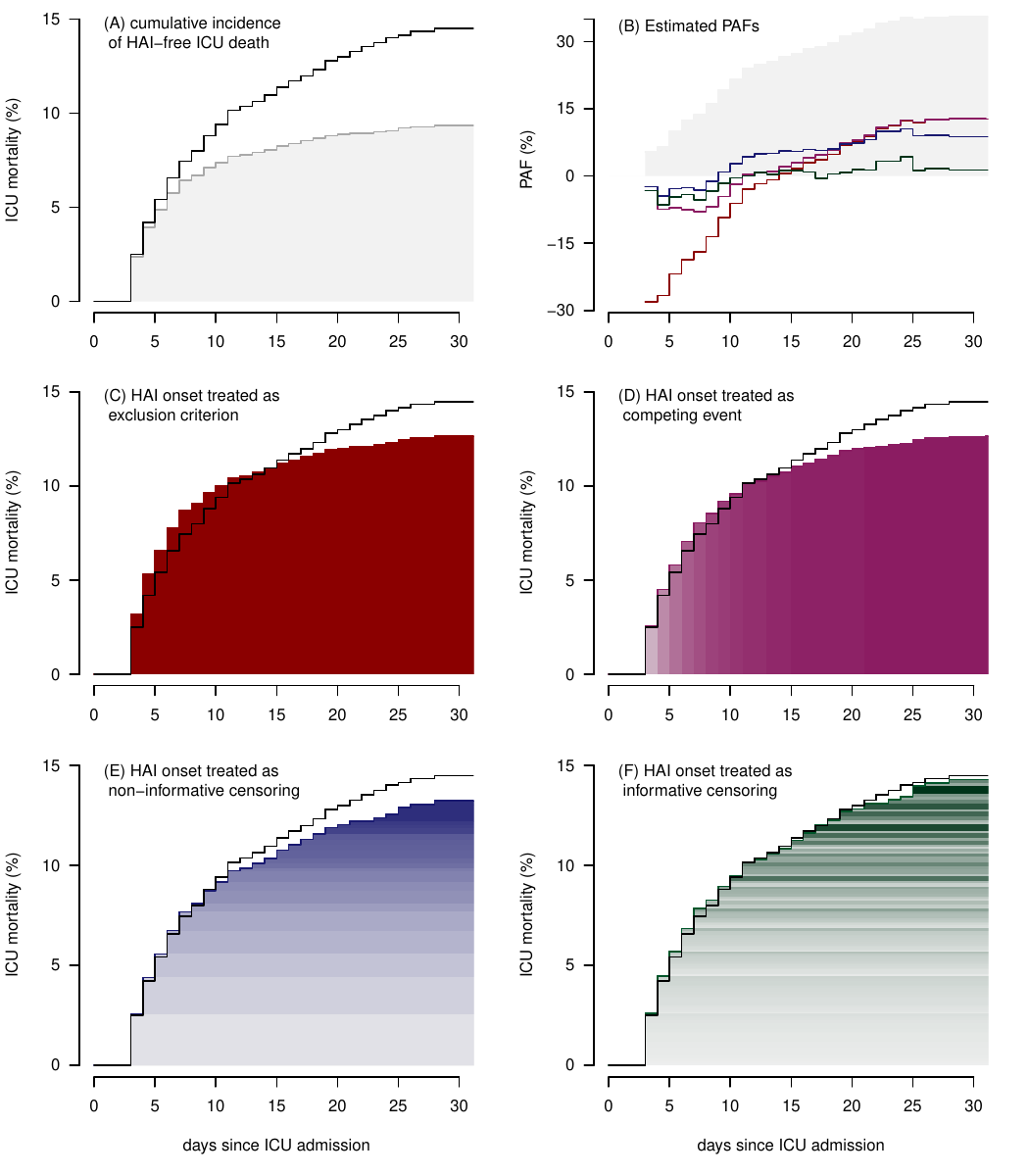}}
\caption{Factual risk of ICU death within the first 30 ICU days as estimated by $\widehat\phi_K^{\text{\tiny AJ}}$ (for $K = 1, ..., 30$) (solid black line in panels A,C-F), counterfactual HAI-free risk of ICU death as estimated by $\widehat\varphi_K^{\star \text{\tiny AJ}}$, $\widehat\varphi_K^{\dagger \text{\tiny AJ}}$, $\widehat\varphi_K^{\circ \text{\tiny AJ}}$ and $\widehat\varphi_K^{\text{\tiny AJ}}$ (solid colored lines in panels C-F), and corresponding estimated population-attributable fractions (PAFs) (panel B; with lower and upper bounds for the PAF under consistency and monotonicity depicted by the gray shaded area). Each estimator has a distinct way of upweighing the increments of the estimated cumulative risk of HAI-free ICU death, as graphically represented by different color shades, with darker shades reflecting larger weights. As a visual reference, $\widehat{\tilde\phi}_K^{\text{\tiny AJ}}$ is depicted with unit weights in light gray (panel A).\label{fig:emp_example2}}
\end{figure}

Figure~\ref{fig:emp_example1} displays the unadjusted state occupation probabilities corresponding to the competing risk models depicted in Figures~\ref{fig:competingrisk1} and~\ref{fig:competingrisk2}, as estimated by their corresponding Aalen-Johansen estimators.
By day 30, 26.07\% of the considered patients had acquired a bacterial infection at the ICU, 81.04\% had been discharged from the ICU (63.91\% without infection), 14.49\% had died at the ICU (9.34\% without infection), and 4.47\%  was still hospitalized (0.67\% without infection).
The contrast between the cumulative incidence of ICU death by day $K$, as estimated by $\widehat\phi_K^{\text{\tiny AJ}}$, and the cumulative incidence of HAI-free ICU death by day $K$, as estimated by $\widehat{\tilde\phi}_K^{\text{\tiny AJ}}$, expresses the proportion of patients who have died \emph{with} an ICU-acquired bacterial infection by day $K$ (the contrast between the darkred curve and the darkred shaded area in Figure~\ref{fig:emp_example1}, as also depicted in panel A of Figure~\ref{fig:emp_example2}).

By day 30, this is about $14.49 - 9.34 = 5.2\%$ of patients. In other words, about $(14.49 - 9.34)/14.49 \approx 36\%$ of patients who have died at the ICU by day 30, have died with an ICU-acquired bacterial infection. Because not all patients who die with infection die from infection, and since infections are considered harmful (rather than beneficial) for all patients, $\tilde\phi_K$ can be considered as a lower bound for the counterfactual HAI-free cumulative incidence $\varphi_K$ (i.e. assuming all patients dying with infection also die from infection), while $\phi_K$ can be considered as an upper bound (i.e. assuming no patient dying with infection dies from infection). The gray shaded area in panel B of Figure~\ref{fig:emp_example2} displays the corresponding lower and upper bounds for the PAF under this monotonicity assumption. To acknowledge that some patients who have died at the ICU with infection within 30 days may not have died had they not acquired infection within that time window, different approaches for estimating the counterfactual risk $\varphi_K$ borrow missing information on counterfactual event times in infected patients from patients who did not (yet) acquire infection. For each of the four considered estimation approaches, this is achieved by upweighing ICU deaths in the latter group in the analysis.
\clearpage
However, each approach applies a different weighting scheme, as displayed in Figure~\ref{fig:emp_example2}. No confidence intervals are displayed to avoid overcrowding. Instead, these are displayed in Figure~\ref{fig:emp_example_app} in Appendix~\ref{sec:app-confint} in the Supplementary Material. Table~\ref{tab:est} displays estimates at ICU day 30, along with the corresponding 95\% confidence intervals.

When HAI onset is treated as a criterion for exclusion from the analytical sample (panel C of Figure~\ref{fig:emp_example2}), exclusion of infected patients is compensated for by uniformly weighing each HAI-free ICU death by the inverse (marginal) probability of remaining without ICU-acquired bacterial infection until ICU death or discharge (about 74\%), which corresponds to a factor $\approx 1.35$. According to this approach it was estimated that, had all patients remained without infection (at least until day 30), 12.65\% would have died at the ICU by day 30, corresponding to an estimated excess mortality due to infection of $14.49-12.65 = 1.84\%$ or an estimated PAF of $(14.49-12.65)/14.49 = 12.70\%$ (panel B of Figure~\ref{fig:emp_example2}). In contrast, during roughly the first two weeks of hospitalization, this approach estimates that certain deaths may have been prevented (or delayed) by infection, as reflected by negative estimated PAFs (panels B--C of Figure~\ref{fig:emp_example2}). This can be explained by the fact that, as every HAI-free ICU death is weighed by the same factor (that can only be determined by the last HAI onset time), early ICU deaths get weighed disproportionally compared to later ICU deaths, resulting in immortal time bias.

When treating HAI onset as a time-dependent exclusion criterion (panel D of Figure~\ref{fig:emp_example2}), progressive depletion of infected patients from the risk set over time is compensated for by weighing each HAI-free ICU death \emph{by} day $K$ by the inverse (marginal) probability of remaining without ICU-acquired bacterial infection until day $K$, a factor that increases over time.
According to this approach it was estimated that, had all patients remained without infection (at least until day 30), 12.64\% would have died at the ICU by day 30, corresponding to an estimated excess mortality due to infection of $14.49-12.64 = 1.85\%$ or an estimated PAF of $(14.49-12.64)/14.49 = 12.78\%$. Even though this approach ensures that smaller weights are apportioned at earlier landmarks, thereby mitigating immortal time bias, its resulting time-dependent PAF estimates may still misleadingly raise the impression that infection may be beneficial during the first 10 days after admission (panel B of Figure~\ref{fig:emp_example2}). This is because all ICU deaths \emph{by} day $K$ are uniformly weighed by a factor that is determined that day. Hence, to estimate the counterfactual risk of ICU death by day $K$ all observed HAI-free ICU deaths \emph{before} day $K$ are weighed by a factor that incorporates information that will only become available \emph{on} day $K$. This approach essentially perpetuates the problem of the first approach, but in a nested (and therefore more subtle) fashion. Instead of uniformly weighing each HAI-free ICU death \emph{across} future landmarks, this approach uniformly weighs each HAI-free ICU death that has occurred \emph{by} day $K$ \emph{on} day $K$. This leads to deaths getting reweighed at each future landmark with incident infections. At landmarks with no future infections estimates of this approach coincide with those of the first approach, which explains why these two approaches converge toward the end follow-up (panel B of Figure~\ref{fig:emp_example2}).

\begin{table}[t]%
\centering
\begin{tabular}{lccr}
\hline\\[0.5ex]%
\shortstack[l]{Estimate\\ (95\% confidence intervals)} & \shortstack[c]{Excess mortality\\ by ICU day 30} & \shortstack[c]{Estimated PAF \\ by ICU day 30} & \shortstack[r]{(Range of) weights\\ for ICU deaths at day 30} \\[0.1cm]\hline\\[-0.3cm]
$\widehat\varphi_{30}^{\star \text{\tiny AJ}}$ = 0.126 (0.106; 0.146) & 1.84\% (0.73; 3.03) & 12.70\% (5.22; 20.74) & $1.35$ \\[0.1cm]%
$\widehat\varphi_{30}^{\dagger \text{\tiny AJ}}$ = 0.126 (0.106; 0.146) & 1.85\% (0.75; 3.04) & 12.78\% (5.23; 20.81) & $1.35$ \\[0.1cm]%
$\widehat\varphi_{30}^{\circ \text{\tiny AJ}}$ = 0.132 (0.110; 0.154) & 1.26\% (0.01; 2.64) & 8.71\% (0.06; 18.16) & $2.52$ \\[0.1cm]%
$\widehat\varphi_{30}^{\text{\tiny AJ}}$ = 0.143 (0.109; 0.206) & 0.20\% (-5.54; 2.90) & 1.35\% (-37.03; 20.10) & $(1.71; 5.21)$ \\[0.1cm]\hline\\[-0.3cm]%
\multicolumn{4}{c}{$\widehat{\tilde\phi}_{30}^{\text{\tiny AJ}}$ = 0.093 (0.078; 0.109)}\\
\multicolumn{4}{c}{$\widehat\phi_{30}^{\text{\tiny AJ}}$ = 0.145 (0.127; 0.162)}\\[0.1cm]\hline%
\end{tabular}
\caption{The counterfactual risk at day 30, as estimated by the different (weight) Aalen-Johansen estimators and corresponding estimated excess mortality and population attributable fractions (PAFs) at ICU day 30. \label{tab:est}}%
\end{table}

When treating HAI onset as (marginally) independent censoring (panel E of Figure~\ref{fig:emp_example2}), progressive depletion of infected patients from the risk set over time is compensated for by recursively weighing each patient at risk on day $K$ on each of the previous days by the inversed complement of the event-specific hazard of HAI onset.
This corresponds to weighing each HAI-free ICU death on day $K$ by the inverse probability of remaining without ICU-acquired bacterial infection (while hospitalized) by day $K$ and weighing each HAI-free ICU death on each day $k$ that \emph{precedes} day $K$ by the inverse probability of remaining without ICU-acquired bacterial infection (while hospitalized) by day $k$.
According to this approach it was estimated that, had all patients remained without infection (at least until day 30), 13.23\% would have died at the ICU by day 30, corresponding to an estimated excess mortality due to infection of $14.49-13.23  = 1.26\%$ or an estimated PAF of $(14.49-13.23)/14.49 = 8.71\%$.
In contrast to the previous two approaches, this approach does not upweigh deceased HAI-free patients based on information that is not yet available at their respective death times, and therefore fully respects the temporal ordering of event times.

When treating HAI onset as conditionally independent censoring (panel F of Figure~\ref{fig:emp_example2}), progressive depletion of infected patients from the risk set over time is compensated for in a similar fashion, while additionally accounting for differential selection of uninfected patients over time to avoid selection bias.
This approach weighs each HAI-free ICU death \emph{on} day $K$ by the inverse probability of remaining without ICU-acquired bacterial infection (while hospitalized) by day $K$ among patients with the same covariate history up to day $K-1$, and weighs each HAI-free ICU death on each day $k$ that \emph{precedes} day $K$ by the inverse probability of remaining without ICU-acquired bacterial infection (while hospitalized) by day $k$ among patients with the same covariate history up to day $k-1$. According to this approach it was estimated that, had all patients remained without infection (at least until day 30), 14.29 would have died at the ICU by day 30, corresponding to an estimated excess deaths due to infection of $14.49-14.29 = 0.20\%$ or an estimated PAF of $(14.49-14.29)/14.49 = 1.35\%$.

\section{Discussion}\label{sec:main-discussion}

Several approaches have been proposed in the epidemiologic literature for estimating the population-attributable fraction (PAF) in the presence of time-dependent exposures and competing events. Recent work by \cite{VonCube2019,VonCube2019a} has compared these approaches and has shed light on conceptual confusion concerning the targeted estimands of these respective approaches. Their work indicated that differences between approaches can be related to whether or not the corresponding target estimand is meant to be interpreted causally or not, leading these authors to suggest a distinction between the directly observable (or `factual') PAF estimand, which corresponds to $\{\phi_K - [\tilde\phi_K/(1-\alpha_K)]\}/\phi_K$, and the causal or `counterfactual' PAF estimand, which corresponds to $(\phi_K - \varphi_K)/\phi_K$. It may be argued, however, that due to clear reference to the term \emph{attribution}, the PAF was meant to be interpreted causally from its outset, or that at least, such interpretation is usually (although perhaps implicitly) evoked by most of the scientific community. The work by von Cube and colleagues clarified that any counterfactual or causal interpretation of an estimate obtained by \citep{Schumacher2007}'s estimator $\widehat\varphi^{\dagger}_K$ (which treats HAI onset as a time-dependent exclusion criterion) implicitly conflates the counterfactual with the factual PAF estimand \citep{Steen2022a}. The latter essentially is a \emph{statistical} estimand that can be expressed as a functional of the observed data and that is directly estimable without assumptions about the underlying causal structure of the observed data. It can be interpreted as the reduction in hospital mortality by day $K$ among patients who had not acquired an infection by that day relative to that in the original population (see Appendix~\ref{sec:app-connest1} in the Supplementary Material). While, arguably, this estimand lacks a causal interpretation, it remains unclear whether this `factually observed burden' of exposure has any relevance or implications for clinical practice. Instead, it could be argued that the fraction of deceased patients that dies \emph{with} exposure $(\phi_K - \tilde\phi_K)/\phi_K$ may be a more relevant factual estimand, because under the aforementioned consistency and monotonicity conditions, this measure sets an upper bound to the counterfactual PAF $(\phi_K - \varphi_K)/\phi_K$ or the fraction of deceased patients that dies \emph{from} exposure.

This work builds upon and extends these recent important contributions by von Cube and colleagues. We highlight that, considering the main goal of estimating the PAF as defined as a causal, counterfactual quantity, proposed estimation approaches can be ordered along a certain hierarchy. By proposing a weighting-based characterization of estimation approaches that aids to better pinpoint different sources of bias within a unified framework, we highlight that each approach implies a refinement with respect to the approach right below it in the hierarchy. Importantly, this characterization may foster deeper and more intuitive understanding both of differences between multistate model based approaches (in terms of how well they respect certain key principles of causal inference) and of the connection between these approaches and g-methods for causal inference, in particular inverse probability of censoring (IPC) weighting.

In this paper, we mainly target estimation of the counterfactual PAF $(\phi_K - \varphi_K)/\phi_K$ in retrospective studies. Recently, there have been increasing calls to cast observational studies for comparative effectiveness research into a more formal causal framework that revolves around explicit `target trial emulation' \citep{Hernan2016,Hernan2016a,Sterne2016,Didelez2016,Labrecque2017,Prosperi2020}. This approach encourages researchers to first define the causal question of interest in terms of a hypothetical (possibly pragmatic) randomized controlled trial and to then either conduct that RCT or to emulate its results from observational data, using its protocol as a guidance to avoid common pitfalls. When trying to endow the PAF with an \emph{explicit} interventional interpretation, as often done \emph{implicitly} by interpreting the PAF as the proportion of preventable death cases, a randomized prevention trial may naturally come to mind. More specifically, under certain assumptions, the PAF can be interpreted as the relative risk reduction in a randomized experiment that compares standard of care with a fully effective prevention strategy that affects the outcome of interest only through eradication of exposure. As such, the estimation approaches under comparison can alternatively be organized hierarchically with respect to how well they emulate this hypothetical target trial.
While we have discussed conditions for identification of the counterfactual risk $\varphi_K$ in Section~\ref{sec:main-ident}, an explicit description of the hypo­thetical target prevention trial (or any attempt thereof) forces researchers to pay particular attention to the consistency assumption (condition~\eqref{main-consistency} in Section~\ref{sec:main-ident}) \citep{Rubin1980,Rubin1986a,Robins1986,Cole2009}.
This assumption links the observable data characterized by \emph{observed} exposures (e.g. under standard of care) to the counterfactual or interventional world, characterized by \emph{intervened upon} exposures (e.g. under fully effective prevention).
For this assumption to be met, uninfected patients' observed outcomes need to be the same as their (counterfactual) outcomes that would have been observed if they had received a preventive intervention that eradicates infection.
To endow PAF estimates in our example with a clear interventional interpretation, in terms of preventable ICU deaths, we would need to be more specific as to how ICU deaths would be prevented: by means of a preventive intervention, or a bundle of interventions (a so-called \emph{compound} treatment \citep{Hernan2011a}), that successfully eradicates HAIs without affecting ICU mortality or discharge other than through elimination of HAIs. In most studies that aim to estimate PAFs, however, such hypothetical preventive intervention bundles are left implicit and are therefore, by definition, ill-defined. This is mostly because fully effective prevention of the exposure is usually impossible given current means and measures and is therefore inherently hypothetical.

This vague and hypothetical nature of interventions doesn't necessarily need to be a major obstacle, as long as one is willing to assume such intervention exists, or may at some point exist. Even if none will ever exist, the result is arguably useful as an upper bound to the impact that imperfect prevention bundles might achieve at the ICU population level. However, without making this (possibly) complex hypothetical intervention more explicit, a few central questions tend to get obscured, even though they seem crucial for certain analytical choices along the way \citep{Hernan2005,Hernan2008a,Hernan2011a}. For instance, does it include all current preventive measures? If so, how does the choice whether or not to adjust for them, change our interpretation? Another important issue is that, if we don’t know which exact measures are included in this prevention bundle, how are we supposed to identify a minimal adjustment set to ensure exchangeability? Finally, apart from the above issues, a fully effective bundle may not be realistic, because not all HAIs may be (easily) preventable. From a decision-making perspective, it may be more sensible to shift focus to so-called `generalized impact fractions' which contrast clinically feasible interventions with varying degrees of prevention effectiveness \citep{Morgenstern1982}.

In conclusion, assessing whether patients die \emph{from} rather than \emph{with} a certain time-dependent exposure not only requires adequate modeling of the timing of events, as has been emphasized in the past \citep{Schumacher2013a}, but also adequate adjustment for the confounding factors affecting these events and their timing.
A weighting based characterization of proposed approaches for estimating the PAF (i) enables more intuitive understanding of these respective approaches, their differences and their connections and (ii) clearly links potential sources of bias to violations of certain key principles of causal inference inherent to some of these approaches.

\section{Software}\label{sec8}

Documented R code for each of the discussed estimation approaches is available from \url{https://github.com/jmpsteen/time-dep-paf}.

\medskip

\section*{Acknowledgments}
Johan Steen, Pawe\l{} Morzywo\l{}ek, Wim Van Biesen and Stijn Vansteelandt were supported by the Flemish Research Council (FWO Research Project 3G068619 --- Grant FWO.OPR.2019.0045.01). The authors would like to thank Maja von Cube and Els Goetghebeur for fruitful discussions on this topic, and Bram Gadeyne, Veerle Brams and Christian Danneels for technical support.

\bibliography{20230119PAFmultistate_vs_counterfactual_modelling}%

\appendix

\section{Estimation of the factual risks $\phi_K$ and $\tilde\phi_K$ from observed data}

Throughout, let $P_n(X=x) \equiv n^{-1} \sum_{i=1}^n I(X_i = x)$ and $E_n(X) \equiv n^{-1} \sum_{i=1}^n X_i$ denote the empirical probability and empirical expectation, respectively, with $I(\cdot)$ the indicator function. Furthermore, let $P_n(X=x|Z=z) \equiv \dfrac{P_n(X=x,Z=z)}{P_n(Z=z)}$ and $E_n(X \vert Z=z) \equiv \dfrac{\sum_{i=1}^n I(Z_i=z) X_i}{\sum_{i=1}^n I(Z_i=z)}$.

\subsection{Cumulative incidence of hospital death $\phi_K$}\label{sec:app-estobsdata1}
Following the definitions from Section~\ref{sec:main-setnot} in the main text, under the assumed temporal ordering $(D_k, Y_k)$, we have
\begin{eqnarray*}
I(T \in (t_{k-1}, t_k], \epsilon = 1) &=& Y_k(1-D_k)(1-Y_{k-1})\\
I(T \in (t_{k-1}, t_k], \epsilon = 2) &=& D_k(1-Y_{k-1})(1-D_{k-1})\\
I(T > t_{k-1}) &=& (1-Y_{k-1})(1-D_{k-1})\\
I(T \in (t_{k-1}, t_k]) &=& I(T > t_{k-1}) - I(T > t_k)\\
&=& (1-Y_{k-1})(1-D_{k-1}) - (1-Y_k)(1-D_k).
\end{eqnarray*}
Define discrete-time event-specific hazards
\begin{eqnarray*}
  h^{(1)}_{k} &\equiv& \Pr\left[T \in (t_{k-1}, t_k], \epsilon = 1 \middle\vert T > t_k \cup \left\{T \in (t_{k-1}, t_k], \epsilon = 1\right\}\right]\\
  &=& \Pr(Y_k = 1 \vert D_{k} = Y_{k-1} = 0) = \dfrac{E[Y_k(1-D_k)(1-Y_{k-1})]}{E[(1-D_{k})(1-Y_{k-1})]}\\
  h^{(2)}_{k} &\equiv& \Pr\left[T \in (t_{k-1}, t_k], \epsilon = 2 \middle\vert T > t_{k-1}\right]\\
  &=& \Pr(D_k = 1 \vert Y_{k-1} = D_{k-1} = 0) = \dfrac{E[D_k(1-Y_{k-1})(1-D_{k-1})]}{E[(1-Y_{k-1})(1-D_{k-1})]},
\end{eqnarray*}
and their respective estimators
\begin{eqnarray*}
  \widehat{h}^{(1)}_{k} &\equiv& \dfrac{E_n[Y_k(1-D_k)(1-Y_{k-1})(1-C_k)]}{E_n[(1-D_{k})(1-Y_{k-1})(1-C_k)]}\\
  \widehat{h}^{(2)}_{k} &\equiv& \dfrac{E_n[D_k(1-Y_{k-1})(1-D_{k-1})(1-C_k)]}{E_n[(1-Y_{k-1})(1-D_{k-1})(1-C_k)]},
\end{eqnarray*}
where $C_k$ is an indicator for censoring by the end of interval $k$.

In the absence of censoring due to loss to follow-up or administrative end of study, the Aalen-Johansen estimator $\widehat\phi_K^{\text{\tiny AJ}}$ (Section~\ref{sec:main-estimands} in the main text) reduces to the empirical cumulative distribution function (ecdf) of hospital mortality $\widehat\phi_K$ (Section~\ref{sec:main-estimands} in the main text):
\begin{eqnarray*}
  \widehat\phi_K^{\text{\tiny AJ}} &=& \sum_{k = 1}^K \widehat{h}^{(1)}_k (1-\widehat{h}^{(2)}_k) \prod_{s = 1}^{k-1} (1-\widehat{h}^{(1)}_s)(1-\widehat{h}^{(2)}_s)\\
  &=& \sum_{k = 1}^K \dfrac{E_n[Y_k(1-D_k)(1-Y_{k-1})]}{E_n[(1-D_k)(1-Y_{k-1})]} \dfrac{E_n[(1-D_k)(1-Y_{k-1})(1-D_{k-1})]}{E_n[(1-Y_{k-1})(1-D_{k-1})]}\\
  &&\qquad\qquad \times \prod_{s = 1}^{k-1} \dfrac{E_n[(1 - Y_s)(1 - D_s)(1 - Y_{s-1})]}{E_n[(1-D_s)(1-Y_{s-1})]} \dfrac{E_n[(1-D_s)(1 - Y_{s-1})(1 - D_{s-1})]}{E_n[(1-Y_{s-1})(1-D_{s-1})]}\\
  &=& \sum_{k = 1}^K \dfrac{E_n[Y_k(1-D_k)(1-Y_{k-1})]}{E_n[(1-Y_{k-1})(1-D_{k-1})]} \prod_{s = 1}^{k-1} \dfrac{E_n[(1 - Y_s)(1 - D_s)]}{E_n[(1-Y_{s-1})(1-D_{s-1})]}\\
  &=& \sum_{k = 1}^K E_n[Y_k(1-D_k)(1-Y_{k-1})] = n^{-1} \sum_{k = 1}^K d_{1k},
\end{eqnarray*}
with $d_{1k} \equiv \sum_{i=1}^n Y_{ik}(1-D_{ik})(1-Y_{i,k-1})$ denoting the number of hospital deaths in interval $k$. The third equality follows from the fact that for any $s \ge s'$, we have $D_s \ge D_{s'}$ and $Y_s \ge Y_{s'}$, and the fourth equality holds because, by definition, for each patient, we have $D_0 \equiv Y_0 \equiv 0$.

\subsection{Cumulative incidence of infection-free hospital death $\tilde\phi_K$}\label{sec:app-estobsdata2}

Following the definitions from Section~\ref{sec:main-setnot} in the main text, under the assumed temporal ordering $(A_k, \tilde D_k, \tilde Y_k)$, we have
\begin{eqnarray*}
  I(\tilde T \in (t_{k-1}, t_k], \tilde \epsilon = 0) &=& A_k(1-A_{k-1})(1-\tilde Y_{k-1})(1-\tilde D_{k-1})\\
  &=& A_k(1-A_{k-1})(1-Y_{k-1}(1-A_{k-1}))(1-D_{k-1}(1-A_{k-1}))\\
  &=& \begin{cases}
  A_k(1-Y_{k-1})(1-D_{k-1}), & \text{if}\ A_{k-1}=0 \\
  0, & \text{if}\ A_{k-1}=1
  \end{cases}\\
  &=& A_k(1-A_{k-1})(1-Y_{k-1})(1-D_{k-1}),\\
  I(\tilde T \in (t_{k-1}, t_k], \tilde \epsilon = 1) &=& \tilde Y_k(1-\tilde Y_{k-1})(1-\tilde D_k)(1-A_k)\\
  &=& Y_k(1-A_k)(1-Y_{k-1}(1-A_{k-1}))(1-D_k(1-A_k))(1-A_k)\\
  &=& \begin{cases}
  Y_k(1-Y_{k-1})(1-D_k), & \text{if}\ A_k=0 \\
  0, & \text{if}\ A_k=1
  \end{cases}\\
  &=& Y_k(1-Y_{k-1})(1-D_k)(1-A_k),\\
  I(\tilde T \in (t_{k-1}, t_k], \tilde \epsilon = 2) &=& \tilde D_k(1-\tilde D_{k-1})(1-A_k)(1-\tilde Y_{k-1})\\
  &=& D_k(1-A_k)(1-D_{k-1}(1-A_{k-1}))(1-A_k)(1-Y_{k-1}(1-A_{k-1}))\\
  &=& \begin{cases}
  D_k(1-D_{k-1})(1-Y_{k-1}), & \text{if}\ A_k=0 \\
  0, & \text{if}\ A_k=1
  \end{cases}\\
  &=& D_k(1-D_{k-1})(1-A_k)(1-Y_{k-1}),\\
  I(\tilde T > t_{k-1}) &=& (1-Y_{k-1})(1-D_{k-1})(1-A_{k-1})\\
  I(\tilde T \in (t_{k-1}, t_k]) &=& I(\tilde T > t_{k-1}) - I(\tilde T > t_k)\\
  &=& (1-Y_{k-1})(1-D_{k-1})(1-A_{k-1}) - (1-Y_k)(1-D_k)(1-A_k).
\end{eqnarray*}
Define discrete-time event-specific hazards
\begin{eqnarray*}
  \tilde h^{(0)}_k &\equiv& \Pr\left[\tilde T \in (t_{k-1}, t_k], \tilde\epsilon = 0 \middle\vert \tilde T > t_{k-1}\right]\\ &=& \Pr(A_k = 1 \vert Y_{k-1} = D_{k-1} = A_{k-1} = 0) = \dfrac{E[A_k(1-Y_{k-1})(1-D_{k-1})(1-A_{k-1})]}{E[(1-Y_{k-1})(1-D_{k-1})(1-A_{k-1})]}\\
  \tilde h^{(1)}_k &\equiv& \Pr\left[\tilde T \in (t_{k-1}, t_k], \tilde\epsilon = 1 \middle\vert \tilde T > t_k \cup \left\{\tilde T \in (t_{k-1}, t_k], \tilde\epsilon = 1\right\} \right]\\ &=& \Pr(Y_k = 1 \vert D_k = A_k = Y_{k-1} = 0) = \dfrac{E[Y_k(1-D_k)(1-A_k)(1-Y_{k-1})]}{E[(1-D_k)(1-A_k)(1-Y_{k-1})]}\\
  \tilde h^{(2)}_k &\equiv& \Pr\left[\tilde T \in (t_{k-1}, t_k], \tilde\epsilon = 2 \middle\vert \tilde T > t_k \cup \left\{\tilde T \in (t_{k-1}, t_k], \tilde\epsilon \in (1,2)\right\} \right]\\ &=& \Pr(D_k = 1 \vert A_k = Y_{k-1} = D_{k-1} = 0) = \dfrac{E[D_k(1-A_k)(1-Y_{k-1})(1-D_{k-1})]}{E[(1-A_k)(1-Y_{k-1})(1-D_{k-1})]},
\end{eqnarray*}
and their respective estimators
\begin{eqnarray*}
  \widehat{\tilde h}^{(0)}_k &\equiv& \dfrac{E_n[A_k(1-Y_{k-1})(1-D_{k-1})(1-A_{k-1})(1-C_k)]}{E_n[(1-Y_{k-1})(1-D_{k-1})(1-A_{k-1})(1-C_k)]}\\
  \widehat{\tilde h}^{(1)}_k &\equiv& \dfrac{E_n[Y_k(1-D_k)(1-A_k)(1-Y_{k-1})(1-C_k)]}{E_n[(1-D_k)(1-A_k)(1-Y_{k-1})(1-C_k)]}\\
  \widehat{\tilde h}^{(2)}_k &\equiv& \dfrac{E_n[D_k(1-A_k)(1-Y_{k-1})(1-D_{k-1})(1-C_k)]}{E_n[(1-A_k)(1-Y_{k-1})(1-D_{k-1})(1-C_k)]}.
\end{eqnarray*}

In the absence of censoring due to loss to follow-up or administrative end of study, the Aalen-Johansen estimator $\widehat{\tilde\phi}_K^{\text{\tiny AJ}}$ (Section~\ref{sec:main-estimands} in the main text) reduces to the ecdf of exposure-free hospital mortality $\widehat{\tilde\phi}_K$ (Section~\ref{sec:main-estimands} in the main text):
\begin{eqnarray*}
&&\sum_{k = 1}^K \widehat{\tilde h}^{(1)}_k (1-\widehat{\tilde h}^{(2)}_k) (1-\widehat{\tilde h}^{(0)}_k) \prod_{s = 1}^{k-1} (1-\widehat{\tilde h}^{(0)}_s)(1-\widehat{\tilde h}^{(1)}_s)(1-\widehat{\tilde h}^{(2)}_s) \\
&=& \sum_{k = 1}^K \dfrac{E_n[Y_k(1-D_k)(1-A_k)(1-Y_{k-1})]}{E_n[(1-D_k)(1-A_k)(1-Y_{k-1})]} \dfrac{E_n[(1-D_k)(1-A_k)(1-Y_{k-1})(1-D_{k-1})]}{E_n[(1-A_k)(1-Y_{k-1})(1-D_{k-1})]}\\
&&\qquad \times \dfrac{E_n[(1-A_k)(1-Y_{k-1})(1-D_{k-1})(1-A_{k-1})]}{E_n[(1-Y_{k-1})(1-D_{k-1})(1-A_{k-1})]}\\
&&\qquad\qquad \times \prod_{s = 1}^{k-1} \dfrac{E_n[(1-A_s)(1-Y_{s-1})(1-D_{s-1})(1-A_{s-1})]}{E_n[(1-Y_{s-1})(1-D_{s-1})(1-A_{s-1})]}
\dfrac{E_n[(1-Y_s)(1-D_s)(1-A_s)(1-Y_{s-1})]}{E_n[(1-D_s)(1-A_s)(1-Y_{s-1})]}\\
&&\qquad\qquad\qquad \times \dfrac{E_n[(1-D_s)(1-A_s)(1-Y_{s-1})(1-D_{s-1})]}{E_n[(1-A_s)(1-Y_{s-1})(1-D_{s-1})]}\\
&=& \sum_{k = 1}^K \dfrac{E_n[Y_k(1-D_k)(1-A_k)(1-Y_{k-1})]}{E_n[(1-Y_{k-1})(1-D_{k-1})(1-A_{k-1})]} \prod_{s = 1}^{k-1} \dfrac{E_n[(1-Y_s)(1-D_s)(1-A_s)]}{E_n[(1-Y_{s-1})(1-D_{s-1})(1-A_{s-1})]}\\
&=& \sum_{k = 1}^K E_n[Y_k(1-D_k)(1-A_k)(1-Y_{k-1})]
= n^{-1} \sum_{k = 1}^K \tilde d_{1k},
\end{eqnarray*}
with $\tilde d_{1k} \equiv \sum_{i=1}^n Y_{ik}(1-D_{ik})(1-A_{ik})(1-Y_{i,k-1})$ denoting the number of exposure-free hospital deaths in interval $k$. The second equality follows from the fact that for any $s \ge s'$, we have $A_s \ge A_{s'}$, $D_s \ge D_{s'}$ and $Y_s \ge Y_{s'}$, and the third equality holds because, by definition, for each patient, we have $A_0 \equiv D_0 \equiv Y_0 \equiv 0$.

Alternatively, define discrete-time subdistribution hazard
\begin{eqnarray*}
  \tilde h^{\text{sd}}_{k} &\equiv& \Pr\left[\tilde T \in (t_{k-1}, t_k], \tilde\epsilon = 1 \middle\vert \tilde T > t_{k-1} \cup \left\{\tilde T \le t_{k-1}, \tilde \epsilon \ne 1\right\}\right]\\
  &=& \Pr(\tilde Y_k = 1 \vert \tilde Y_{k-1} = 0) = \dfrac{E[\tilde Y_k(1-\tilde Y_{k-1})]}{E[1-\tilde Y_{k-1}]},
\end{eqnarray*}
and its estimator
\begin{eqnarray*}
  \widehat{\tilde h}^{\text{sd}}_{k} &\equiv& \dfrac{E_n[\tilde Y_k(1-\tilde Y_{k-1})(1-C_k)]}{E_n[(1-\tilde Y_{k-1})(1-C_k)]}.
\end{eqnarray*}
An alternative estimator for $\tilde\phi_K$ then corresponds to the complement of a Kaplan-Meier like estimator
\begin{eqnarray*}
\widehat{\tilde \phi}_K^{\text{\tiny KM}} &=& 1 - \prod_{k = 1}^{K} (1-\widehat{\tilde h}^{\text{sd}}_k)\\
  &=& 1 - \left\{\prod_{k = 1}^{K-1} (1-\widehat{\tilde h}^{\text{sd}}_k) - \widehat{\tilde h}^{\text{sd}}_K \prod_{k = 1}^{K-1} (1-\widehat{\tilde h}^{\text{sd}}_k)\right\}\\
  &=& 1 - \left\{\prod_{k = 1}^{K-2} (1-\widehat{\tilde h}^{\text{sd}}_k) - \widehat{\tilde h}^{\text{sd}}_{K-1} \prod_{k = 1}^{K-2} (1-\widehat{\tilde h}^{\text{sd}}_k) - \widehat{\tilde h}^{\text{sd}}_K \prod_{k = 1}^{K-1} (1-\widehat{\tilde h}^{\text{sd}}_k)\right\}\\
  &=& 1 - \left\{(1 - \widehat{\tilde h}^{\text{sd}}_{1}) - \widehat{\tilde h}^{\text{sd}}_{2}(1-\widehat{\tilde h}^{\text{sd}}_{1}) - ... - \widehat{\tilde h}^{\text{sd}}_{K-1} \prod_{k = 1}^{K-2} (1-\widehat{\tilde h}^{\text{sd}}_k) - \widehat{\tilde h}^{\text{sd}}_K \prod_{k = 1}^{K-1} (1-\widehat{\tilde h}^{\text{sd}}_k)\right\}\\
  &=& \sum_{k = 1}^K \widehat{\tilde h}^{\text{sd}}_k \prod_{s = 1}^{k-1} (1-\widehat{\tilde h}^{\text{sd}}_s).
\end{eqnarray*}
In the absence of censoring due to loss to follow-up or administrative end of study, $\widehat{\tilde \phi}_K^{\text{\tiny KM}}$ can again be shown to be algebraically equivalent to both $\widehat{\tilde \phi}_K^{\text{\tiny AJ}}$ and $\widehat{\tilde \phi}_K$ because then we have
\begin{eqnarray*}
  \sum_{k = 1}^K \widehat{\tilde h}^{\text{sd}}_k \prod_{s = 1}^{k-1} (1-\widehat{\tilde h}^{\text{sd}}_s)
  &=& \sum_{k = 1}^K \dfrac{E_n[\tilde Y_k(1-\tilde Y_{k-1})]}{E_n[1-\tilde Y_{k-1}]} \prod_{s = 1}^{k-1} \dfrac{E_n[(1 - \tilde Y_s)(1 - \tilde Y_{s-1})]}{E_n[1-\tilde Y_{s-1}]}\\
  &=& \sum_{k = 1}^K E_n[\tilde Y_k(1-\tilde Y_{k-1})] = \sum_{k = 1}^K E_n[Y_k(1-Y_{k-1})(1-A_k)]\\
  &=& \sum_{k = 1}^K E_n[Y_k(1-Y_{k-1})(1-D_k)(1-A_k)]
= n^{-1} \sum_{k = 1}^K \tilde d_{1k},
\end{eqnarray*}
where the second equality holds because for any $s \ge s'$, we have $\tilde Y_s \ge \tilde Y_{s'}$, and, by definition, for each patient, we have $\tilde Y_0 \equiv 0$, the third equality holds because, for every $k$, we have
\begin{eqnarray*}
  \tilde Y_k(1-\tilde Y_{k-1}) &=& Y_k(1-A_k)(1-Y_{k-1}(1-A_{k-1})) = \begin{cases}
  Y_k(1-Y_{k-1}) & \text{if}\ A_k=0 \\
  0 & \text{if}\ A_k=1
  \end{cases}\\
  &=& Y_k(1-Y_{k-1})(1-A_k),
\end{eqnarray*}
and the fourth equality holds because, for every $k$, we have $Y_k = Y_k(1-D_k)$ due to the assumed temporal order of competing events within interval $k$ and the deterministic relation between the competing events.

Furthermore, note that $\tilde h^{\text{sd}}_{k}$ can be rewritten as
\begin{eqnarray*}
  &&\dfrac{\Pr(\tilde Y_k = 1,\tilde Y_{k-1} = 0)}{1-\Pr(\tilde Y_{k-1} = 1)} = \dfrac{\Pr(Y_k = 1, A_k = 0, Y_{k-1} = 0)}{1-\Pr(Y_{k-1} = 1, A_{k-1} = 0)}
  = \dfrac{\Pr(Y_k = 1, D_k = A_k = Y_{k-1} = 0)}{1-\Pr(Y_{k-1} = 1, D_{k-1} = A_{k-1} = 0)}\\
  &&= \dfrac{\Pr(D_k = A_k = Y_{k-1} = 0)}{1-\Pr(Y_{k-1} = 1, D_{k-1} = A_{k-1} = 0)} \Pr(Y_k = 1 \vert D_k = A_k = Y_{k-1} = 0) = c^{\text{sd}}_k \tilde h^{(1)}_{k},
\end{eqnarray*}
with
\begin{eqnarray}
  c^{\text{sd}}_k &&\equiv \dfrac{\Pr(D_k = A_k = Y_{k-1} = 0)}{1-\Pr(Y_{k-1} = 1, D_{k-1} = A_{k-1} = 0)} = \dfrac{\Pr(D_k = A_k = Y_{k-1} = 0)}{1-\Pr(Y_{k-1} = 1, D_k = A_k = 0)}\nonumber\\
  &&= \dfrac{\Pr(D_k = A_k = Y_{k-1} = 0)}{\Pr(D_k = A_k = Y_{k-1} = 0) + \Pr(A_k = 1) + \Pr(D_k = 1, A_k = 0)}\nonumber 
\end{eqnarray}
representing the proportion of patients in the subdistribution hazard risk set of interval $k$ that has not yet experienced a competing event (exposure onset or exposure-free hospital discharge). \cite{Putter2020} referred to $c^{\text{sd}}_k$ as the `reduction factor', as it expresses by how much the subdistribution hazard risk set in interval $k$ is reduced to arrive at the event-specific hazard risk set. Conversely, the reduction factor illustrates that, to arrive at the subdistribution hazard risk set, the event-specific hazard risk set is `extended' by also including patients that have experienced a competing event.

\section{Estimation of the counterfactual risk $\varphi_K$ from counterfactual data}\label{sec:app-estcfdata}

Suppose, hypothetically speaking, that for each subject, we could observe their counterfactual data.
Following the definitions from Section~\ref{sec:main-setnot} in the main text, under the assumed temporal ordering $(D^{\overline a = 0}_k, Y^{\overline a = 0}_k)$, we would then have
\begin{eqnarray*}
I(T^{\overline a = 0} \in (t_{k-1}, t_k], \epsilon^{\overline a = 0} = 1) &=& Y^{\overline a = 0}_k(1-D^{\overline a = 0}_k)(1-Y^{\overline a = 0}_{k-1})\\
I(T^{\overline a = 0} \in (t_{k-1}, t_k], \epsilon^{\overline a = 0} = 2) &=& D^{\overline a = 0}_k(1-Y^{\overline a = 0}_{k-1})(1-D^{\overline a = 0}_{k-1})\\
I(T^{\overline a = 0} > t_{k-1}) &=& (1-Y^{\overline a = 0}_{k-1})(1-D^{\overline a = 0}_{k-1})\\
I(T^{\overline a = 0} \in (t_{k-1}, t_k]) &=& I(T^{\overline a = 0}> t_{k-1}) - I(T^{\overline a = 0}> t_k)\\
&=& (1-Y^{\overline a = 0}_{k-1})(1-D^{\overline a = 0}_{k-1}) - (1-Y^{\overline a = 0}_k)(1-D^{\overline a = 0}_k).
\end{eqnarray*}
Define counterfactual discrete-time event-specific hazards
\begin{eqnarray*}
  h^{(1), \overline a = 0}_k &\equiv& \Pr\left[T^{\overline a = 0} \in (t_{k-1}, t_k], \epsilon^{\overline a = 0} = 1 \middle\vert T^{\overline a = 0} > t_k \cup \left\{T^{\overline a = 0} \in (t_{k-1}, t_k], \epsilon^{\overline a = 0} = 1\right\}\right]\\&=& \Pr(Y^{\overline a = 0}_k = 1 \vert D^{\overline a = 0}_k = Y^{\overline a = 0}_{k-1} = 0) = \dfrac{E[Y^{\overline a = 0}_k(1-D^{\overline a = 0}_k)(1-Y^{\overline a = 0}_{k-1})]}{E[(1-D^{\overline a = 0}_{k})(1-Y^{\overline a = 0}_{k-1})]}\\
  h^{(2), \overline a = 0}_k &\equiv& \Pr\left[T^{\overline a = 0} \in (t_{k-1}, t_k], \epsilon^{\overline a = 0} = 2 \middle\vert T^{\overline a = 0} > t_{k-1}\right]\\&=& \Pr(D^{\overline a = 0}_k = 1 \vert Y^{\overline a = 0}_{k-1} = D^{\overline a = 0}_{k-1} = 0)  = \dfrac{E[D^{\overline a = 0}_k(1-Y^{\overline a = 0}_{k-1})(1-D^{\overline a = 0}_{k-1})]}{E[(1-Y^{\overline a = 0}_{k-1})(1-D^{\overline a = 0}_{k-1})]},
\end{eqnarray*}
and their respective estimators
\begin{eqnarray*}
  \widehat{h}^{(1), \overline a = 0}_k &\equiv& \dfrac{E_n[Y^{\overline a = 0}_k(1-D^{\overline a = 0}_k)(1-Y^{\overline a = 0}_{k-1})]}{E_n[(1-D^{\overline a = 0}_{k})(1-Y^{\overline a = 0}_{k-1})]}\\
  \widehat{h}^{(2), \overline a = 0}_k &\equiv& \dfrac{E_n[D^{\overline a = 0}_k(1-Y^{\overline a = 0}_{k-1})(1-D^{\overline a = 0}_{k-1})]}{E_n[(1-Y^{\overline a = 0}_{k-1})(1-D^{\overline a = 0}_{k-1})]}.
\end{eqnarray*}

The Aalen-Johansen estimator again reduces to the ecdf of counterfactual exposure-free hospital mortality:
\begin{eqnarray*}
  &&\sum_{k = 1}^K \widehat{h}^{(1),\overline a = 0}_k \left(1-\widehat{h}^{(2),\overline a = 0}_k\right) \prod_{s = 1}^{k-1} \left(1-\widehat{h}^{(1),\overline a = 0}_s\right)\left(1-\widehat{h}^{(2),\overline a = 0}_s\right)\\
  &=& \sum_{k = 1}^K \dfrac{E_n[Y^{\overline a = 0}_k(1-D^{\overline a = 0}_k)(1-Y^{\overline a = 0}_{k-1})]}{E_n[(1-D^{\overline a = 0}_k)(1-Y^{\overline a = 0}_{k-1})]} \dfrac{E_n[(1-D^{\overline a = 0}_k)(1-Y^{\overline a = 0}_{k-1})(1-D^{\overline a = 0}_{k-1})]}{E_n[(1-Y^{\overline a = 0}_{k-1})(1-D^{\overline a = 0}_{k-1})]}\\
  &&\qquad\qquad \times \prod_{s = 1}^{k-1} \dfrac{E_n[(1 - Y^{\overline a = 0}_s)(1 - D^{\overline a = 0}_s)(1 - Y^{\overline a = 0}_{s-1})]}{E_n[(1-D^{\overline a = 0}_s)(1-Y^{\overline a = 0}_{s-1})]} \dfrac{E_n[(1-D^{\overline a = 0}_s)(1 - Y^{\overline a = 0}_{s-1})(1 - D^{\overline a = 0}_{s-1})]}{E_n[(1-Y^{\overline a = 0}_{s-1})(1-D^{\overline a = 0}_{s-1})]}\\
  &=& \sum_{k = 1}^K \dfrac{E_n[Y^{\overline a = 0}_k(1-D^{\overline a = 0}_k)(1-Y^{\overline a = 0}_{k-1})]}{E_n[(1-Y^{\overline a = 0}_{k-1})(1-D^{\overline a = 0}_{k-1})]} \prod_{s = 1}^{k-1} \dfrac{E_n[(1 - Y^{\overline a = 0}_s)(1 - D^{\overline a = 0}_s)]}{E_n[(1-Y^{\overline a = 0}_{s-1})(1-D^{\overline a = 0}_{s-1})]}\\
  &=& \sum_{k = 1}^K E_n[Y^{\overline a = 0}_k(1-D^{\overline a = 0}_k)(1-Y^{\overline a = 0}_{k-1})]
  = n^{-1} \sum_{k = 1}^K d^{\overline a = 0}_{1k},
\end{eqnarray*}
with $d^{\overline a = 0}_{1k} \equiv \sum_{i=1}^n Y^{\overline a = 0}_{ik}(1-D^{\overline a = 0}_{ik})(1-Y^{\overline a = 0}_{i,k-1})$ denoting the number of counterfactual (exposure-free) hospital deaths in interval $k$ and where the third equality holds because, by definition, for each patient, we have $D^{\overline a = 0}_0 \equiv Y^{\overline a = 0}_0 \equiv 0$.

Alternatively, define counterfactual discrete-time subdistribution hazard
\begin{eqnarray*}
  h^{\text{sd},\overline a = 0}_{k} &\equiv& \Pr\left[T^{\overline a = 0} \in (t_{k-1}, t_k], \epsilon^{\overline a = 0} = 1 \middle\vert T^{\overline a = 0} > t_{k-1} \cup \left\{T^{\overline a = 0} \le t_{k-1}, \epsilon^{\overline a = 0} \ne 1\right\}\right]\\
  &=& \Pr(Y^{\overline a = 0}_k = 1 \vert Y^{\overline a = 0}_{k-1} = 0) = \dfrac{E[Y^{\overline a = 0}_k(1-Y^{\overline a = 0}_{k-1})]}{E[1-Y^{\overline a = 0}_{k-1}]},
\end{eqnarray*}
and its estimator
\begin{eqnarray*}
  \widehat{h}^{\text{sd},\overline a = 0}_{k} &\equiv& \dfrac{E_n[Y^{\overline a = 0}_k(1-Y^{\overline a = 0}_{k-1})]}{E_n[(1-Y^{\overline a = 0}_{k-1})]}.
\end{eqnarray*}
An alternative estimator for $\varphi_K$ then corresponds to the complement of a Kaplan-Meier like estimator
\begin{eqnarray}
  1 - \prod_{k = 1}^{K} (1-\widehat{h}^{\text{sd},\overline a = 0}_k) = \sum_{k = 1}^K \widehat{h}^{\text{sd},\overline a = 0}_k \prod_{s = 1}^{k-1} (1-\widehat{h}^{\text{sd},\overline a = 0}_s).\label{app-kaplanmeierlike_cf}
\end{eqnarray}
This estimator can be shown to be algebraically equivalent to the aforementioned Aalen-Johansen estimator and the ecdf of counterfactual exposure-free hospital mortality
\begin{eqnarray*}
  \sum_{k = 1}^K \widehat{h}^{\text{sd},\overline a = 0}_k \prod_{s = 1}^{k-1} (1-\widehat{h}^{\text{sd},\overline a = 0}_s)
  &=& \sum_{k = 1}^K \dfrac{E_n[Y^{\overline a = 0}_k(1-Y^{\overline a = 0}_{k-1})]}{E_n[1-Y^{\overline a = 0}_{k-1}]} \prod_{s = 1}^{k-1} \dfrac{E_n[(1 - Y^{\overline a = 0}_s)(1 - Y^{\overline a = 0}_{s-1})]}{E_n[1-Y^{\overline a = 0}_{s-1}]}\\
  &=& \sum_{k = 1}^K E_n[Y^{\overline a = 0}_k(1-Y^{\overline a = 0}_{k-1})]\\
  &=& \sum_{k = 1}^K E_n[Y^{\overline a = 0}_k(1-D^{\overline a = 0}_k)(1-Y^{\overline a = 0}_{k-1})]
= n^{-1} \sum_{k = 1}^K d^{\overline a = 0}_{1k},
\end{eqnarray*}
where the second equality holds because, by definition, for each patient, we have $Y^{\overline a = 0}_0 \equiv 0$, and the third equality holds because, for every $k$, we have $Y^{\overline a = 0}_k = Y^{\overline a = 0}_k(1-D^{\overline a = 0}_k)$ due to the assumed temporal order of competing events within interval $k$ and the deterministic relation between the competing events.

\section{Assumptions for non-parametric identification of counterfactual quantities} 

\subsection{Non-parametric identification of the counterfactual risk $\varphi_K$} \label{sec:app-ident1}

Under the following structural assumptions, for every $k = 1, ..., K$,
\begin{itemize}
  \item Sequential exchangeability
  \begin{eqnarray}\label{app-seqexch}
  \underline{Y}_k^{\overline{a} = 0} \cip  A_k \vert \overline{L}_{k-1}, \overline{D}_{k-1}, Y_{k-1} = A_{k-1} = 0
  \end{eqnarray}
  \item Positivity
  \begin{eqnarray}\label{app-positivity}
  &\Pr\left(A_k = 0 \middle\vert \overline{L}_{k-1}, A_{k-1} = D_{k-1} = Y_{k-1} = 0\right) > 0 \text{ w.p.1}
  \end{eqnarray}
  \item Consistency
  \begin{eqnarray}\label{app-consistency}
  \text{If } A_k = 0 \text{ then } \overline{L}_k = \overline{L}_k^{\overline{a} = 0} \text{, } \overline{D}_k = \overline{D}_k^{\overline{a} = 0} \text{ and } \overline{Y}_k = \overline{Y}_k^{\overline{a} = 0},
  \end{eqnarray}
\end{itemize}
the counterfactual risk $\varphi_K$ for $K = 0,1,...,\tau$ can be identified from the observed data $(\overline A_K, \overline D_K, \overline Y_K, \overline L_K)$ as follows
\begin{eqnarray}
  \Pr(Y_K^{\overline a = 0} = 1) &=& \sum_{k=1}^K \Pr(Y_k^{\overline a = 0} = 1, \overline Y_{k-1}^{\overline a = 0} = \overline 0)\nonumber\\
  &=& \sum_{k=1}^K \sum_{l_0} \Pr(Y_k^{\overline a = 0} = 1, Y_{k-1}^{\overline a = 0} = \cdots = Y_1^{\overline a = 0} = 0 \vert L_0 = l_0) f(L_0 = l_0)\nonumber\\
  \eqref{app-seqexch},\eqref{app-positivity} &=& \sum_{k=1}^K \sum_{l_0} \Pr(Y_k^{\overline a = 0} = 1, Y_{k-1}^{\overline a = 0} = \cdots = Y_1^{\overline a = 0} = 0 \vert L_0 = l_0, A_1 = 0) f(L_0 = l_0)\nonumber\\
  &=& \sum_{k=1}^K \sum_{l_0, \overline d_1} \Pr(Y_k^{\overline a = 0} = 1, Y_{k-1}^{\overline a = 0} = \cdots = Y_1^{\overline a = 0} = 0 \vert L_0 = l_0, \overline D_1 = \overline d_1, A_1 = 0)\nonumber\\[-5pt]
  &&\qquad \qquad \times \Pr(D_1 = d_1 \vert L_0 = l_0, A_1 = 0) f(L_0 = l_0)\nonumber\\
  &=& \sum_{k=1}^K \sum_{l_0, \overline d_1} \Pr(Y_1^{\overline a = 0} = 0 \vert L_0 = l_0, \overline D_1 = \overline d_1, A_1 = 0)\nonumber\\[-10pt]
  &&\qquad \qquad \times \Pr(Y_k^{\overline a = 0} = 1, Y_{k-1}^{\overline a = 0} = \cdots = Y_2^{\overline a = 0} = 0 \vert L_0 = l_0, \overline D_1 = \overline d_1, A_1 = Y_1^{\overline a = 0} = 0)\nonumber\\
  &&\qquad \qquad \times \Pr(D_1 = d_1 \vert L_0 = l_0, A_1 = 0) f(L_0 = l_0)\nonumber\\
  \eqref{app-consistency} &=& \sum_{k=1}^K \sum_{l_0, \overline d_1} \Pr(Y_1 = 0 \vert L_0 = l_0, \overline D_1 = \overline d_1, A_1 = 0)\nonumber\\[-10pt]
  &&\qquad \qquad \times \Pr(Y_k^{\overline a = 0} = 1, Y_{k-1}^{\overline a = 0} = \cdots = Y_2^{\overline a = 0} = 0 \vert L_0 = l_0, \overline D_1 = \overline d_1, A_1 = Y_1 = 0)\nonumber\\
  &&\qquad \qquad \times \Pr(D_1 = d_1 \vert L_0 = l_0, A_1 = 0) f(L_0 = l_0)\nonumber\\
  &=& \sum_{k=1}^K \sum_{\overline l_1, \overline d_1} \Pr(Y_1 = 0 \vert L_0 = l_0, \overline D_1 = \overline d_1, A_1 = 0)\nonumber\\[-10pt]
  &&\qquad \qquad \times \Pr(Y_k^{\overline a = 0} = 1, Y_{k-1}^{\overline a = 0} = \cdots = Y_2^{\overline a = 0} = 0 \vert \overline L_1 = \overline l_1, \overline D_1 = \overline d_1, A_1 = Y_1 = 0)\nonumber\\
  &&\qquad \qquad \times  f(L_1 = l_1 \vert L_0 = l_0, \overline D_1 = \overline d_1, A_1 = Y_1 = 0) \Pr(D_1 = d_1 \vert L_0 = l_0, A_1 = 0) f(L_0 = l_0)\nonumber\\
  \eqref{app-seqexch},\eqref{app-positivity} &=& \sum_{k=1}^K \sum_{\overline l_1, \overline d_1} \Pr(Y_1 = 0 \vert L_0 = l_0, \overline D_1 = \overline d_1, A_1 = 0)\nonumber\\[-10pt]
  &&\qquad \qquad \times \Pr(Y_k^{\overline a = 0} = 1, Y_{k-1}^{\overline a = 0} = \cdots = Y_2^{\overline a = 0} = 0 \vert \overline L_1 = \overline l_1, D_1 = d_1, A_2 = Y_1 = 0)\nonumber\\ &&\qquad \qquad \times f(L_1 = l_1 \vert L_0 = l_0, \overline D_1 = \overline d_1, A_1 = Y_1 = 0) \Pr(D_1 = d_1 \vert L_0 = l_0, A_1 = 0) f(L_0 = l_0)\nonumber\\
  &=& \sum_{k=1}^K \sum_{\overline l_1, \overline d_2} \Pr(Y_1 = 0 \vert L_0 = l_0, \overline D_1 = \overline d_1, A_1 = 0)\nonumber\\[-10pt]
  &&\qquad \qquad \times \Pr(Y_k^{\overline a = 0} = 1, Y_{k-1}^{\overline a = 0} = \cdots = Y_2^{\overline a = 0} = 0 \vert \overline L_1 = \overline l_1, \overline D_2 = \overline d_2, A_2 = Y_1 = 0)\nonumber\\
  &&\qquad \qquad \times \Pr(D_2 = d_2 \vert \overline L_1 = \overline l_1, \overline D_1 = \overline d_1, A_1 = Y_1 = 0) f(L_1 = l_1 \vert L_0 = l_0, \overline D_1 = \overline d_1, A_1 = Y_1 = 0)\nonumber\\
  &&\qquad \qquad \times \Pr(D_1 = d_1 \vert L_0 = l_0, A_1 = 0) f(L_0 = l_0)\nonumber\\
  \eqref{app-consistency} &=& \sum_{k=1}^K \sum_{\overline l_1, \overline d_2} \Pr(Y_1 = 0 \vert L_0 = l_0, \overline D_1 = \overline d_1, A_1 = 0)\nonumber\\[-10pt]
  &&\qquad \qquad \times \Pr(Y_2 = 0 \vert \overline L_1 = \overline l_1, \overline D_2 = \overline d_2, A_2 = Y_1 = 0)\nonumber\\
  &&\qquad \qquad \times \Pr(Y_k^{\overline a = 0} = 1, Y_{k-1}^{\overline a = 0} = \cdots = Y_3^{\overline a = 0} = 0 \vert \overline L_1 = \overline l_1, \overline D_2 = \overline d_2, A_2 = Y_2 = 0)\nonumber\\
  &&\qquad \qquad \times \Pr(D_2 = d_2 \vert \overline L_1 = \overline l_1, \overline D_1 = \overline d_1, A_1 = Y_1 = 0) f(L_1 = l_1 \vert L_0 = l_0, \overline D_1 = \overline d_1, A_1 = Y_1 = 0)\nonumber\\
  &&\qquad \qquad \times \Pr(D_1 = d_1 \vert L_0 = l_0, A_1 = 0) f(L_0 = l_0)\nonumber\\
  \eqref{app-seqexch},\eqref{app-consistency}&=& \cdots\nonumber\\
  &=& \sum_{k=1}^K \sum_{\overline l_{k-1}, \overline d_k} \Pr(Y_k = 1 \vert \overline L_{k-1} = \overline l_{k-1}, \overline D_k = \overline d_k, A_k = Y_{k-1} = 0)\nonumber\\[-10pt]
  &&\qquad \qquad \times \prod_{s = 1}^{k} \Pr(D_s = d_s \vert \overline L_{s-1} = \overline l_{s-1}, \overline D_{s-1} = \overline d_{s-1}, A_s = Y_{s-1} = 0)\nonumber\\[-10pt]
  &&\qquad \qquad \qquad \quad \times f(L_{s-1} = l_{s-1} \vert \overline L_{s-2} = \overline l_{s-2}, \overline D_{s-1} = \overline d_{s-1}, A_{s-1} = Y_{s-1} = 0)\nonumber\\
  &&\qquad \qquad \qquad \quad \times \Pr(Y_{s-1} = 0 \vert \overline L_{s-2} = \overline l_{s-2}, \overline D_{s-1} = \overline d_{s-1}, A_{s-1} = Y_{s-2} = 0)\label{gformula}\\
  (*) &=& \sum_{k=1}^K \sum_{\overline l_{k-1}} \Pr(Y_k = 1 \vert \overline L_{k-1} = \overline l_{k-1}, A_k = D_k = Y_{k-1} = 0)\nonumber\\[-10pt]
  &&\qquad \qquad\times \prod_{s = 1}^{k} \Pr(D_s =0 \vert \overline L_{s-1} = \overline l_{s-1}, A_s = D_{s-1} = Y_{s-1} = 0)\nonumber\\[-10pt]
  &&\qquad \qquad \qquad \quad \times f(L_{s-1} = l_{s-1} \vert \overline L_{s-2} = \overline l_{s-2}, A_{s-1} = D_{s-1} = Y_{s-1} = 0)\nonumber\\
  &&\qquad \qquad \qquad \quad \times \Pr(Y_{s-1} = 0 \vert \overline L_{s-2} = \overline l_{s-2}, A_{s-1} = D_{s-1} = Y_{s-2} = 0)\nonumber\\
  &=& \sum_{k=1}^K \sum_{\overline l_{k-1}} \Pr(Y_k = 1 \vert \overline L_{k-1} = \overline l_{k-1}, A_k = D_k = Y_{k-1} = 0)\nonumber\\[-10pt]
  &&\qquad \qquad \times \prod_{s = 1}^{k} \Pr(D_s =0 \vert \overline L_{s-1} = \overline l_{s-1}, A_s = D_{s-1} = Y_{s-1} = 0) \nonumber\\[-10pt]
  &&\qquad \qquad \qquad \quad \times \dfrac{\Pr(A_s =0 \vert \overline L_{s-1} = \overline l_{s-1}, A_{s-1} = D_{s-1} = Y_{s-1} = 0)}{\Pr(A_s =0 \vert \overline L_{s-1} = \overline l_{s-1}, A_{s-1} = D_{s-1} = Y_{s-1} = 0)}\nonumber\\
  &&\qquad \qquad \qquad \quad \times f(L_{s-1} = l_{s-1} \vert \overline L_{s-2} = \overline l_{s-2}, A_{s-1} = D_{s-1} = Y_{s-1} = 0)\nonumber\\
  &&\qquad \qquad \qquad \quad \times \Pr(Y_{s-1} = 0 \vert \overline L_{s-2} = \overline l_{s-2}, A_{s-1} = D_{s-1} = Y_{s-2} = 0)\nonumber\\[5pt]
  &=& \sum_{k=1}^K \sum_{\overline l_{k-1}} \dfrac{\Pr(Y_k = 1, \overline Y_{k-1} = \overline D_k = \overline A_k = \overline 0, \overline L_{k-1} = \overline l_{k-1})}{\prod_{s = 1}^{k} \Pr(A_s = 0 \vert \overline L_{s-1} = \overline l_{s-1}, A_{s-1} = D_{s-1} = Y_{s-1} = 0)}\nonumber\\
  &=& \sum_{k=1}^K E\left[ \dfrac{Y_k(1-D_k)(1-A_k)(1-Y_{k-1})}{\prod_{s = 1}^{k} \Pr(A_s = 0 \vert \overline L_{s-1}, A_{s-1} = D_{s-1} = Y_{s-1} = 0)} \right]\nonumber\\
  &=& \sum_{k=1}^K E\left[ Y_k(1-D_k)(1-A_k)(1-Y_{k-1})W_k \right],\nonumber
\end{eqnarray}
with
\begin{eqnarray*}
  W_{k} &\equiv& \prod_{s=1}^k \dfrac{1}{\Pr\left(A_s=0 \middle\vert \overline L_{s-1}, A_{s-1} = D_{s-1} = Y_{s-1} = 0\right)},
\end{eqnarray*}
and where the equality indicated by $(*)$ follows from the deterministic relation between the competing events.

\subsection{Non-parametric identification of counterfactual subdistribution hazard $h^{\text{sd},\overline a = 0}_k$}\label{sec:app-ident2}

Following \cite{Young2020}, under assumptions~\eqref{app-seqexch}-\eqref{app-consistency},
the counterfactual subdistribution hazard $h^{\text{sd},\overline a = 0}_k$ (defined in Appendix~\ref{sec:app-estcfdata}) is non-parametrically identified.
Under these assumptions, we have
\begin{enumerate}
  \item \begin{eqnarray*}
     \Pr(Y_k^{\overline a = 0} = 1, \overline Y_{k-1}^{\overline a = 0} = \overline 0) &=& E\left[Y_k(1-D_k)(1-A_k)(1-Y_{k-1}) W_k \right]\\
     &=& E\left[Y_k(1-A_k)(1-Y_{k-1}) W^{\text{sd}}_k \right],
   \end{eqnarray*}
   with
   \begin{eqnarray*}
     W^{\text{sd}}_{k} &\equiv& \prod_{s=1}^k \dfrac{1}{\Pr\left(A_s=0 \middle\vert \overline L_{s-1}, \overline D_{s-1}, A_{s-1} = Y_{s-1} = 0\right)},
   \end{eqnarray*}
  and where the first equality is demonstrated in Appendix~\ref{sec:app-ident1} and the second equality follows from the fact that, for every $k$, we have $Y_k = Y_k(1-D_k)$, because of the deterministic relation between the competing events, and that when $D_k = 0$, we have $W^{\text{sd}}_{k} = W_{k}$,
  \item \begin{eqnarray*}
     \Pr(\overline Y_{k-1}^{\overline a = 0} = \overline 0) &=& \sum_{\overline l_{k-2}, \overline d_{k-1}} \Pr(Y_{k-1} = 0 \vert \overline L_{k-2} = \overline l_{k-2}, \overline D_{k-1} = \overline d_{k-1}, A_{k-1} = Y_{k-2} = 0)\nonumber\\[-10pt]
     &&\qquad \qquad \times \prod_{s = 1}^{k-1} \Pr(D_s = d_s \vert \overline L_{s-1} = \overline l_{s-1}, \overline D_{s-1} = \overline d_{s-1}, A_s = Y_{s-1} = 0)\nonumber\\[-10pt]
     &&\qquad \qquad \qquad \quad \times f(L_{s-1} = l_{s-1} \vert \overline L_{s-2} = \overline l_{s-2}, \overline D_{s-1} = \overline d_{s-1}, A_{s-1} = Y_{s-1} = 0)\nonumber\\
     &&\qquad \qquad \qquad \quad \times \Pr(Y_{s-1} = 0 \vert \overline L_{s-2} = \overline l_{s-2}, \overline D_{s-1} = \overline d_{s-1}, A_{s-1} = Y_{s-2} = 0)\nonumber\\[5pt]
     &=& \sum_{\overline l_{k-2}, \overline d_{k-1}} \dfrac{\Pr(\overline Y_{k-1} = \overline A_{k-1} = \overline 0, \overline D_{k-1} = \overline d_{k-1}, \overline L_{k-2} = \overline l_{k-2})}{\prod_{s = 1}^{k-1} \Pr(A_s = 0 \vert \overline L_{s-1} = \overline l_{s-1}, \overline D_{s-1} = \overline d_{s-1}, A_{s-1} = Y_{s-1} = 0)}\nonumber\\
     &=& E\left[ \dfrac{(1-Y_{k-1})(1-A_{k-1})}{\prod_{s = 1}^{k-1} \Pr(A_s = 0 \vert \overline L_{s-1}, \overline D_{s-1}, A_{s-1} = Y_{s-1} = 0)} \right]\nonumber\\
     &=& E\left[ (1-Y_{k-1}) (1-A_{k-1}) W^{\text{sd}}_{k-1} \right] = E\left[ (1-Y_{k-1}) (1-A_{k}) W^{\text{sd}}_{k} \right],
   \end{eqnarray*}
   where the first equality follows along the lines of Appendix~\ref{sec:app-ident1} and the last equality holds because \citep[also see Lemma 2 in][]{Young2020}
   \begin{eqnarray*}
     &&E\left[(1-Y_k)(1-A_{k+1}) W^{\text{sd}}_{k+1}\right] = E\left[\dfrac{(1-Y_k)(1-A_{k+1}) W^{\text{sd}}_k}{\Pr(A_{k+1} = 0 \vert \overline{L}_k, \overline D_k, Y_k = A_k = 0)}\right]\nonumber\\
     &&=\sum_{\overline{l}_k, \overline{d}_k} E\left[\dfrac{(1-Y_k)(1-A_{k+1}) W^{\text{sd}}_k}{\Pr(A_{k+1} = 0 \vert \overline{L}_k = \overline{l}_k, \overline{D}_k = \overline{d}_k, Y_k = A_k = 0)}\middle\vert \overline{L}_k = \overline{l}_k, \overline{D}_k = \overline{d}_k, Y_k = A_k = 0 \right]\nonumber\\
     &&\qquad\qquad \times \Pr(\overline{L}_k = \overline{l}_k, \overline{D}_k = \overline{d}_k, Y_k = A_k = 0)\nonumber\\[5pt]
     &&=\sum_{\overline{l}_k, \overline{d}_k} \dfrac{\Pr(A_{k+1} = 0 \vert \overline{L}_k = \overline{l}_k, \overline{D}_k = \overline{d}_k, Y_k = A_k = 0)}{\Pr(A_{k+1} = 0 \vert \overline{L}_k = \overline{l}_k, \overline{D}_k = \overline{d}_k, Y_k = A_k = 0)} E\left[(1-Y_k)(1-A_k) W^{\text{sd}}_k\middle\vert \overline{L}_k = \overline{l}_k, \overline{D}_k = \overline{d}_k, Y_k = A_k = 0 \right]\nonumber\\
     &&\qquad\qquad \times \Pr(\overline{L}_k = \overline{l}_k, \overline{D}_k = \overline{d}_k, Y_k = A_k = 0)\nonumber\\[5pt]
     &&=\sum_{\overline{l}_k, \overline{d}_k} E\left[(1-Y_k)(1-A_k) W^{\text{sd}}_k\middle\vert \overline{L}_k = \overline{l}_k, \overline{D}_k = \overline{d}_k, Y_k = A_k = 0 \right] \Pr(\overline{L}_k = \overline{l}_k, \overline{D}_k = \overline{d}_k, \overline{Y}_k = \overline{A}_k = 0)\nonumber\\
     &&= E\left[(1-Y_k)(1-A_k) W^{\text{sd}}_k\right].\nonumber
   \end{eqnarray*}
\end{enumerate}

Taken altogether, under assumptions~\eqref{app-seqexch}-\eqref{app-consistency} we have
\begin{eqnarray*}
  h^{\text{sd},\overline a = 0}_k &\equiv& \Pr(Y^{\overline a = 0}_k = 1 \vert Y^{\overline a = 0}_{k-1} = 0) = \dfrac{E\left[Y_k(1-A_k)(1-Y_{k-1}) W^{\text{sd}}_k \right]}{E\left[ (1-A_{k})(1-Y_{k-1}) W^{\text{sd}}_{k} \right]}.
\end{eqnarray*}

\subsection{Non-parametric identification of counterfactual event-specific hazards $h^{(1),\overline a = 0}_k$ and $h^{(2),\overline a = 0}_k$}\label{sec:app-ident3}

Following \cite{Young2020}, under assumptions~\eqref{app-seqexch}-\eqref{app-consistency} and an additional sequential exchangeability assumption
\begin{eqnarray}\label{app-seqexch2}
\underline{D}_k^{\overline{a} = 0} \cip  A_k \vert \overline{L}_{k-1}, \overline{Y}_{k-1}, D_{k-1} = A_{k-1} = 0,
\end{eqnarray}
the counterfactual event-specific hazards $h^{(1),\overline a = 0}_k$ and $h^{(2),\overline a = 0}_k$ (defined in Appendix~\ref{sec:app-estcfdata}) are non-parametrically identified. This can be understood upon noting that sequential exchangeability assumptions~\eqref{app-seqexch} and~\eqref{app-seqexch2} together imply
\begin{eqnarray}
(\underline{Y}_k^{\overline{a} = 0}, \underline{D}_k^{\overline{a} = 0}) \cip  A_k \vert \overline{L}_{k-1}, Y_{k-1} = D_{k-1} = A_{k-1} = 0,\label{app-seqexch3}
\end{eqnarray}
such that under assumptions~\eqref{app-seqexch}-\eqref{app-consistency} and~\eqref{app-seqexch2} we have
\begin{enumerate}
  \item \begin{eqnarray*}
     \Pr(Y_k^{\overline a = 0} = 1, \overline D_k^{\overline a = 0} = \overline Y_{k-1}^{\overline a = 0} = \overline 0) &=& E\left[Y_k(1-D_k)(1-A_k)(1-Y_{k-1}) W_k \right],
   \end{eqnarray*}
  as demonstrated in Appendix~\ref{sec:app-ident1},
  \item \begin{eqnarray*}
    \Pr(\overline D_k^{\overline a = 0} = \overline Y_{k-1}^{\overline a = 0} = \overline 0) &=& \Pr(D_k^{\overline a = 0} = Y_{k-1}^{\overline a = 0} = D_{k-1}^{\overline a = 0} = \cdots = Y_1^{\overline a = 0} = D_1^{\overline a = 0} = 0)\\
    &=& \sum_{l_0} \Pr(D_k^{\overline a = 0} = Y_{k-1}^{\overline a = 0} = D_{k-1}^{\overline a = 0} = \cdots = Y_1^{\overline a = 0} = D_1^{\overline a = 0} = 0 \vert L_0 = l_0) f(L_0 = l_0)\\
    \eqref{app-positivity},\eqref{app-seqexch3} &=& \sum_{l_0} \Pr(D_k^{\overline a = 0} = Y_{k-1}^{\overline a = 0} = D_{k-1}^{\overline a = 0} = \cdots = Y_1^{\overline a = 0} = D_1^{\overline a = 0} = 0 \vert L_0 = l_0, A_1 = 0) f(L_0 = l_0)\\
    \eqref{app-consistency} &=& \sum_{l_0} \Pr(Y_1 = D_1 = 0 \vert L_0 = l_0, A_1 = 0)\\[-10pt]
    &&\qquad \times \Pr(D_k^{\overline a = 0} = Y_{k-1}^{\overline a = 0} = D_{k-1}^{\overline a = 0} = \cdots = Y_2^{\overline a = 0} = D_2^{\overline a = 0} = 0 \vert L_0 = l_0, A_1 = D_1 = Y_1 = 0)\\
    &&\qquad \times f(L_0 = l_0)\\
    &=& \sum_{\overline l_1} \Pr(Y_1 = D_1 = 0 \vert L_0 = l_0, A_1 = 0)\\[-10pt]
    && \qquad \times \Pr(D_k^{\overline a = 0} = Y_{k-1}^{\overline a = 0} = D_{k-1}^{\overline a = 0} = \cdots = Y_2^{\overline a = 0} = D_2^{\overline a = 0} = 0 \vert \overline L_1 = \overline l_1, A_1 = D_1 = Y_1 = 0)\\
    && \qquad \times f(L_1 = l_1 \vert L_0 = l_0, A_1 = D_1 = Y_1 = 0) f(L_0 = l_0)\\
    \eqref{app-positivity},\eqref{app-seqexch3} &=& \sum_{\overline l_1} \Pr(Y_1 = D_1 = 0 \vert L_0 = l_0, A_1 = 0)\\[-10pt]
    &&\qquad \times \Pr(D_k^{\overline a = 0} = Y_{k-1}^{\overline a = 0} = D_{k-1}^{\overline a = 0} = \cdots = Y_2^{\overline a = 0} = D_2^{\overline a = 0} = 0 \vert \overline L_1 = \overline l_1, A_2 = D_1 = Y_1 = 0)\\
    &&\qquad \times f(L_1 = l_1 \vert L_0 = l_0, A_1 = D_1 = Y_1 = 0) f(L_0 = l_0)\\
    \eqref{app-consistency} &=& \sum_{\overline l_1} \Pr(Y_1 = D_1 = 0 \vert L_0 = l_0, A_1 = 0)\\[-10pt]
    &&\qquad \times \Pr(Y_2 = D_2 = 0 \vert \overline L_1 = \overline l_1, A_2 = D_1 = Y_1 = 0)\\
    &&\qquad \times \Pr(D_k^{\overline a = 0} = Y_{k-1}^{\overline a = 0} = D_{k-1}^{\overline a = 0} = \cdots = Y_3^{\overline a = 0} = D_3^{\overline a = 0} = 0 \vert \overline L_1 = \overline l_1, A_2 = D_2 = Y_2 = 0)\\
    &&\qquad \times f(L_1 = l_1 \vert L_0 = l_0, A_1 = D_1 = Y_1 = 0) f(L_0 = l_0)\\
    \eqref{app-positivity},\eqref{app-consistency},\eqref{app-seqexch3} &=& \cdots\\
    &=& \sum_{\overline l_{k-1}} \Pr(D_k = 0 \vert \overline L_{k-1} = \overline l_{k-1}, A_k = D_{k-1} = Y_{k-1} = 0)\\[-10pt]
    &&\qquad \times \prod_{s = 1}^{k} \Pr(Y_{s-1} = D_{s-1} = 0 \vert \overline L_{s-2} = \overline l_{s-2}, A_{s-1} = D_{s-2} = Y_{s-2} = 0)\\[-5pt]
    &&\qquad \qquad \qquad \times f(L_{s-1} = l_{s-1} \vert \overline L_{s-2} = \overline l_{s-2}, A_{s-1} = D_{s-1} = Y_{s-1} = 0)\\[5pt]
    &=& \sum_{\overline l_{k-1}} \dfrac{\Pr(\overline Y_{k-1} = \overline D_k = \overline A_k = \overline 0, \overline L_{k-1} = \overline l_{k-1})}{\prod_{s = 1}^{k} \Pr(A_s = 0 \vert \overline L_{s-1} = \overline l_{s-1}, A_{s-1} = D_{s-1} = Y_{s-1} = 0)} \\
    &=& E\left[ \dfrac{(1-Y_{k-1})(1-D_k)(1-A_k)}{\prod_{s = 1}^{k} \Pr(A_s = 0 \vert \overline L_{s-1}, A_{s-1} = D_{s-1} = Y_{s-1} = 0)} \right]\\
    &=& E\left[(1-Y_{k-1})(1-D_k)(1-A_k) W_k \right],
  \end{eqnarray*}
  \item \begin{eqnarray*}
    \Pr(\overline Y_{k-1}^{\overline a = 0} = \overline D_{k-1}^{\overline a = 0} = \overline 0) &&= E\left[(1-Y_{k-1})(1-D_{k-1})(1-A_{k-1}) W_{k-1} \right]\\
    &&= E\left[(1-Y_{k-1})(1-D_{k-1})(1-A_k) W_k \right],
  \end{eqnarray*}
  where the last equality holds because \cite[also see Lemma 5 in][]{Young2020}
  \begin{eqnarray*}
    &&E[(1-Y_k)(1-D_k)(1-A_{k+1}) W_{k+1}]\nonumber\\
    &&=E\left[\dfrac{(1-Y_k)(1-D_k)(1-A_{k+1}) W_k}{\Pr(A_{k+1} = 0 \vert \overline{L}_k, Y_k = D_k = A_k = 0)}\right]\nonumber\\
    &&=\sum_{\overline{l}_k} E\left[\dfrac{(1-Y_k)(1-D_k)(1-A_{k+1}) W_k}{\Pr(A_{k+1} = 0 \vert \overline{L}_k = \overline{l}_k, Y_k = D_k = A_k = 0)}\middle\vert \overline{L}_k = \overline{l}_k, Y_k = D_k = A_k = 0 \right]\nonumber\\
    &&\qquad\qquad \times \Pr(\overline{L}_k = \overline{l}_k, \overline{Y}_k = \overline{D}_k = \overline{A}_k = 0)\nonumber\\[5pt]
    &&=\sum_{\overline{l}_k} \dfrac{\Pr(A_{k+1} = 0 \vert \overline{L}_k = \overline{l}_k, Y_k = D_k = A_k = 0)}{\Pr(A_{k+1} = 0 \vert \overline{L}_k = \overline{l}_k, Y_k = D_k = A_k = 0)} E\left[(1-Y_k)(1-D_k)(1-A_k) W_k\middle\vert \overline{L}_k = \overline{l}_k, Y_k = D_k = A_k = 0 \right]\nonumber\\
    &&\qquad\qquad \times \Pr(\overline{L}_k = \overline{l}_k, \overline{Y}_k = \overline{D}_k = \overline{A}_k = 0)\nonumber\\[5pt]
    &&=\sum_{\overline{l}_k} E\left[(1-Y_k)(1-D_k)(1-A_k) W_k\middle\vert \overline{L}_k = \overline{l}_k, Y_k = D_k = A_k = 0 \right]\nonumber\\[-10pt]
    &&\qquad\qquad \times  \Pr(\overline{L}_k = \overline{l}_k, \overline{Y}_k = \overline{D}_k = \overline{A}_k = \overline 0)\nonumber\\
    &&= E[(1-Y_k)(1-D_k)(1-A_k) W_k].
  \end{eqnarray*}
\end{enumerate}

Taken altogether, under assumptions~\eqref{app-seqexch}-\eqref{app-consistency} and~\eqref{app-seqexch2} we have
\begin{eqnarray*}
  h^{(1),\overline a = 0}_k &\equiv& \Pr(Y_k^{\overline a = 0} = 1 \vert Y_{k-1}^{\overline a = 0} = D_k^{\overline a = 0} = 0) = \dfrac{E\left[Y_k(1-D_k)(1-A_k)(1-Y_{k-1}) W_k \right]}{E\left[(1-Y_{k-1})(1-D_k)(1-A_k) W_k \right]}\\ 
  h^{(2),\overline a = 0}_k &\equiv& \Pr(D_k^{\overline a = 0} = 1 \vert Y_{k-1}^{\overline a = 0} = D_{k-1}^{\overline a = 0} = 0) = 1-\dfrac{E\left[(1-Y_{k-1})(1-D_k)(1-A_k) W_k \right]}{E\left[(1-Y_{k-1})(1-D_{k-1})(1-A_k) W_k \right]}\\
  &=& \dfrac{E\left[(1-Y_{k-1})(1-D_{k-1})(1-A_k) W_k \right] - E\left[(1-Y_{k-1})(1-D_k)(1-A_k) W_k \right]}{E\left[(1-Y_{k-1})(1-D_{k-1})(1-A_k) W_k \right]}\\
  &=& \dfrac{E\left[D_k(1-Y_{k-1})(1-D_{k-1})(1-A_k) W_k \right]}{E\left[(1-Y_{k-1})(1-D_{k-1})(1-A_k) W_k \right]}.
\end{eqnarray*}

\subsection{Single world intervention graphs \label{sec:app-swigs}}

Although sequential exchangeability conditions~\eqref{app-seqexch} and~\eqref{app-seqexch2} may be quite difficult to interpret, these conditional independencies involving counterfactuals can be graphically interrogated from single world intervention graphs (SWIGs) \citep{Richardson2013}. Intuitively, a SWIG can be thought of as a causal diagram that explicitly depicts a hypothetical world in which certain nodes have been intervened upon according a pre-specified hypothetical intervention. For more details about SWIGs and how to construct them, we refer the reader to \cite{Richardson2011,Richardson2013} or Appendix A in \cite{Young2020}, for a shorter and more targeted description in the context of competing and censoring events.

\begin{figure}[!htbp]
  \center
  \includegraphics{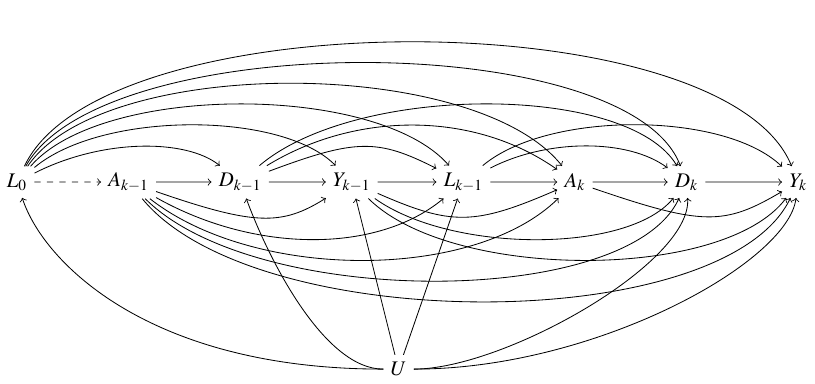}
  \caption{Causal diagram template (dagitty link: \url{http://dagitty.net/dags.html?id=DekFi_})\label{fig:dag}}
\end{figure}

\begin{figure}[!htbp]
  \center
  \includegraphics{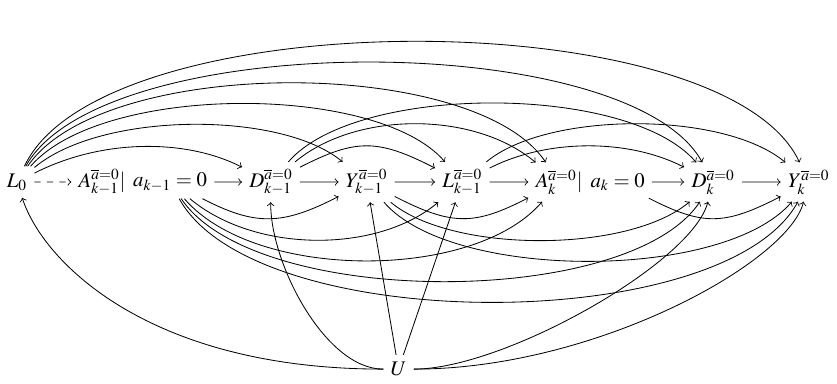}
  \caption{SWIG transformation of the causal diagram template under a hypothetical intervention that eliminates exposure (dagitty link: \url{http://dagitty.net/dags.html?id=aZSqjm})\label{fig:swig}}
\end{figure}

In this section, we specifically focus on a causal diagram that graphically represents a set of assumptions under which sequential exchangeability conditions~\eqref{app-seqexch} and~\eqref{app-seqexch2} hold, as can be evaluated from its corresponding SWIG obtained under a hypothetical intervention that eliminates exposure or, in other words, sets $\overline A_K$ to 0. Consider the causal diagram template depicted in Figure~\ref{fig:dag}. We refer to it as a template because it is meant as a simplification that only explicitly depicts two time waves $k-1$ and $k$, but that suffices for illustrative purposes. The node $U$ represents a (possibly vector-valued) set of unmeasured confounders of the relation between measured confounders $\overline L_k$, on the one hand, and hospital death and discharge, on the other hand. In time-to-event settings, $U$ is often conceived of an unobserved `frailty' component. The edge from $L_0$ to $A_{k-1}$ is dashed to imply the existence of previous time waves between baseline and interval $k-1$. Figure~\ref{fig:swig} depicts the SWIG template derived from Figure~\ref{fig:dag} obtained under a hypothetical intervention that eliminates exposure.

Under consistency~\eqref{app-consistency}, the causal diagram in Figure~\ref{fig:dag} implies sequential exchangeability~\eqref{app-seqexch} for any $k \in \{0,...,K\}$ by the absence of unblocked backdoor paths on the SWIG in Figure~\ref{fig:swig} between the observed value $A_k$ and future counterfactual outcomes $\underline Y^{\overline a = 0}_k$ conditional on $(\overline L_{k-1}, \overline D_{k-1}, A_{k-1} = Y_{k-1} = 0)$. Furthermore, under consistency~\eqref{app-consistency}, also sequential exchangeability~\eqref{app-seqexch2} is implied for any $k \in \{0,...,K\}$ by the absence of unblocked backdoor paths on the SWIG between the observed value $A_k$ and future counterfactual outcomes $\underline D^{\overline a = 0}_k$ conditional on $(\overline L_{k-1}, \overline Y_{k-1}, A_{k-1} = D_{k-1} = 0)$.

Importantly, this graphical representation illustrates that exchangeability conditions~\eqref{app-seqexch} and~\eqref{app-seqexch2} do not preclude unmeasured confounding of the relation between the outcome of interest and its competing event, nor of the relation between the outcome (competing event) at time interval $k-1$ and the outcome (competing event) at later intervals $\ge k$.

\section{Weighting-based characterization of proposed estimators\\ for the counterfactual risk $\varphi_K$}

\subsection{Treating exposure as an exclusion criterion}\label{sec:app-wcharest1}
A naive multistate model based approach ignores HAI onset and (implicitly) conflates the counterfactual risk $\varphi_K$ with
$$\Pr(T \le t_K, \epsilon = 1 \vert \tilde \epsilon \ne 0) = \Pr(\tilde T \le t_K, \tilde \epsilon = 1 \vert \tilde \epsilon \ne 0) = \Pr\left(\tilde Y_K = 1 \middle\vert A_{\tau} = 0\right) = \tilde\phi_K/(1-\alpha_{\tau}).$$ This quantity can, in turn, be estimated using the estimators $\widehat\phi_K$, $\widehat{\tilde\phi}_K$, $\widehat\phi^{\text{\tiny AJ}}_K$ and $\widehat{\tilde\phi}^{\text{\tiny AJ}}_K$ after exclusion of ever exposed patients (for whom $\tilde \epsilon = 0$ or $A_{\tau} = 1$) from the analysis.

For instance, define discrete-time event-specific hazards $h^{\star (1)}_k$ and $h^{\star (2)}_k$ as
\begin{eqnarray*}
  h^{\star (1)}_{k} &\equiv& \Pr\left[T \in (t_{k-1}, t_k], \epsilon = 1 \middle\vert \tilde\epsilon \ne 0, \left(T > t_k \cup \left\{T \in (t_{k-1}, t_k], \epsilon = 1\right\}\right) \right]\\
  &=& \Pr(Y_k = 1 \vert A_{\tau} = D_{k} = Y_{k-1} = 0) = \dfrac{E[Y_k(1-D_k)(1-Y_{k-1})(1-A_{\tau})]}{E[(1-D_{k})(1-Y_{k-1})(1-A_{\tau})]}\\
  h^{\star (2)}_{k} &\equiv& \Pr\left[T \in (t_{k-1}, t_k], \epsilon = 2 \middle\vert \tilde\epsilon \ne 0, T > t_{k-1} \right]\\
  &=& \Pr(D_k = 1 \vert A_{\tau} = Y_{k-1} = D_{k-1} = 0) = \dfrac{E[D_k(1-Y_{k-1})(1-D_{k-1})(1-A_{\tau})]}{E[(1-Y_{k-1})(1-D_{k-1})(1-A_{\tau})]}.
\end{eqnarray*}
The quantity $\tilde\phi_K/\left(1-\alpha_{\tau}\right)$ can then be estimated using the following modified Aalen-Johansen estimator
\begin{eqnarray*}
 \widehat{\varphi}^{\star \text{\tiny AJ}}_K &=& \sum_{k = 1}^K \widehat{h}^{\star (1)}_k (1-\widehat{h}^{\star (2)}_k) \prod_{s = 1}^{k-1} (1-\widehat{h}^{\star (1)}_s)(1-\widehat{h}^{\star (2)}_s).
\end{eqnarray*}
with $\widehat{h}^{\star (1)}_k$ and $\widehat{h}^{\star (2)}_k$ corresponding sample proportions.

In the absence of censoring due to loss to follow-up or administrative end of study, $\widehat{\varphi}^{\star \text{\tiny AJ}}_K$ reduces to a weighted empirical cumulative distribution function
\begin{eqnarray}
  &&\sum_{k = 1}^K \widehat{h}^{\star (1)}_k (1-\widehat{h}^{\star (2)}_k) \prod_{s = 1}^{k-1} (1-\widehat{h}^{\star (1)}_s)(1-\widehat{h}^{\star (2)}_s)\nonumber\\
&=& \sum_{k = 1}^K \dfrac{E_n[Y_k(1-D_k)(1-Y_{k-1})(1-A_{\tau})]}{E_n[(1-D_k)(1-Y_{k-1})(1-A_{\tau})]} \dfrac{E_n[(1-D_k)(1-Y_{k-1})(1-D_{k-1})(1-A_{\tau})]}{E_n[(1-Y_{k-1})(1-D_{k-1})(1-A_{\tau})]}\nonumber\\
&&\qquad\qquad \times \prod_{s = 1}^{k-1} \dfrac{E_n[(1 - Y_s)(1 - D_s)(1 - Y_{s-1})(1-A_{\tau})]}{E_n[(1-D_s)(1-Y_{s-1})(1-A_{\tau})]} \dfrac{E_n[(1-D_s)(1 - Y_{s-1})(1 - D_{s-1})(1-A_{\tau})]}{E_n[(1-Y_{s-1})(1-D_{s-1})(1-A_{\tau})]}\nonumber\\
&=& \sum_{k = 1}^K \dfrac{E_n[Y_k(1-D_k)(1-Y_{k-1})(1-A_{\tau})]}{E_n[(1-Y_{k-1})(1-D_{k-1})(1-A_{\tau})]} \prod_{s = 1}^{k-1} \dfrac{E_n[(1 - Y_s)(1 - D_s)(1-A_{\tau})]}{E_n[(1-Y_{s-1})(1-D_{s-1})(1-A_{\tau})]}\nonumber\\
&=& \sum_{k = 1}^K \dfrac{E_n[Y_k(1-Y_{k-1})(1-D_k)(1-A_{\tau})]}{E_n[(1-Y_0)(1-D_0)(1-A_{\tau})]} = \sum_{k = 1}^K \dfrac{E_n[Y_k(1-Y_{k-1})(1-D_k)(1-A_{\tau})]}{P_n(A_{\tau}=0)}\nonumber\\
&=& \sum_{k = 1}^K E_n\left[Y_k(1-Y_{k-1})(1-D_k)(1-A_k)\widehat{W}^{\star}\right] = n^{-1}\sum_{k = 1}^K \tilde d^{w}_{1k} = \widehat{\varphi}^{\star}_K,\nonumber
\end{eqnarray}
with $\tilde d^{w}_{1k} \equiv \sum_{i=1}^n Y_{ik}(1-D_{ik})(1-A_{ik})(1-Y_{i,k-1})\widehat W^{\star}_i$ denoting the weighted number of HAI-free hospital deaths in interval $k$ and
\begin{eqnarray*}
  \widehat{W}^{\star}_i \equiv \widehat{W}^{\star} &\equiv& \dfrac{1}{1-P_n(\tilde \epsilon = 0)} = \dfrac{1}{P_n(A_{\tau} = 0)} = 1 + \dfrac{P_n(A_{\tau} = 1)}{P_n(A_{\tau} = 0)}.
\end{eqnarray*}
The fifth equality holds because, for each $k \le \tau$, we have
\begin{eqnarray*}
  Y_k(1-A_{\tau}) &=& \begin{cases}
  1 & \text{iff}\ Y_k(1-A_k) = 1; \\
  0 & \text{otherwise}.
  \end{cases}
\end{eqnarray*}
This holds because
\begin{enumerate}
  \item $Y_k(1-A_k) = 1$ iff $Y_k = 1$ and $A_k = 0$, in which case we have $\underline A_{k+1} = \underline 0$, and so $Y_k(1-A_{\tau}) = 1$, due to the deterministic relation between the competing events.
  \item $Y_k(1-A_{\tau}) = 1$ iff $Y_k = 1$ and $A_{\tau} = 0$, in which case we have $A_k = 0$, and so $Y_k(1-A_k) = 1$, due to the fact that by definition, for any $k \le \tau$, we have $A_k \le A_{\tau}$.
\end{enumerate}

Note that, even though a subscript $i$ is used in the summation, the weights $W^{\star}_{i}$ are not subject-dependent, as clarified by their definition $W^{\star}_i \equiv W^{\star}$. The estimator $\widehat{\varphi}^{\star}_K$ is again a weighted version of the empirical cumulative distribution of infection-free hospital death $\widehat{\tilde\phi}_K$ with weights $\widehat{W}^{\star}_i$ equal to the inverse probability of remaining without HAI until hospital death or discharge (or, equivalently, being included in the final sample for this naive analysis), but also a weighted version of the empirical cumulative distribution of hospital death $\widehat{\phi}_K$ with weights $(1-A_{ik}) \widehat{W}^{\star}_i$.

\subsection{Treating exposure onset as a time-dependent exclusion criterion}\label{sec:app-wcharest2}
\begin{figure}[!htbp]
\centerline{\includegraphics[width=300pt]{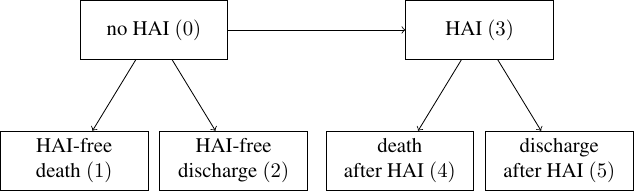}}
\caption{The extended (or progressive) illness death model.\label{fig:extillnessdeath}}
\end{figure}

\cite{Schumacher2007} describe a method to estimate the counterfactual risk $\varphi_K$ from an extended or progressive illness death model, as depicted in Figure~\ref{fig:extillnessdeath}.

Let $P_{ij}(s,t)$ denote the transition probability $\Pr(X_t = j\vert X_s = i)$ for $s \le t$, with $X_s$ the observed state at time $s$ and $X_t$ the observed state at time $t$. In the considered setting, we have that $\Pr(X_0 = 0) = 1$, such that, for each pair $(j,t)$, $P_{0j}(0,t)$ corresponds to a so-called state occupation probability (or cumulative incidence) $\Pr(X_t = j)$. Schumacher and colleagues target the following quantity
\begin{eqnarray*}
   \dfrac{P_{01}(0,t)}{P_{00}(0,t)+P_{01}(0,t)+P_{02}(0,t)} = \dfrac{P_{01}(0,t)}{1-P_{03}(0,t)-P_{04}(0,t)-P_{05}(0,t)},
\end{eqnarray*}
which, given that
\begin{eqnarray*}
  \Pr(\tilde T > t) &=& P_{00}(0,t)\\
  \Pr(\tilde T \le t, \tilde \epsilon = 1) &=& P_{01}(0,t)\\
  \Pr(\tilde T \le t, \tilde \epsilon = 2) &=& P_{02}(0,t)\\
  \Pr(\tilde T \le t, \tilde \epsilon = 0) &=& 1-P_{00}(0,t)-P_{01}(0,t)-P_{02}(0,t),
\end{eqnarray*}
equals
\begin{eqnarray*}
  \dfrac{\Pr(\tilde T \le t, \tilde \epsilon = 1)}{1-\Pr(\tilde T \le t, \tilde \epsilon = 0)} = \dfrac{\Pr(\tilde T \le t, \tilde \epsilon = 1)}{\Pr\left[\tilde T > t \cup (\tilde T \le t, \tilde \epsilon \ne 0) \right]} = \Pr\left[\tilde T \le t, \tilde \epsilon = 1 \middle\vert \tilde T > t \cup (\tilde T \le t, \tilde \epsilon \ne 0) \right].
\end{eqnarray*}
In other words, interpreting estimates of this quantity targeted by Schumacher and colleagues implicitly corresponds to conflating $\varphi_K$ with
$$\Pr\left[\tilde T \le t_K, \tilde \epsilon = 1 \middle\vert \tilde T > t_K \cup (\tilde T \le t_K, \tilde \epsilon \ne 0)\right]  = \Pr\left(\tilde Y_K = 1 \middle\vert A_K = 0\right) = \tilde\phi_K/(1-\alpha_K).$$

The relevant state occupation probabilities for estimating this targeted observable quantity can also be obtained from a simpler competing risk model with three absorbing states (and no transient states), as depicted in Figure~\ref{fig:competingrisk2}, which is obtained by collapsing HAI onset and subsequent hospital death or discharge into a single absorbing state.
This corresponds to the competing risk model with observed event time $\tilde T$ and event types $\tilde \epsilon \in \{0,1,2\}$, discussed in the main text.
Let $P^{\prime}_{0k}(0,t)$ denote the state occupation probability $\Pr(X'_t = k\vert X'_0 = 0)$ for $0 \le t$, with $X'_t$ the observed state at time $t$, as defined in the competing risk model depicted in Figure~\ref{fig:competingrisk2}, under the careful consideration that event types $\tilde \epsilon = 1$ and $\tilde \epsilon = 2$ correspond to states 1 and 2, respectively, but event type $\tilde \epsilon = 0$ corresponds to state 3.
The targeted quantity can then be expressed as
\begin{eqnarray*}
   \dfrac{P^{\prime}_{01}(0,t)}{P^{\prime}_{00}(0,t)+P^{\prime}_{01}(0,t)+P^{\prime}_{02}(0,t)} = \dfrac{P^{\prime}_{01}(0,t)}{1-P^{\prime}_{03}(0,t)}.
\end{eqnarray*}

In other words, each probability (the cumulative risk of HAI-free death and the cumulative risk of HAI onset, respectively) in the targeted functional
\begin{eqnarray*}
  \dfrac{\Pr(\tilde T \le t_K, \tilde \epsilon = 1)}{1-\Pr(\tilde T \le t_K, \tilde \epsilon = 0)} = \dfrac{\tilde\phi_K}{1-\alpha_K}
\end{eqnarray*}
can be estimated by their corresponding Aalen-Johansen estimator, such that the estimator proposed by Schumacher and colleagues can be expressed as
\begin{eqnarray*}
  \dfrac{\widehat{\tilde\phi}^{\text{\tiny AJ}}_K}{\left(1-\widehat{\alpha}^{\text{\tiny AJ}}_K\right)}
\end{eqnarray*}
with
\begin{eqnarray*}
  \widehat{\alpha}^{\text{\tiny AJ}}_K = \sum_{k = 1}^K \widehat{\tilde h}^{(0)}_k \prod_{s = 1}^{k-1} (1-\widehat{\tilde h}^{(0)}_s)(1-\widehat{\tilde h}^{(1)}_s)(1-\widehat{\tilde h}^{(2)}_s).
\end{eqnarray*}

Alternatively, since the quantity $\tilde\phi_K/(1-\alpha_K)$ targeted by Schumacher and colleagues equals the risk of infection-free hospital death in patients that have not acquired infection by interval $K$, $\widehat{\tilde\phi}^{\text{\tiny AJ}}_K/\left(1-\widehat{\alpha}^{\text{\tiny AJ}}_K\right)$ can be shown to be algebraically equivalent with $\widehat\phi^{\text{\tiny AJ}}_K$ and $\widehat{\tilde\phi}^{\text{\tiny AJ}}_K$ after exclusion of patients infected by the end of interval $K$ (for whom $T \le t_K, \epsilon = 0$ or $A_K = 1$) from the analytical sample (or equivalently, application of weights $1-A_{iK}$ for each patient $i$ at each interval $k$).

For instance, define $h^{\dagger (1)}_{k,K}$ and $h^{\dagger (2)}_{k,K}$ as
\begin{eqnarray*}
  h^{\dagger (1)}_{k,K} &\equiv& \Pr\left[T \in (t_{k-1}, t_k], \epsilon = 1 \middle\vert \left(T > t_k \cup \left\{T \in (t_{k-1}, t_k], \epsilon = 1\right\}\right), \left( \tilde T > t_K \cup \left\{\tilde T \le t_K, \tilde\epsilon \ne 0 \right\}\right) \right]\\
  &=& \Pr(Y_k = 1 \vert A_K = D_{k} = Y_{k-1} = 0) = \dfrac{E[Y_k(1-D_k)(1-Y_{k-1})(1-A_K)]}{E[(1-D_{k})(1-Y_{k-1})(1-A_K)]}\\
  h^{\dagger (2)}_{k,K} &\equiv& \Pr\left[T \in (t_{k-1}, t_k], \epsilon = 2 \middle\vert T > t_{k-1}, \left( \tilde T > t_K \cup \left\{\tilde T \le t_K, \tilde\epsilon \ne 0 \right\}\right) \right]\\
  &=& \Pr(D_k = 1 \vert A_K = Y_{k-1} = D_{k-1} = 0) = \dfrac{E[D_k(1-Y_{k-1})(1-D_{k-1})(1-A_K)]}{E[(1-Y_{k-1})(1-D_{k-1})(1-A_K)]}.
\end{eqnarray*}
The quantity $\tilde\phi_K/\left(1-\alpha_K\right)$ can then be estimated using the following modified Aalen-Johansen estimator
\begin{eqnarray*}
 \widehat{\varphi}^{\dagger \text{\tiny AJ}}_K &=& \sum_{k = 1}^K \widehat{h}^{\dagger (1)}_{k,K} (1-\widehat{h}^{\dagger (2)}_{k,K}) \prod_{s = 1}^{k-1} (1-\widehat{h}^{\dagger (1)}_{s,K})(1-\widehat{h}^{\dagger (2)}_{s,K})
\end{eqnarray*}
with $\widehat{h}^{\dagger (1)}_{k,K}$ and $\widehat{h}^{\dagger (2)}_{k,K}$ corresponding sample proportions.

In the absence of censoring due to loss to follow-up or administrative end of study, $\widehat{\varphi}^{\dagger \text{\tiny AJ}}_K$ reduces to a weighted empirical cumulative distribution function
\begin{eqnarray}
  &&\sum_{k = 1}^K \widehat{h}^{\dagger (1)}_{k,K} (1-\widehat{h}^{\dagger (2)}_{k,K}) \prod_{s = 1}^{k-1} (1-\widehat{h}^{\dagger (1)}_{s,K})(1-\widehat{h}^{\dagger (2)}_{s,K})\nonumber\\
  &=& \sum_{k = 1}^K \dfrac{E_n[Y_k(1-D_k)(1-Y_{k-1})(1-A_K)]}{E_n[(1-D_k)(1-Y_{k-1})(1-A_K)]} \dfrac{E_n[(1-D_k)(1-Y_{k-1})(1-D_{k-1})(1-A_K)]}{E_n[(1-Y_{k-1})(1-D_{k-1})(1-A_K)]}\nonumber\\
  &&\qquad\qquad \times \prod_{s = 1}^{k-1} \dfrac{E_n[(1 - Y_s)(1 - D_s)(1 - Y_{s-1})(1-A_K)]}{E_n[(1-D_s)(1-Y_{s-1})(1-A_K)]} \dfrac{E_n[(1-D_s)(1 - Y_{s-1})(1 - D_{s-1})(1-A_K)]}{E_n[(1-Y_{s-1})(1-D_{s-1})(1-A_K)]}\nonumber\\
  &=& \sum_{k = 1}^K \dfrac{E_n[Y_k(1-D_k)(1-Y_{k-1})(1-A_K)]}{E_n[(1-Y_{k-1})(1-D_{k-1})(1-A_K)]} \prod_{s = 1}^{k-1} \dfrac{E_n[(1 - Y_s)(1 - D_s)(1-A_K)]}{E_n[(1-Y_{s-1})(1-D_{s-1})(1-A_K)]}\nonumber\\
  &=& \sum_{k = 1}^K \dfrac{E_n[Y_k(1-Y_{k-1})(1-D_k)(1-A_K)]}{E_n[(1-Y_0)(1-D_0)(1-A_K)]} = \sum_{k = 1}^K \dfrac{E_n[Y_k(1-Y_{k-1})(1-D_k)(1-A_K)]}{P_n(A_K=0)}\nonumber\\
  &=& \sum_{k = 1}^K E_n\left[Y_k(1-Y_{k-1})(1-D_k)(1-A_k)\widehat{W}^{\dagger}\right] = n^{-1}\sum_{k = 1}^K \tilde d^{w_K}_{1k} = \widehat{\varphi}^{\dagger}_K,\nonumber
\end{eqnarray}
with $\tilde d^{w_K}_{1k} \equiv \sum_{i=1}^n Y_{ik}(1-D_{ik})(1-A_{ik})(1-Y_{i,k-1})\widehat W^{\dagger}_{iK}$ denoting the weighted number of HAI-free hospital deaths in interval $k$ and
\begin{eqnarray*}
\widehat{W}^{\dagger}_{iK} \equiv \widehat{W}^{\dagger}_K &\equiv& \dfrac{1}{1-P_n(\tilde T \le t_K, \tilde \epsilon = 0)} = \dfrac{1}{P_n(A_K = 0)} = 1 + \dfrac{P_n(A_K = 1)}{P_n(A_K = 0)}.
\end{eqnarray*}
The fifth equality holds because, for each $k \le K$, we have
\begin{eqnarray*}
  Y_k(1-A_K) &=& \begin{cases}
  1 & \text{iff}\ Y_k(1-A_k) = 1; \\
  0 & \text{otherwise}.
  \end{cases}
\end{eqnarray*}
This holds because
\begin{enumerate}
  \item $Y_k(1-A_k) = 1$ iff $Y_k = 1$ and $A_k = 0$, in which case we have $\underline A_{k+1} = 0$, and so $Y_k(1-A_K) = 1$, due to the deterministic relation between the competing events.
  \item $Y_k(1-A_K) = 1$ iff $Y_k = 1$ and $A_K = 0$, in which case we have $A_k = 0$, and so $Y_k(1-A_k) = 1$, due to the fact that by definition, for any $k \le K$, we have $A_k \le A_K$.
\end{enumerate}

Note again that, even though a subscript $i$ is used in the summation, the weights $W^{\dagger}_{iK}$ are not subject-dependent, as clarified by their definition $W^{\dagger}_{iK} \equiv W^{\dagger}_K$. Also important to note is that events (in interval $k$) are not weighed according to a function that depends on the time of the event, but according to a function that depends on the landmark interval $K$ at which the cumulative incidence function is evaluated. This implies that the same event is weighed differently according to the chosen landmark.
The estimator $\widehat{\varphi}^{\dagger}_K$ is again a weighted version of the empirical cumulative distribution of infection-free hospital death $\widehat{\tilde\phi}_K$ with weights $\widehat{W}^{\dagger}_{iK}$ equal to the inverse probability of remaining without HAI up to (and including) interval $K$ (irrespective of hospitalization status), but also a weighted version of the empirical cumulative distribution of hospital death $\widehat{\phi}_K$ with weights $(1-A_{ik}) \widehat{W}^{\dagger}_{iK}$.

Finally, given that in the absence of censoring due to loss to follow-up or administrative end of study, $\widehat{\tilde\phi}^{\text{\tiny AJ}}_K = \widehat{\tilde\phi}_K$ and that, similarly, it can be shown that $\widehat{\alpha}^{\text{\tiny AJ}}_K = \widehat{\alpha}_K$, with
\begin{eqnarray*}
  \widehat{\alpha}_K &=& \sum_{k=1}^K E_n[A_k(1-Y_{k-1})(1-D_{k-1})(1-A_{k-1})] = P_n(A_K = 1),
\end{eqnarray*}
it is then trivial to see that Schumacher and colleagues' estimator $\widehat{\tilde\phi}^{\text{\tiny AJ}}_K/\left(1-\widehat{\alpha}^{\text{\tiny AJ}}_K\right)$ is algebraically equivalent to both $\widehat{\varphi}^{\dagger}_K$ and $\widehat{\varphi}^{\dagger \text{\tiny AJ}}_K$.

\subsubsection{Implicit hypothetical intervention when treating exposure onset as a (time-dependent) exclusion criterion}

Upon noting that $h^{\dagger (1)}_{k,K}$ can be expressed as a weighted function of $\tilde h^{(1)}_k$
\begin{eqnarray*}
  h^{\dagger (1)}_{k,K} &&\equiv \Pr(Y_k = 1 \vert A_K = D_k = Y_{k-1} = 0)\\
  &&= \dfrac{\Pr(Y_k = 1, Y_{k-1} = D_k = A_k = 0, A_K = 0)}{\Pr(Y_{k-1} = D_k = A_k = 0, A_K = 0)}\\
  &&= \dfrac{\Pr(A_K = 0 \vert Y_k = 1, Y_{k-1} = D_k = A_k = 0)}{\Pr(A_K = 0 \vert Y_{k-1} = D_k = A_k = 0)}
  \dfrac{\Pr(Y_k = 1, Y_{k-1} = D_k = A_k = 0)}{\Pr(Y_{k-1} = D_k = A_k = 0)}\\
  &&= \dfrac{\Pr(A_K = 0 \vert Y_k = 1, Y_{k-1} = D_k = A_k = 0)}{\Pr(A_K = 0 \vert Y_{k-1} = D_k = A_k = 0) } \tilde h^{(1)}_k = c^{\dagger (1)}_{k,K} \tilde h^{(1)}_k,
\end{eqnarray*}
with expansion factor
\begin{eqnarray*}
  c^{\dagger (1)}_{k,K} \equiv \dfrac{1}{\Pr(A_K = 0 \vert D_k = A_k = Y_{k-1} = 0)},
\end{eqnarray*}
and that, similarly, $h^{\dagger (2)}_{k,K}$ can be expressed as a weighted function of $\tilde h^{(2)}_k$
\begin{eqnarray*}
  h^{\dagger (2)}_{k,K} &&= \Pr(D_k = 1 \vert A_K = Y_{k-1} = D_{k-1} = 0) \\
  &&= \dfrac{\Pr(D_k = 1, A_k = Y_{k-1} = D_{k-1} = 0, A_K = 0)}{\Pr(A_k = Y_{k-1} = D_{k-1} = 0, A_K = 0)}\\
  &&= \dfrac{\Pr(A_K = 0 \vert D_k = 1, A_k = Y_{k-1} = D_{k-1} = 0)}{\Pr(A_K = 0 \vert A_k = Y_{k-1} = D_{k-1} = 0)}
  \dfrac{\Pr(D_k = 1, A_k = Y_{k-1} = D_{k-1} = 0)}{\Pr(A_k = Y_{k-1} = D_{k-1} = 0)}\\
  &&= \dfrac{\Pr(A_K = 0 \vert D_k = 1, A_k = Y_{k-1} = D_{k-1} = 0)}{\Pr(A_K = 0 \vert A_k = Y_{k-1} = D_{k-1} = 0)} \tilde h^{(2)}_k = c^{\dagger (2)}_{k,K} \tilde h^{(2)}_k,
\end{eqnarray*}
with expansion factor
\begin{eqnarray*}
  c^{\dagger (2)}_{k,K} \equiv \dfrac{1}{\Pr(A_K = 0 \vert A_k = Y_{k-1} = D_{k-1} = 0)},
\end{eqnarray*}
$\widehat{\varphi}^{\dagger \text{\tiny AJ}}_K$ can alternatively be expressed as
\begin{eqnarray*}
  \widehat\varphi_K^{\dagger \text{\tiny AJ}} &&= \sum_{k=1}^K \widehat{h}^{\dagger (1)}_{k,K} \left(1 - \widehat{h}^{\dagger (2)}_{k,K}\right) \prod_{s=1}^{k-1} \left(1 - \widehat{h}^{\dagger (1)}_{s,K}\right) \left(1-\widehat{h}^{\dagger (2)}_{s,K}\right)\\
  &&= \sum_{k=1}^K \widehat{c}^{\dagger (1)}_{k,K} \widehat{\tilde h}^{(1)}_k \left(1 - \widehat{c}^{\dagger (2)}_{k,K} \widehat{\tilde h}^{(2)}_k \right) \prod_{s=1}^{k-1} \left(1 - \widehat{c}^{\dagger (1)}_{s,K} \widehat{\tilde h}^{(1)}_s \right) \left(1 - \widehat{c}^{\dagger (2)}_{s,K} \widehat{\tilde h}^{(2)}_s \right)\\
  &&= \sum_{k=1}^K \widehat{c}^{\dagger (1)}_{k,K} \widehat{\tilde h}^{(1)}_k \left(1 - 0 \widehat{\tilde h}^{(0)}_k \right) \left(1 - \widehat{c}^{\dagger (2)}_{k,K} \widehat{\tilde h}^{(2)}_k \right) \prod_{s=1}^{k-1} \left(1 - 0 \widehat{\tilde h}^{(0)}_s \right) \left(1 - \widehat{c}^{\dagger (1)}_{s,K} \widehat{\tilde h}^{(1)}_s \right) \left(1 - \widehat{c}^{\dagger (2)}_{s,K} \widehat{\tilde h}^{(2)}_s \right).
\end{eqnarray*}
This illustrates that, from a counterfactual point of view, the estimation approach of Schumacher and colleagues envisages a hypothetical intervention that sets the hazard of exposure or infection to zero, but that also increases the hazard of both in-hospital death and discharge in interval $k \le K$ based on information available by interval $K$, thereby somehow allowing patients to die or get discharged multiple times (as if these were recurrent events).

Furthermore, upon noting that
\begin{eqnarray*}
  h^{\star (1)}_{k} &&= h^{\dagger (1)}_{k,\tau}\\
  h^{\star (2)}_{k} &&= h^{\dagger (2)}_{k,\tau},
\end{eqnarray*}
it is clear that the more naive estimator $\widehat\varphi_K^{\star \text{\tiny AJ}}$ that treats exposure onset as an exclusion criterion can be rewritten as
\begin{eqnarray*}
  \widehat\varphi_K^{\star \text{\tiny AJ}} &&= \sum_{k=1}^K \widehat{h}^{\star (1)}_{k} \left(1 - \widehat{h}^{\star (2)}_{k}\right) \prod_{s=1}^{k-1} \left(1 - \widehat{h}^{\star (1)}_{s}\right) \left(1-\widehat{h}^{\star (2)}_{s}\right)\\
  &&= \sum_{k=1}^K \widehat{h}^{\dagger (1)}_{k,\tau} \left(1 - \widehat{h}^{\dagger (2)}_{k,\tau}\right) \prod_{s=1}^{k-1} \left(1 - \widehat{h}^{\dagger (1)}_{s,\tau}\right) \left(1-\widehat{h}^{\dagger (2)}_{s,\tau}\right)\\
  &&= \sum_{k=1}^K \widehat{c}^{\dagger (1)}_{k,\tau} \widehat{\tilde h}^{(1)}_k \left(1 - \widehat{c}^{\dagger (2)}_{k,\tau} \widehat{\tilde h}^{(2)}_k \right) \prod_{s=1}^{k-1} \left(1 - \widehat{c}^{\dagger (1)}_{s,\tau} \widehat{\tilde h}^{(1)}_s \right) \left(1 - \widehat{c}^{\dagger (2)}_{s,\tau} \widehat{\tilde h}^{(2)}_s \right)\\
  &&= \sum_{k=1}^K \widehat{c}^{\dagger (1)}_{k,\tau} \widehat{\tilde h}^{(1)}_k \left(1 - 0 \widehat{\tilde h}^{(0)}_k \right) \left(1 - \widehat{c}^{\dagger (2)}_{k,\tau} \widehat{\tilde h}^{(2)}_k \right) \prod_{s=1}^{k-1} \left(1 - 0 \widehat{\tilde h}^{(0)}_s \right) \left(1 - \widehat{c}^{\dagger (1)}_{s,\tau} \widehat{\tilde h}^{(1)}_s \right) \left(1 - \widehat{c}^{\dagger (2)}_{s,\tau} \widehat{\tilde h}^{(2)}_s \right),
\end{eqnarray*}
illustrating that, from a counterfactual point of view, also $\widehat\varphi_K^{\star \text{\tiny AJ}}$ envisages a hypothetical intervention that sets the hazard of exposure or infection to zero, but that increases the hazard of both in-hospital death and discharge in interval $k \le K$ based on information that is only available by the maximal follow-up interval $\tau$ (instead of landmark interval $K$), thereby also somehow treating death and discharge as recurrent events. The main difference, is that the weighting or expansion factor that determines the extent to which deceased or discharged patients may artificially `re-enter' the risk set, is not determined by the landmark interval $K$ for which $\varphi_K$ is estimated, but by the maximal follow-up interval $\tau$.

Finally, note that the notion of an extended risk set is also characteristic for the subdistribution hazard $\tilde h^{\text{sd}}_k$ (defined in Appendix~\ref{sec:app-estobsdata2}). However, a main difference is that the extended risk set of the subdistribution hazard implies $\tilde h^{\text{sd}}_k = c^{\text{sd}}_k \tilde h^{(1)}_k \le \tilde h^{(1)}_k$, while the extended risk set of the hazard-like quantity $h^{\dagger (1)}_{k,K}$ implies $h^{\dagger (1)}_{k,K} = c^{\dagger (1)}_{k,K} \tilde h^{(1)}_k \ge \tilde h^{(1)}_k$. In other words, the subdistribution hazard $\tilde h^{\text{sd}}_k$ is smaller than the corresponding event-specific hazard $\tilde h^{(1)}_k$ because the subdistribution hazard risk set includes patients that are technically no longer at risk for event $\tilde\epsilon = 1$ because they have experienced a competing event. This reduction is expressed by the reduction factor $c^{\text{sd}}_k$ (defined in Appendix~\ref{sec:app-estobsdata2}). In contrast, the extended risk set of $h^{\dagger (1)}_{k,K}$ for $k < K$ somehow artificially considers patients that are (in reality) no longer at risk for event $\tilde\epsilon = 1$ in interval $K$ because they have already experienced \emph{this} event (instead of a \emph{competing} event) in interval $k$ to remain at risk. Conceptually, it seems as if, right before their death in interval $k$, the hypothetical intervention creates a parallel world in which some patients can time travel to interval $K$ and then go back in time, to interval $k$, to experience the event in this parallel world. In the calculation of the hazard-like quantity $h^{\dagger (1)}_{k,K}$, these parallel worlds are then also taken into account. Arguably, this type of intervention may be hard to conceive, to say the least.

\subsection{Treating exposure onset as (marginally) independent censoring}\label{sec:app-wcharest3}

As noted in Section~\ref{sec:main-timedepexp-cens} in the main text, censoring patients that have gotten infected at their infection onset time\footnote{Under the aforementioned assumed temporal order of event types within an interval $k$, this implies that throughout, we assume artificial censoring times (infection onset times) to occur right before possibly tied event times (infection-free hospital death or discharge times) within interval $k$.}, leads to the following Aalen-Johansen estimator, which is equivalent to $\widehat\phi^{\text{\tiny AJ}}_K$ upon excluding patients that have gotten infected by the end of interval $k$ (for whom $T \le t_k, \epsilon = 0$ or $A_k = 1$) from the risk set of interval $k$ for each $k = 1, ..., K$:
\begin{eqnarray*}
  \widehat{\varphi}^{\circ \text{\tiny AJ}}_K &=&\sum_{k = 1}^K \widehat{\tilde h}^{(1)}_k (1-\widehat{\tilde h}^{(2)}_k) \prod_{s = 1}^{k-1} (1-\widehat{\tilde h}^{(1)}_s)(1-\widehat{\tilde h}^{(2)}_s).
\end{eqnarray*}
with $\widehat{\tilde h}^{(1)}_k$ and $\widehat{\tilde h}^{(2)}_k$ corresponding sample proportions.
In the absence of censoring due to loss to follow-up or administrative end of study $\widehat{\varphi}^{\circ \text{\tiny AJ}}_K$ again reduces to a weighted empirical cumulative distribution function
\begin{eqnarray*}
  \widehat{\varphi}^{\circ \text{\tiny AJ}}_K &&= \sum_{k = 1}^K \dfrac{E_n[Y_k(1-D_k)(1-Y_{k-1})(1-A_k)]}{E_n[(1-D_k)(1-Y_{k-1})(1-A_k)]} \dfrac{E_n[(1-D_k)(1-Y_{k-1})(1-D_{k-1})(1-A_k)]}{E_n[(1-Y_{k-1})(1-D_{k-1})(1-A_k)]}\\
  &&\qquad\qquad \times \prod_{s = 1}^{k-1} \dfrac{E_n[(1 - Y_s)(1 - D_s)(1 - Y_{s-1})(1-A_s)]}{E_n[(1-D_s)(1-Y_{s-1})(1-A_s)]} \dfrac{E_n[(1-D_s)(1 - Y_{s-1})(1 - D_{s-1})(1-A_s)]}{E_n[(1-Y_{s-1})(1-D_{s-1})(1-A_s)]}\\
  &&= \sum_{k = 1}^K \dfrac{E_n[Y_k(1-Y_{k-1})(1-D_k)(1-A_k)]}{E_n[(1-Y_{k-1})(1-D_{k-1})(1-A_k)]} \prod_{s = 1}^{k-1} \dfrac{E_n[(1-Y_s)(1-D_s)(1-A_s)]}{E_n[(1-Y_{s-1})(1-D_{s-1})(1-A_s)]}\nonumber\\
  &&= \sum_{k = 1}^K E_n[Y_k(1-Y_{k-1})(1-D_k)(1-A_k)] \prod_{s = 1}^{k} \dfrac{E_n[(1-Y_{s-1})(1-D_{s-1})(1-A_{s-1})]}{E_n[(1-Y_{s-1})(1-D_{s-1})(1-A_s)]}\nonumber\\
  &&= \sum_{k = 1}^K \dfrac{E_n[Y_k(1-Y_{k-1})(1-D_k)(1-A_k)]}{\prod_{s = 1}^{k} \widehat{\Pr}(A_s=0 | Y_{s-1}=D_{s-1}=A_{s-1} = 0)}\nonumber\\
  &&= \sum_{k = 1}^K E_n[Y_k(1-Y_{k-1})(1-D_k)(1-A_k)\widehat W^{\circ}_k] = n^{-1} \sum_{k = 1}^K \tilde d^{w_k}_{1k} = \widehat{\varphi}^{\circ}_K,\nonumber
\end{eqnarray*}
with $\tilde d^{w_k}_{1k} \equiv \sum_{i=1}^n Y_{ik}(1-D_{ik})(1-A_{ik})(1-Y_{i,k-1})\widehat W^{\circ}_{ik}$ denoting the weighted number of HAI-free hospital deaths in interval $k$ and
\begin{eqnarray*}
  \widehat W^{\circ}_{ik} \equiv \widehat W^{\circ}_k &\equiv& \prod_{s = 1}^k \dfrac{1}{1-\widehat{\Pr}(\tilde T \in (t_{s-1}, t_s], \tilde \epsilon = 0 \vert \tilde T > t_{s-1})} = \prod_{s = 1}^k \dfrac{1}{\widehat{\Pr}(A_s=0 | Y_{s-1} = D_{s-1} = A_{s-1} = 0)}\nonumber\\
  &=& 1+\sum_{s = 1}^k \dfrac{\widehat{\Pr}(A_s =1 | Y_{s-1} = D_{s-1} = A_{s-1} = 0)}{\widehat{\Pr}(A_{s} =0 | Y_{s-1} = D_{s-1} = A_{s-1} = 0)} \prod_{s' = 1}^{s-1} \left\{ 1+ \dfrac{\widehat{\Pr}(A_{s'} =1 | Y_{s'-1} = D_{s'-1} = A_{s'-1} = 0)}{\widehat{\Pr}(A_{s'} =0 | Y_{s'-1} = D_{s'-1} = A_{s'-1} = 0)} \right\}.
\end{eqnarray*}
The third equality follows from the fact that we have $A_0 \equiv D_0 \equiv Y_0 \equiv 0$. The fourth equality only holds if for each $s = 1,...,k$, we have
\begin{eqnarray*}
\widehat{\Pr}(A_s=0 | Y_{s-1}=D_{s-1}=A_{s-1} = 0) &=& P_n(A_s=0 | Y_{s-1}=D_{s-1}=A_{s-1} = 0)\\
&=& \dfrac{E_n[(1-Y_{s-1})(1-D_{s-1})(1-A_s)]}{E_n[(1-Y_{s-1})(1-D_{s-1})(1-A_{s-1})]},
\end{eqnarray*}
with $P_n(X=x \vert Z=z)$ as defined in Appendix~\ref{sec:app-estobsdata1}, which implies non-parametric estimation of the weights $W^{\circ}_k$. The last equality in the expression of $\widehat W^{\circ}_k$ is proven in Appendix~\ref{sec:app-decompw}. Note that even though a subscript $i$ is used in the summation, the weights $W^{\circ}_{ik}$ are not subject-dependent, as clarified by their definition $W^{\circ}_{ik} \equiv W^{\circ}_k$.

The algebraic equivalence of $\widehat{\varphi}^{\circ \text{\tiny AJ}}_K$ and $\widehat{\varphi}^{\circ}_K$ (under non-parametric estimation of the weights $W^{\circ}_k$) illustrates that, in accordance with \cite{Satten2001}, in the presence of censoring, the Aalen-Johansen estimator for the cumulative incidence function corresponds to an inverse probability of censoring (IPC) weighted empirical cumulative distribution function with weights $\widehat{W}^{\circ}_k$ equal to the inverse of Kaplan-Meier estimates for the censoring event by interval $k$ (upon reversing the roles of censoring events and events of interest).
Note that $\widehat{\varphi}^{\circ}_K$ is a weighted version of the ecdf of infection-free hospital death $\widehat{\tilde\phi}_K$ with weights $\widehat{W}^{\circ}_{ik}$ and a weighted version of the ecdf of hospital death $\widehat{\phi}_K$ with weights $\widehat{W}^{\bullet}_{ik} = (1-A_{ik}) \widehat{W}^{\circ}_{ik}$.

Furthermore, as mentioned before, $\widehat{\varphi}^{\circ \text{\tiny AJ}}_K$ can be considered a weighted version of $\widehat{\phi}^{\text{\tiny AJ}}_K$ with weight $1-A_{ik}$ for subject $i$ at interval $k$.
Upon substituting this weight by $\widehat{W}^{\bullet}_{ik}$, which is licensed by the fact that $\widehat{W}^{\circ}_{ik}$ only depends on $k$, we obtain
\begin{eqnarray*}
  \widehat{\varphi}^{\bullet \text{\tiny AJ}}_K &=& \sum_{k = 1}^K \dfrac{E_n[Y_k(1-D_k)(1-Y_{k-1})\widehat{W}^{\bullet}_{k}]}{E_n[(1-D_k)(1-Y_{k-1})\widehat{W}^{\bullet}_{k}]} \dfrac{E_n[(1-D_k)(1-Y_{k-1})(1-D_{k-1})\widehat{W}^{\bullet}_{k}]}{E_n[(1-Y_{k-1})(1-D_{k-1})\widehat{W}^{\bullet}_{k}]}\\
  &&\qquad\qquad \times \prod_{s = 1}^{k-1} \dfrac{E_n[(1 - Y_s)(1 - D_s)(1 - Y_{s-1})\widehat{W}^{\bullet}_{s}]}{E_n[(1-D_s)(1-Y_{s-1})\widehat{W}^{\bullet}_{s}]} \dfrac{E_n[(1-D_s)(1 - Y_{s-1})(1 - D_{s-1})\widehat{W}^{\bullet}_{s}]}{E_n[(1-Y_{s-1})(1-D_{s-1})\widehat{W}^{\bullet}_{s}]},
\end{eqnarray*}
an IPC weighted version of $\widehat{\phi}^{\text{\tiny AJ}}_K$ that is algebraically equivalent to $\widehat{\varphi}^{\circ \text{\tiny AJ}}_K$. Algebraic equivalence of $\widehat{\varphi}^{\circ \text{\tiny AJ}}_K$ and $\widehat{\varphi}^{\circ}_K$ then alternatively follows from
\begin{eqnarray*}
  &&\sum_{k = 1}^K \dfrac{E_n[Y_k(1-D_k)(1-Y_{k-1})\widehat{W}^{\bullet}_{k}]}{E_n[(1-D_k)(1-Y_{k-1})\widehat{W}^{\bullet}_{k}]} \dfrac{E_n[(1-D_k)(1-Y_{k-1})(1-D_{k-1})\widehat{W}^{\bullet}_{k}]}{E_n[(1-Y_{k-1})(1-D_{k-1})\widehat{W}^{\bullet}_{k}]}\\
  &&\qquad\qquad \times \prod_{s = 1}^{k-1} \dfrac{E_n[(1 - Y_s)(1 - D_s)(1 - Y_{s-1})\widehat{W}^{\bullet}_{s}]}{E_n[(1-D_s)(1-Y_{s-1})\widehat{W}^{\bullet}_{s}]} \dfrac{E_n[(1-D_s)(1 - Y_{s-1})(1 - D_{s-1})\widehat{W}^{\bullet}_{s}]}{E_n[(1-Y_{s-1})(1-D_{s-1})\widehat{W}^{\bullet}_{s}]}\\
  &&= \sum_{k = 1}^K \dfrac{E_n[Y_k(1-Y_{k-1})(1-D_k)\widehat W^{\bullet}_k]}{E_n[(1-Y_{k-1})(1-D_{k-1})\widehat W^{\bullet}_k]} \prod_{s = 1}^{k-1} \dfrac{E_n[(1-Y_s)(1-D_s)\widehat W^{\bullet}_s]}{E_n[(1-Y_{s-1})(1-D_{s-1})\widehat W^{\bullet}_s]}\\
  &&= \sum_{k = 1}^K E_n[Y_k(1-Y_{k-1})(1-D_k)\widehat W^{\bullet}_k] = \sum_{k = 1}^K E_n[Y_k(1-Y_{k-1})(1-D_k)(1-A_k)\widehat W^{\circ}_k],
\end{eqnarray*}
where the first equality again follows from the fact that we have $A_0 \equiv D_0 \equiv Y_0 \equiv 0$, and from
\begin{eqnarray*}
  && E_n[(1-Y_k)(1-D_k)\widehat W^{\bullet}_{k+1}]\nonumber\\
  &&= E_n \left[\dfrac{(1-Y_k)(1-D_k)(1-A_{k+1})\widehat W^{\bullet}_k}{P_n(A_{k+1}=0 | Y_k = D_k = A_k = 0)}\middle\vert Y_k = D_k = A_k = 0 \right] P_n(\overline Y_k = \overline D_k = \overline A_k = \overline 0)\nonumber\\
  &&= E_n \left[(1-Y_k)(1-D_k)\widehat W^{\bullet}_k\middle\vert Y_k = D_k = A_k = 0 \right] P_n(\overline Y_k = \overline D_k = \overline A_k = \overline 0)\nonumber\\
  &&= E_n \left[(1-Y_k)(1-D_k)\widehat W^{\bullet}_k \right],
\end{eqnarray*}
with $E_n(X \vert Z=z)$ as defined in Appendix~\ref{sec:app-estobsdata1}. The proof of this equality is also given in Lemma 5 in \cite{Young2020}.

Following the results of Appendix~\ref{sec:app-estcfdata}, $\widehat{\varphi}^{\circ \text{\tiny AJ}}_K$ and $\widehat{\varphi}^{\circ}_K$ are consistent estimators of the counterfactual risk $\varphi_K$ under the assumption that (artificial) censoring is (marginally) \emph{independent} of the counterfactual event times or, equivalently, under the assumption that the observable event-specific hazards of hospital death and discharge among patients who have remained uninfected up until (and including) interval $k$ equal the counterfactual infection-free event-specific hazards of hospital death and discharge, respectively. That is, for every $k$, we have
\begin{eqnarray*}
  h^{(1), \overline a = 0}_k &=& \tilde h^{(1)}_k\\
  h^{(2), \overline a = 0}_k &=& \tilde h^{(2)}_k.
\end{eqnarray*}

Note that these equalities also follow from conditions~\eqref{app-seqexch}-\eqref{app-consistency} and-\eqref{app-seqexch2} with $\overline L_{k-1} = \emptyset$. Under these assumptions, we have weights $W_{ik} = W^{\circ}_k$ that only depend on interval $k$, such that the identification results from Appendix~\ref{sec:app-ident3} reduce to $\tilde h^{(1)}_k$ and $\tilde h^{(2)}_k$, respectively.

\subsubsection{Implicit hypothetical intervention when treating exposure onset as censoring event}

Finally, note that $\widehat\varphi_K^{\circ \text{\tiny AJ}}$ corresponds to $\widehat{\tilde\phi}^{\text{\tiny AJ}}_K$ upon setting the event-specific hazards of the competing event HAI onset $\tilde h^{(0)}_k$ to zero for each $k = 1, ..., K$ \citep{Arjas1993,Keiding2001}

\begin{eqnarray*}
  \widehat\varphi_K^{\circ \text{\tiny AJ}} &&= \sum_{k=1}^K \widehat{\tilde h}^{(1)}_k \left(1 - 0 \widehat{\tilde h}^{(0)}_k \right) \left(1 - \widehat{\tilde h}^{(2)}_k \right) \prod_{s=1}^{k-1} \left(1 - 0 \widehat{\tilde h}^{(0)}_s \right) \left(1 - \widehat{\tilde h}^{(1)}_s \right) \left(1 - \widehat{\tilde h}^{(2)}_s \right).
\end{eqnarray*}
This illustrates that, from a counterfactual point of view, censoring a time-to-event at a competing event time envisages a
hypothetical intervention that sets the hazard of that competing event (infection onset) to zero. In contrast to the approach by \cite{Schumacher2007}, this hypothetical intervention does not retrospectively increase the hazards of in-hospital death or discharge.

\subsection{Treating exposure onset as conditionally independent censoring}\label{sec:app-wcharest4}

Define weights
\begin{eqnarray*}
  W_{ik} &\equiv& \prod_{s=1}^k \dfrac{1}{1-\Pr(\tilde T \in (t_{s-1}, t_s], \tilde \epsilon = 0 \vert \tilde T > t_{s-1}, \overline L_{i,s-1})} = \prod_{s=1}^k \dfrac{1}{\Pr(A_{is}=0 | \overline L_{i,s-1}, A_{i,s-1} = D_{i,s-1} = Y_{i,s-1} = 0)}\nonumber\\
  &=& 1+\sum_{s=1}^k \dfrac{\Pr(A_{is} =1 | \overline L_{i,s-1}, A_{i,s-1} = D_{i,s-1} = Y_{i,s-1} = 0)}{\Pr(A_{is} =0 | \overline L_{i,s-1}, A_{i,s-1} = D_{i,s-1} = Y_{i,s-1} = 0)} \prod_{s'=1}^{s-1} \left\{ 1+ \dfrac{\Pr(A_{i,s'} =1 | \overline L_{i,s'-1}, A_{i,s'-1} = D_{i,s'-1} = Y_{i,s'-1} = 0)}{\Pr(A_{i,s'} =0 | \overline L_{i,s'-1}, A_{i,s'-1} = D_{i,s'-1} = Y_{i,s'-1} = 0)} \right\},
\end{eqnarray*}
and weights $W^{\prime}_{ik} \equiv (1-A_{ik}) W_{ik}$, with $\overline L_k$ a time-dependent covariate set under which sequential exchangeability~\eqref{app-seqexch} holds. The last equality follows from the results in Appendix~\ref{sec:app-decompw}.

Furthermore, define
\begin{eqnarray}
\tilde h^{(1),w_{ik}}_k &\equiv& \dfrac{E[Y_k(1-D_k)(1-A_k)(1-Y_{k-1}) W_k]}{E[(1-D_k)(1-A_k)(1-Y_{k-1}) W_k]}\nonumber\\
\tilde h^{(2),w_{ik}}_k &\equiv& \dfrac{E[D_k(1-A_k)(1-Y_{k-1})(1-D_{k-1}) W_k]}{E[(1-A_k)(1-Y_{k-1})(1-D_{k-1}) W_k]},\nonumber
\end{eqnarray}
and their respective estimators
\begin{eqnarray}
\widehat{\tilde h}^{(1),w_{ik}}_k(\hat \theta) &\equiv& \dfrac{E_n[Y_k(1-D_k)(1-A_k)(1-Y_{k-1})(1-C_k) W_k(\hat \theta)]}{E_n[(1-D_k)(1-A_k)(1-Y_{k-1}) (1-C_k) W_k(\hat \theta)]}\nonumber\\
\widehat{\tilde h}^{(2),w_{ik}}_k(\hat \theta) &\equiv& \dfrac{E_n[D_k(1-A_k)(1-Y_{k-1})(1-D_{k-1}) (1-C_k) W_k(\hat \theta)]}{E_n[(1-A_k)(1-Y_{k-1})(1-D_{k-1}) (1-C_k) W_k(\hat \theta)]},\nonumber
\end{eqnarray}
with
\begin{eqnarray*}
  W_{ik}(\hat \theta) &\equiv& \prod_{s=1}^k \dfrac{1}{1-p(s, \overline L_{i,s-1}; \widehat \theta)},
\end{eqnarray*}
where $p(k, \overline L_{i,k-1}; \theta)$ is a pre-specified model (e.g. a pooled logistic regression model or a Cox proportional hazard model), indexed by (vector-valued) parameter $\theta$, for the discrete-time event-specific hazard of exposure in interval $k$ conditional on the confounder history $\Pr(A_{k}=1 | \overline L_{k-1}, A_{k-1} = D_{k-1} = Y_{k-1} = C_{k} = 0)$. Whenever a fully saturated (i.e. non-parametric) model is used, we will suppress the dependence on $\widehat \theta$ in the notation and use the simplified notation $\widehat{\tilde h}^{(1),w_{ik}}_k$, $\widehat{\tilde h}^{(2),w_{ik}}_k$ and $\widehat W_{ik}$ instead.

Under conditions~\eqref{app-seqexch}-\eqref{app-consistency}, correct specification of $p(k, \overline L_{i,k-1}; \theta)$, and in the presence of independent censoring due to loss to follow-up or administrative end of study, $\varphi_K$ can be consistently estimated by the following IPC weighted Aalen-Johansen estimator
\begin{eqnarray}
  \widehat\varphi^{\text{\tiny AJ}}_K &&= \sum_{k = 1}^K \widehat{\tilde h}^{(1),w_{ik}}_k(\hat \theta) \left(1-\widehat{\tilde h}^{(2),w_{ik}}_k(\hat \theta)\right) \prod_{s = 1}^{k-1} \left(1-\widehat{\tilde h}^{(1),w_{is}}_s(\hat \theta)\right) \left(1-\widehat{\tilde h}^{(2),w_{is}}_s(\hat \theta)\right). \nonumber
\end{eqnarray}
In the absence of censoring due to loss to follow-up or administrative end of study, and under non-parametric estimation of the weights $W_k$, $\widehat\varphi^{\text{\tiny AJ}}_K$ reduces to an IPC weighted empirical cumulative distribution function of exposure-free hospital death
\begin{eqnarray}
  \widehat\varphi^{\text{\tiny AJ}}_K &=& \sum_{k = 1}^K \dfrac{E_n[Y_k(1-D_k)(1-Y_{k-1})(1-A_k)\widehat W_k]}{E_n[(1-D_k)(1-Y_{k-1})(1-A_k)\widehat W_k]} \dfrac{E_n[(1-D_k)(1-Y_{k-1})(1-D_{k-1})(1-A_k)\widehat W_k]}{E_n[(1-Y_{k-1})(1-D_{k-1})(1-A_k)\widehat W_k]}\nonumber\\
  &&\qquad\qquad \times \prod_{s = 1}^{k-1} \dfrac{E_n[(1 - Y_s)(1 - D_s)(1 - Y_{s-1})(1-A_s)\widehat W_s]}{E_n[(1-D_s)(1-Y_{s-1})(1-A_s)\widehat W_s]} \dfrac{E_n[(1-D_s)(1 - Y_{s-1})(1 - D_{s-1})(1-A_s)\widehat W_s]}{E_n[(1-Y_{s-1})(1-D_{s-1})(1-A_s)\widehat W_s]}\nonumber\\
  &&= \sum_{k = 1}^K \dfrac{E_n[Y_k(1-D_k)(1- A_k)(1-Y_{k-1})\widehat W_k]}{E_n[(1-A_k)(1-Y_{k-1})(1-D_{k-1})\widehat W_k]} \prod_{s = 1}^{k-1} \dfrac{E_n[(1-Y_s)(1-D_s)(1-A_s)\widehat W_s]}{E_n[(1-A_s)(1-Y_{s-1})(1-D_{s-1})\widehat W_s]}\nonumber\\
  &&= \sum_{k = 1}^K E_n[Y_k(1-D_k)(1-A_k)(1-Y_{k-1})\widehat W_k] = n^{-1} \sum_{k = 1}^K \tilde d^{w_{ik}}_{1k} = \widehat\varphi_K,\nonumber
\end{eqnarray}
with $\tilde d^{w_{ik}}_{1k} \equiv \sum_{i=1}^n Y_{ik}(1-D_{ik})(1-A_{ik})(1-Y_{i,k-1})\widehat W_{ik}$ denoting the weighted number of HAI-free hospital deaths in interval $k$. The third equality follows from the fact that we have $A_0 \equiv D_0 \equiv Y_0 \equiv 0$ and that, under non-parametric estimation of the weights, we have \citep[also see Lemma 5 in][]{Young2020}
\begin{eqnarray}
  &&E_n[(1-Y_k)(1-D_k)(1-A_{k+1}) \widehat W_{k+1}]\nonumber\\
  &&=E_n\left[\dfrac{(1-Y_k)(1-D_k)(1-A_{k+1}) \widehat W_k}{P_n(A_{k+1} = 0 \vert \overline{L}_k, Y_k = D_k = A_k = 0)}\right]\nonumber\\
  &&=\sum_{\overline{l}_k} E_n\left[\dfrac{(1-Y_k)(1-D_k)(1-A_{k+1}) \widehat W_k}{P_n(A_{k+1} = 0 \vert \overline{L}_k = \overline{l}_k, Y_k = D_k = A_k = 0)}\middle\vert \overline{L}_k = \overline{l}_k, Y_k = D_k = A_k = 0 \right]\nonumber\\[-5pt]
  &&\qquad\qquad \times P_n(\overline{L}_k = \overline{l}_k, \overline{Y}_k = \overline{D}_k = \overline{A}_k = 0)\nonumber\\
  &&=\sum_{\overline{l}_k} E_n\left[(1-Y_k)(1-D_k)(1-A_k) \widehat W_k\middle\vert \overline{L}_k = \overline{l}_k, Y_k = D_k = A_k = 0 \right]\nonumber\\[-10pt]
  &&\qquad\qquad \times  P_n(\overline{L}_k = \overline{l}_k, \overline{Y}_k = \overline{D}_k = \overline{A}_k = \overline 0)\nonumber\\
  &&= E_n[(1-Y_k)(1-D_k)(1-A_k) \widehat W_k].\nonumber
\end{eqnarray}
However, if $\overline L_k$ is high-dimensional, or contains non-categorical covariates, non-parametric estimation of the weights becomes practically infeasible due to the curse of dimensionality, and estimation based on a (semi)parametric model for $p(k, \overline L_{i,k-1}; \theta)$ is warranted. Under the imposed (parametric) modeling assumptions, the aforementioned algebraic equivalence no longer holds and this gives rise to distinct estimators depending on the model specification.
Again note that $\widehat{\varphi}_K$ is a weighted version of the ecdf of infection-free hospital death $\widehat{\tilde\phi}_K$ with weights $\widehat{W}_{k}$ and a weighted version of the ecdf of hospital death $\widehat{\phi}_K$ with weights $(1-A_{k}) \widehat{W}_{k}$.

Following the results of Appendix~\ref{sec:app-estcfdata}, under correct specification of $p(k, \overline L_{i,k-1}; \theta)$, $\widehat{\varphi}^{\text{\tiny AJ}}_K$ and $\widehat{\varphi}_K$ are consistent estimators of the counterfactual risk $\varphi_K$ under the assumption that (artificial) censoring is independent of the counterfactual event times \emph{conditional} on a set of time-dependent covariates or, equivalently, under the assumption that, for every $k$,
\begin{eqnarray*}
  h^{(1), \overline a = 0}_k &=& \tilde h^{(1),w_{ik}}_k\\
  h^{(2), \overline a = 0}_k &=& \tilde h^{(2),w_{ik}}_k.
\end{eqnarray*}
Note that these equalities follow from conditions~\eqref{app-seqexch}-\eqref{app-consistency} and~\eqref{app-seqexch2}. However, building on recent work by \cite{Young2020}, and as indicated in Appendix~\ref{sec:app-ident1} (or Section~\ref{sec:main-ident} in the main text), non-parametric identification of the counterfactual risk $\varphi_K$ only requires conditions~\eqref{app-seqexch}-\eqref{app-consistency} to be met. This implies that the additional sequential exchangeability condition~\eqref{app-seqexch2} (see Appendix~\ref{sec:app-ident3}) is required to warrant interpretation of $\widehat{\tilde h}^{(1),w_{ik}}_k(\widehat \theta)$ and $\widehat{\tilde h}^{(2),w_{ik}}_k(\widehat \theta)$ in terms of counterfactual event-specific hazards. Conditions~\eqref{app-seqexch}-\eqref{app-consistency} and~\eqref{app-seqexch2} hence permit to reconstruct the hypothetical population (in which no one were exposed) from the observed data in such a way that each factor of the Aalen-Johansen estimator that involves unobserved counterfactual data $(\overline D_k^{\overline a = 0}, \overline Y_k^{\overline a = 0})$ (for $k \le K$) (see Appendix~\ref{sec:app-estcfdata}) can be estimated from the observed factual data $(\overline L_k, \overline A_k, \overline D_k, \overline Y_k)$ by an IPC weighted equivalent.

Moreover, following the results from Appendices~\ref{sec:app-ident1} and~\ref{sec:app-ident3}, under conditions~\eqref{app-seqexch}-\eqref{app-consistency} and~\eqref{app-seqexch2}, the (marginal) counterfactual event-specific hazards can be expressed as weighted sums of the corresponding (conditional) observable event-specific hazards within strata defined by the confounder history:
\begin{eqnarray*}
  h^{(1),\overline a = 0}_k &=& \dfrac{\Pr(Y_k^{\overline a = 0} = 1, \overline D_k^{\overline a = 0} = \overline Y_{k-1}^{\overline a = 0} = \overline 0)}{\Pr(\overline D_k^{\overline a = 0} =  \overline Y_{k-1}^{\overline a = 0} = 0)} = \dfrac{\sum_{\overline l_{k-1}} \tilde h^{(1)}_k(\overline l_{k-1}) g^{(1)}(\overline l_{k-1})}{\sum_{\overline l_{k-1}} g^{(1)}(\overline l_{k-1})}\\
  h^{(2),\overline a = 0}_k &=& \dfrac{\Pr(D_k^{\overline a = 0} = 1, \overline Y_{k-1}^{\overline a = 0} = \overline D_{k-1}^{\overline a = 0} = 0)}{\Pr(\overline Y_{k-1}^{\overline a = 0} = \overline D_{k-1}^{\overline a = 0} = 0)} = \dfrac{\sum_{\overline l_{k-1}} \tilde h^{(2)}_k(\overline l_{k-1}) g^{(2)}(\overline l_{k-1})}{\sum_{\overline l_{k-1}} g^{(2)}(\overline l_{k-1})},
\end{eqnarray*}
with
\begin{eqnarray*}
  \tilde h^{(1)}_k(\overline l_{k-1}) &\equiv& \Pr(Y_k = 1 \vert Y_{k-1} = D_k = A_k = 0, \overline L_{k-1} = \overline l_{k-1})\\
  \tilde h^{(2)}_k(\overline l_{k-1}) &\equiv& \Pr(D_k = 1 \vert Y_{k-1} = D_{k-1} = A_k = 0, \overline L_{k-1} = \overline l_{k-1})\\
  g^{(1)}(\overline l_{k-1}) &\equiv& \dfrac{\Pr(\overline Y_{k-1} = \overline D_k = \overline A_k = \overline 0, \overline L_{k-1} = \overline l_{k-1})}{\prod_{s = 1}^{k} \Pr(A_s = 0 \vert \overline L_{s-1} = \overline l_{s-1}, A_{s-1} = D_{s-1} = Y_{s-1} = 0)}\\
  g^{(2)}(\overline l_{k-1}) &\equiv& \dfrac{\Pr(\overline Y_{k-1} = \overline D_{k-1} = \overline A_k = \overline 0, \overline L_{k-1} = \overline l_{k-1})}{\prod_{s = 1}^{k} \Pr(A_s = 0 \vert \overline L_{s-1} = \overline l_{s-1}, A_{s-1} = D_{s-1} = Y_{s-1} = 0)}.
\end{eqnarray*}
This characterization clarifies that $\widehat{\varphi}^{\circ \text{\tiny AJ}}_K$ is a special case of $\widehat{\varphi}^{\text{\tiny AJ}}_K$ under the stronger assumption that the \emph{marginal} counterfactual hazards equal the \emph{marginal} factual observable hazards (as clarified at the end of Appendix~\ref{sec:app-wcharest3}) or, in other words that, for every $k$, $\overline L_k = \emptyset$, such that the hazards $\tilde h^{(1)}_k(\overline l_{k-1})$ and $\tilde h^{(2)}_k(\overline l_{k-1})$, and weights $g^{(1)}(\overline l_{k-1})$ and $g^{(2)}(\overline l_{k-1})$ are rendered dependent of only $k$.

More generally, estimation of the counterfactual risk $\varphi_K$ by $\widehat{\varphi}^{\text{\tiny AJ}}_K$, rests on the assumption that the counterfactual hazards equal the factual observable hazards \emph{conditional} on a sufficient set $\overline L_k$:
\begin{eqnarray*}
  h^{(1),\overline a = 0}_k(\overline l_{k-1}) \equiv \Pr(Y_k^{\overline a = 0} = 1\vert \overline L_{k-1}^{\overline a = 0} = \overline l_{k-1}, D_k^{\overline a = 0} = Y_{k-1}^{\overline a = 0} = 0) &=& \tilde h^{(1)}_k(\overline l_{k-1})\\
  h^{(2),\overline a = 0}_k(\overline l_{k-1}) \equiv \Pr(D_k^{\overline a = 0} = 1\vert \overline L_{k-1}^{\overline a = 0} = \overline l_{k-1}, Y_{k-1}^{\overline a = 0} = D_{k-1}^{\overline a = 0} = 0) &=& \tilde h^{(2)}_k(\overline l_{k-1}).
\end{eqnarray*}

\section{Decomposition of accumulated weight}\label{sec:app-decompw}

Censoring leads to weight being transferred in the analysis from censored to uncensored individuals. In this section we illustrate more formally how the weight that gets transferred can be expressed as part of the total weight accumulated by individuals that have remained uncensored until interval $k$. Following Appendix~\ref{sec:app-wcharest3}, we have
\begin{eqnarray*}
  W^{\circ}_k &\equiv&
  \prod_{s = 1}^k \dfrac{1}{1-\Pr(A_s=1 | Y_{s-1} = D_{s-1} = A_{s-1} = 0)} = 1 + \dfrac{1-\prod_{s = 1}^k\Pr(A_s=1 | Y_{s-1} = D_{s-1} = A_{s-1} = 0)}{\prod_{s = 1}^k \Pr(A_s=1 | Y_{s-1} = D_{s-1} = A_{s-1} = 0)}.
\end{eqnarray*}

For instance, in the special case where $k = 3$, we have
\begin{eqnarray*}
  W^{\circ}_3 &=& 1+\dfrac{1 - (1-P_1)(1-P_2)(1-P_3)}{(1-P_1)(1-P_2)(1-P_3)}\\
  &=& 1+\dfrac{1 - (1-P_3) + (1-P_3) - (1-P_2)(1-P_3) + (1-P_2)(1-P_3) - (1-P_1)(1-P_2)(1-P_3)}{(1-P_1)(1-P_2)(1-P_3)}\\
  &=& 1+\dfrac{P_3 + P_2 (1-P_3) + P_1(1-P_2)(1-P_3)}{(1-P_1)(1-P_2)(1-P_3)} = 1+\dfrac{P_1}{1-P_1} + \dfrac{P_2}{(1-P_2)(1-P_1)} + \dfrac{P_3}{(1-P_3)(1-P_2)(1-P_1)},
\end{eqnarray*}
with $P_s$ shorthand notation for $\Pr(A_s =1 | Y_{s-1} = D_{s-1} = A_{s-1} = 0)$.

Alternatively, we can obtain
\begin{eqnarray*}
  W^{\circ}_3 &=& 1 + \dfrac{1 - (1-P_1) + (1-P_1) - (1-P_1)(1-P_2) + (1-P_1)(1-P_2) - (1-P_1)(1-P_2)(1-P_3)}{(1-P_1)(1-P_2)(1-P_3)}\\
  &=& 1 + \dfrac{P_1 + P_2 (1-P_1) + P_3 (1-P_2)(1-P_1)}{(1-P_1)(1-P_2)(1-P_3)} = 1+\dfrac{P_1}{(1-P_1)(1-P_2)(1-P_3)} + \dfrac{P_2}{(1-P_2)(1-P_3)} + \dfrac{P_3}{1-P_3}.
\end{eqnarray*}

Generalizing from the special case where $k=3$, we can see that in the general case, upon re-writing the numerator of the fraction as a telescoping sum, we obtain the following (retrospective) decomposition in terms of the weight \emph{directly} received from patients censored in each interval $s \le k$ and the \emph{indirect} weight received through the same patients who have, in turn, also accumulated/received from patients censored in earlier intervals $s' < s$:
\begin{eqnarray*}
  W^{\circ}_k &=& 1+\dfrac{1 - \sum_{s=2}^k \prod_{s'=s}^k (1-P_{s'}) + \sum_{s=2}^k \prod_{s'=s}^k (1-P_{s'}) - \prod_{s = 1}^{k} (1-P_s)}{\prod_{s = 1}^{k} (1-P_s)}\\
  &=& 1+\dfrac{\sum_{s=1}^k P_s \prod_{s' = s+1}^{k} (1-P_{s'})}{\prod_{s = 1}^{k} (1-P_s)} = 1+\sum_{s=1}^k \dfrac{P_s}{1-P_s} \prod_{s'=1}^{s-1} \dfrac{1}{(1-P_{s'})} = 1+\sum_{s=1}^k \dfrac{P_s}{1-P_s} \prod_{s'=1}^{s-1} \left\{ 1+ \dfrac{P_{s'}}{1-P_{s'}} \right\}\\
  &=& 1+\sum_{s=1}^k \underbrace{\dfrac{\Pr(A_s =1 | Y_{s-1} = D_{s-1} = A_{s-1} = 0)}{1-\Pr(A_s =1 | Y_{s-1} = D_{s-1} = A_{s-1} = 0)}}_{\substack{\text{weight directly received from patients}\\ \text{censored in interval $s$}}} \underbrace{\prod_{s'=1}^{s-1} \left\{ 1+ \dfrac{\Pr(A_{s'} =1 | Y_{s'-1} = D_{s'-1} = A_{s'-1} = 0)}{1-\Pr(A_{s'} =1 | Y_{s'-1} = D_{s'-1} = A_{s'-1} = 0)} \right\}}_{\substack{\text{factor by which this weight is increased}\\ \text{to also include the weight patients censored in interval $s$}\\ \text{have accumulated from patients censored in earlier intervals $s' < s$}}}.
\end{eqnarray*}

An alternative (prospective) decomposition in terms of the weight patients censored in interval $s \le k$ transfer forward in time until interval $k$ can be obtained upon re-writing the numerator of the fraction as an alternative telescoping sum
\begin{eqnarray*}
  W^{\circ}_k &=& 1+\dfrac{1 - \sum_{s=1}^{k-1} \prod_{s'=1}^s (1-P_{s'}) + \sum_{s=1}^{k-1} \prod_{s'=1}^s (1-P_{s'}) - \prod_{s = 1}^{k} (1-P_s)}{\prod_{s = 1}^{k} (1-P_s)}\\
  &=& 1+\dfrac{\sum_{s=1}^k P_s \prod_{s'=1}^{s-1} (1-P_{s'})}{\prod_{s = 1}^{k} (1-P_s)} = 1+\sum_{s=1}^k \dfrac{P_s}{1-P_s} \prod_{s'=s+1}^k \dfrac{1}{(1-P_{s'})} = 1+\sum_{s=1}^k \dfrac{P_s}{1-P_s} \prod_{s'=s+1}^k \left\{ 1+ \dfrac{P_{s'}}{1-P_{s'}} \right\}\\
  &=& 1+\sum_{s=1}^k \underbrace{\dfrac{\Pr(A_s =1 | Y_{s-1} = D_{s-1} = A_{s-1} = 0)}{1-\Pr(A_s =1 | Y_{s-1} = D_{s-1} = A_{s-1} = 0)}}_{\substack{\text{weight directly transferred by patients}\\ \text{censored in interval $s$}}} \underbrace{\prod_{s'=s+1}^k \left\{ 1+ \dfrac{\Pr(A_{s'} =1 | Y_{s'-1} = D_{s'-1} = A_{s'-1} = 0)}{1-\Pr(A_{s'} =1 | Y_{s'-1} = D_{s'-1} = A_{s'-1} = 0)} \right\}}_{\substack{\text{factor by which this weight}\\ \text{is cumulatively carried forward over time}\\ \text{to patients censored in later intervals $s' > s$}}}.
\end{eqnarray*}

\section{Connection with other estimation approaches}\label{sec:app-connest}

In this section, we illustrate that the estimators $\widehat{\varphi}^{\text{\tiny AJ}}_K$ and $\widehat{\varphi}_K$ have a clear connection with other estimators and estimation approaches proposed in the statistical literature.

\subsection{Connection with estimation approach of Schumacher \emph{et al} (2007)}\label{sec:app-connest1}

Recall that the estimator originally proposed by \cite{Schumacher2007} can be expressed as (see Appendix~\ref{sec:app-wcharest2})
\begin{eqnarray*}
  \dfrac{\widehat{\tilde\phi}^{\text{\tiny AJ}}_K}{1-\widehat{\alpha}^{\text{\tiny AJ}}_K}.
\end{eqnarray*}
In the absence of censoring due to loss to follow-up or administrative end of the study, this estimator corresponds to (see Appendix~\ref{sec:app-estobsdata2})
\begin{eqnarray*}
  \dfrac{\sum_{k=1}^K \sum_{i=1}^n Y_{ik} (1-Y_{i,k-1}) (1-D_{ik}) (1-A_{ik})}{n-\sum_{k=1}^K \sum_{i=1}^n A_{ik} (1-A_{i,k-1}) (1-Y_{i,k-1}) (1-D_{i,k-1})},
\end{eqnarray*}
which expresses the fraction of deceased patients by landmark interval $K$ among patients who have not acquired an infection by interval $K$. In other words, at each successive landmark interval $K$, the considered population (from which $n$ patients are randomly sampled into the study sample) is implicitly reduced to those patients who either remained hospitalized and uninfected until the end of interval $K$ or have died or been discharged by interval $K$ while being uninfected, as indicated by the denominator.

Upon weighing the contribution of each patient $i$ to the increments of the cumulative incidence functions of exposure-free hospital death and of exposure onset at each interval $k$ by $(1-A_{ik}) \widehat W_{ik}$ this estimator can be shown to correspond to $\widehat \varphi_K$:
\begin{eqnarray*}
  &&\dfrac{\sum_{k=1}^K \sum_{i=1}^n Y_{ik} (1-Y_{i,k-1}) (1-D_{ik}) (1-A_{ik}) (1-A_{ik}) \widehat W_{ik}}{n-\sum_{k=1}^K \sum_{i=1}^n A_{ik} (1-A_{i,k-1}) (1-Y_{i,k-1}) (1-D_{i,k-1})(1-A_{ik}) \widehat W_{ik}}\\
  &&= n^{-1} \sum_{k=1}^K \sum_{i=1}^n Y_{ik} (1-Y_{i,k-1}) (1-D_{ik}) (1-A_{ik}) (1-A_{ik}) \widehat W_{ik} = n^{-1} \sum_{k=1}^K d^{w_{ik}}_k = \widehat \varphi_K.
\end{eqnarray*}

After IPC weighing by $\widehat{W}^{\prime}_k$, the considered (pseudo)population at each landmark interval $K$ now corresponds to the original population under the hypothetical scenario where \emph{no} patient would have acquired an infection by interval $K$ or, equivalently, \emph{all} patients would have remained uninfected until interval $K$ or would have died or been discharged by interval $K$ while being uninfected (i.e. the numerator $= n$), so as to evaluate which fraction of this hypothetical cohort of patients would have died at the hospital by interval $K$.

\subsection{Connection with alternative estimators in Bekaert \emph{et al} (2010)}\label{sec:app-connest3}

To illustrate the connection with the estimators for $\varphi_K$ from \cite{Bekaert2010}, we first introduce some additional notation. Let $A^{\prime}_k = A_k$ for every $k:t_k\le T$. For every $k:t_k>T$, $A^{\prime}_k$ is undefined. Furthermore, let $\epsilon^{\prime}_{ik} = j, j\in \{0,1,2\},$ denote the event status of patient $i$ by interval $k$, where $j = 0$ in the absence of an event, $j = 1$ in case of hospital death, or $j = 2$ in case of hospital discharge.

The naive estimator from Bekaert and colleagues (expression (2) in their paper) for the counterfactual exposure path $\overline a = 0$ then corresponds with
\begin{eqnarray*}
  &&\dfrac{\sum_{i=1}^n \mathbf{1}_{\epsilon^{\prime}_{iK} = 1} \prod_{k=1}^K \{ \mathbf{1}_{A^{\prime}_{ik} = 0} \mathbf{1}_{\epsilon^{\prime}_{i,k-1} = 0} + \mathbf{1}_{\epsilon^{\prime}_{i,k-1} \ne 0} \}}{\sum_{i=1}^n \prod_{k=1}^K \{ \mathbf{1}_{A^{\prime}_{ik} = 0} \mathbf{1}_{\epsilon^{\prime}_{i,k-1} = 0} + \mathbf{1}_{\epsilon^{\prime}_{i,k-1} \ne 0} \}},
\end{eqnarray*}
which, in our discrete time process notation, can be re-expressed as
\begin{eqnarray*}
  && = \dfrac{\sum_{i=1}^n \mathbf{1}_{\epsilon^{\prime}_{iK} = 1} \prod_{k=1}^K \mathbf{1}_{A_{ik} = 0}}{\sum_{i=1}^n \prod_{k=1}^K \mathbf{1}_{A_{ik} = 0}} = \dfrac{\sum_{i=1}^n \mathbf{1}_{T_i \le t_K, \epsilon_{i} = 1} \mathbf{1}_{A_{iK} = 0}}{\sum_{i=1}^n \mathbf{1}_{A_{iK} = 0}} = \dfrac{E_n[Y_K(1-A_K)]}{E_n[1-A_K]} = \dfrac{E_n[\tilde Y_K]}{E_n[1-A_K]}.
\end{eqnarray*}
In the absence of censoring due to loss to follow-up or administrative end of study, this corresponds to
\begin{eqnarray*}
  \dfrac{\widehat{\tilde\phi}^{\text{\tiny AJ}}_K}{1-\widehat{\alpha}^{\text{\tiny AJ}}_K} = \widehat{\varphi}^{\dagger \text{\tiny AJ}}_K,
\end{eqnarray*}
the estimator proposed by \cite{Schumacher2007} (see Appendix~\ref{sec:app-wcharest2}).
This equivalence has also been highlighted by \cite{VonCube2019a} and is demonstrated in their Appendix B.2.2.

The IP weighted estimator from Bekaert and colleagues for the counterfactual exposure path $\overline a = 0$ (expression (3) in their paper) corresponds with
\begin{eqnarray*}
  \widehat{\varphi}^{\text{B}}_K &&= \dfrac{\sum_{i=1}^n \mathbf{1}_{\epsilon^{\prime}_{iK} = 1} \prod_{k=1}^K \dfrac{\mathbf{1}_{A^{\prime}_{ik} = 0} \mathbf{1}_{\epsilon^{\prime}_{i,k-1} = 0} + \mathbf{1}_{\epsilon^{\prime}_{i,k-1} \ne 0}}{\widehat{\Pr}(A^{\prime}_k = 0 \vert \epsilon^{\prime}_{i,k-1}, \overline A^{\prime}_{i,k-1}, \overline L_{i,k-1})} }{\sum_{i=1}^n \prod_{k=1}^K \dfrac{\mathbf{1}_{A^{\prime}_{ik} = 0} \mathbf{1}_{\epsilon^{\prime}_{i,k-1} = 0} + \mathbf{1}_{\epsilon^{\prime}_{i,k-1} \ne 0}}{\widehat{\Pr}(A^{\prime}_k = 0 \vert \epsilon^{\prime}_{i,k-1}, \overline A^{\prime}_{i,k-1}, \overline L_{i,k-1})}},
\end{eqnarray*}
with $\widehat{\Pr}(A^{\prime}_k = 0 \vert \epsilon^{\prime}_{i,k-1}, \overline A^{\prime}_{i,k-1}, \overline L_{i,k-1})$ defined to equal 1 whenever $\epsilon^{\prime}_{i,k-1} \ne 0$ and $\overline A^{\prime}_k$ and $\overline L_{i,k-1}$ are consequently not defined.
In our discrete time process notation, this IP weighted estimator can be re-expressed as
\begin{eqnarray*}
  &&\dfrac{\sum_{i=1}^n \mathbf{1}_{\epsilon^{\prime}_{iK} = 1} \prod_{k=1}^K \dfrac{\mathbf{1}_{A_{ik} = 0}}{\widehat{\Pr}(A_k = 0 \vert \epsilon^{\prime}_{i,k-1}, \overline A_{i,k-1}, \overline L_{i,k-1})} }{\sum_{i=1}^n \prod_{k=1}^K \dfrac{\mathbf{1}_{A_{ik} = 0}}{\widehat{\Pr}(A_k = 0 \vert \epsilon^{\prime}_{i,k-1}, \overline A_{i,k-1}, \overline L_{i,k-1})}}\\
  &&= \dfrac{\sum_{i=1}^n \dfrac{\mathbf{1}_{\epsilon^{\prime}_{iK} = 1} \mathbf{1}_{A_{iK} = 0}}{\prod_{k=1}^K \widehat{\Pr}(A_k = 0 \vert \epsilon^{\prime}_{i,k-1}, \overline A_{i,k-1}, \overline L_{i,k-1})} }{\sum_{i=1}^n \dfrac{\mathbf{1}_{A_{iK} = 0}}{\prod_{k=1}^K \widehat{\Pr}(A_k = 0 \vert \epsilon^{\prime}_{i,k-1}, \overline A_{i,k-1}, \overline L_{i,k-1})}} = \dfrac{E_n\left[\dfrac{Y_K(1-A_K)}{\prod_{k=1}^K \widehat{\Pr}(A_k = 0\vert \overline L_{k-1}, \overline A_{k-1}, \overline D_{k-1}, \overline Y_{k-1})}\right]}{E_n\left[\dfrac{1-A_K}{\prod_{k=1}^K \widehat{\Pr}(A_k = 0\vert \overline L_{k-1}, \overline A_{k-1}, \overline D_{k-1}, \overline Y_{k-1})}\right]}.
\end{eqnarray*}
From this alternative expression, it is clear that, under non-parametric estimation of the weights (and in the absence of censoring due to loss to follow-up or administrative end of study), this IP weighted estimator of Bekaert and colleagues is again algebraically equivalent with the IPC weighted estimators $\widehat{\varphi}^{\text{\tiny AJ}}_K$, $\widehat{\varphi}^{\text{\tiny KM}}_K$ and $\widehat{\varphi}_K$:
\begin{eqnarray*}
  &&\dfrac{E_n\left[\dfrac{Y_K(1-A_K)}{\prod_{k=1}^K P_n(A_k = 0\vert \overline L_{k-1}, \overline A_{k-1}, \overline D_{k-1}, \overline Y_{k-1})}\right]}{E_n\left[\dfrac{1-A_K}{\prod_{k=1}^K P_n(A_k = 0\vert \overline L_{k-1}, \overline A_{k-1}, \overline D_{k-1}, \overline Y_{k-1})}\right]}\\
  &&= E_n\left[\dfrac{Y_K(1-A_K)}{\prod_{k=1}^K P_n(A_k = 0\vert \overline L_{k-1}, \overline A_{k-1}, \overline D_{k-1}, \overline Y_{k-1})}\right] = \sum_{k=1}^K E_n\left[\dfrac{Y_k(1- Y_{k-1})(1-A_k)}{\prod_{s = 1}^{k} P_n(A_s = 0\vert \overline L_{s-1}, \overline A_{s-1}, \overline D_{s-1}, \overline Y_{s-1})}\right]\\
  &&= \sum_{k=1}^K E_n\left[\dfrac{Y_k(1- Y_{k-1})(1-D_k)(1-A_k)}{\prod_{s = 1}^{k} P_n(A_s = 0\vert \overline L_{s-1}, \overline A_{s-1}, \overline D_{s-1}, \overline Y_{s-1})}\right] = \sum_{k=1}^K E_n\left[\dfrac{Y_k(1- Y_{k-1})(1-D_k)(1-A_k)}{\prod_{s = 1}^{k} P_n(A_s = 0\vert \overline L_{s-1}, A_{s-1} = D_{s-1} = Y_{s-1} = 0)}\right]\\[5pt]
  &&= \sum_{k=1}^K E_n\left[Y_k(1- Y_{k-1})(1-D_k)(1-A_k) \widehat W_k\right] = \widehat{\varphi}_K,
\end{eqnarray*}
where the second equality again holds because, for each $k\le K$, we have $Y_k(1-A_K) = Y_k(1-A_k)$, and the third equality follows from the deterministic relation between the competing events.

\subsection{Connection with alternative estimator in Young \emph{et al} (2020)}

Define weights
\begin{eqnarray*}
  W^{\text{sd}}_{ik} &\equiv& \prod_{s=1}^k \dfrac{1}{1-\Pr(\tilde T \in (t_{s-1}, t_s], \tilde \epsilon = 0 \vert \tilde T > t_{s-1} \cup \{ \tilde T \le t_{s-1}, \tilde \epsilon = 2 \}, \overline L_{i,s-1})}\\
  &=& \prod_{s=1}^k \dfrac{1}{\Pr(A_{is} = 0 \vert \overline L_{i,s-1}, \overline D_{i,s-1}, A_{i,s-1} = Y_{i,s-1} = 0)}.
\end{eqnarray*}
Furthermore, define
\begin{eqnarray}
  h^{\text{sd},w_{ik}}_k &\equiv& \dfrac{E\left[Y_k(1-A_k)(1-Y_{k-1}) W^{\text{sd}}_k\right]}{E\left[(1-A_k)(1-Y_{k-1}) W^{\text{sd}}_k\right]},\nonumber
\end{eqnarray}
and its estimator
\begin{eqnarray*}
  \widehat{h}^{\text{sd},w_{ik}}_k(\widehat \theta) &\equiv& \dfrac{E_n\left[Y_k(1-A_k)(1-Y_{k-1})(1-C_k) W^{\text{sd}}_k(\widehat \theta)\right]}{E_n\left[(1-A_k)(1-Y_{k-1})(1-C_k) W^{\text{sd}}_k(\widehat \theta)\right]},
\end{eqnarray*}
with
\begin{eqnarray*}
  W^{\text{sd}}_{ik}(\widehat \theta) &\equiv& \prod_{s=1}^k \dfrac{1}{\left\{ 1-p(s,\overline L_{i,s-1}; \widehat\theta) \right\}^{1-D_{i,s-1}}},
\end{eqnarray*}
where, as before, $p(k, \overline L_{k-1};\theta)$ is a pre-specified model (e.g. a pooled logistic regression model or a Cox proportional hazards model), indexed by (vector-valued) parameter $\theta$, for the discrete-time event-specific hazard of exposure in interval $k$ conditional on the confounder history $\Pr(A_{k}=1 | \overline L_{k-1}, A_{k-1} = D_{k-1} = Y_{k-1} = 0)$. Whenever a fully saturated (i.e. non-parametric) model is used, we will suppress the dependence on $\widehat \theta$ in the notation and use the simplified notation $\widehat{h}^{\text{sd},w_{ik}}_k$ and $\widehat W^{\text{sd}}_{ik}$ instead.

Following \cite{Young2020} and the results in Appendix~\ref{sec:app-ident2} it follows that, under conditions~\eqref{app-seqexch}-\eqref{app-consistency} the counterfactual subdistribution hazard $h^{\text{sd},\overline a = 0}_k$ (defined in Appendix~\ref{sec:app-estcfdata}) is non-parametrically identified from the observed data as
\begin{eqnarray*}
  h^{\text{sd},\overline a = 0}_k = h^{\text{sd},w_{ik}}_k.
\end{eqnarray*}
As a result, under conditions~\eqref{app-seqexch}-\eqref{app-consistency}, correct specification of $p(k, \overline L_{i,k-1}; \theta)$, and in the presence of independent censoring due to loss to follow-up or administrative end of study, $\varphi_K$ can be consistently estimated by the following IPC weighted Kaplan-Meier like estimator:
\begin{eqnarray*}
  \widehat{\varphi}^{\text{\tiny KM}}_K &=& \sum_{k=1}^K \widehat{h}^{\text{sd},w_{ik}}_k(\widehat \theta) \prod_{s=1}^{k-1} \left(1 - \widehat{h}^{\text{sd},w_{is}}_s(\widehat \theta)\right).
\end{eqnarray*}
Moreover, in the absence of censoring due to loss to follow-up or administrative end of the study and under non-parametric estimation of the weights $W^{\text{sd}}_{k}$, $\widehat{\varphi}^{\text{\tiny KM}}_K$ again reduces to an IPC weighted empirical cumulative distribution function of exposure-free hospital death
\begin{eqnarray*}
  \widehat{\varphi}^{\text{\tiny KM}}_K &=& \sum_{k=1}^K \dfrac{E_n\left[ Y_k(1-Y_{k-1})(1-A_k) \widehat W^{\text{sd}}_k \right]}{E_n\left[ (1-Y_{k-1})(1-A_k) \widehat W^{\text{sd}}_k \right]} \prod_{s = 1}^{k-1} \dfrac{E_n\left[ (1-Y_s)(1-A_s) \widehat W^{\text{sd}}_s \right]}{E_n\left[ (1-Y_{s-1})(1-A_s)  \widehat W^{\text{sd}}_s \right]} = \sum_{k=1}^K E_n\left[ Y_k(1-Y_{k-1})(1-A_k)  \widehat W^{\text{sd}}_k \right]\\
  &=& \sum_{k=1}^K E_n\left[ Y_k(1-Y_{k-1})(1-D_k)(1-A_k)  \widehat W^{\text{sd}}_k \right] = \sum_{k=1}^K E_n\left[ Y_k(1-Y_{k-1})(1-D_k)(1-A_k) \widehat W_k \right]\\
  &=& n^{-1} \sum_{k=1}^K \tilde d^{w_{ik}}_{1k} = \widehat \varphi_K,
\end{eqnarray*}
with $\tilde d^{w_{ik}}_{1k} \equiv \sum_{i=1}^n Y_{ik} (1-D_{ik}) (1-A_{ik}) (1-Y_{i,k-1}) \widehat W_{ik}$ denoting the weighted number of HAI-free hospital deaths in interval $k$. The second equality follows from the fact that we have $A_0 \equiv Y_0 \equiv 0$ and that, under non-parametric estimation of $W^{\text{sd}}_{k}$, we have \citep[also see Lemma 2 in][]{Young2020}
\begin{eqnarray}
  &&E_n\left[(1-Y_k)(1-A_{k+1}) \widehat W^{\text{sd}}_{k+1}\right] =E_n\left[\dfrac{(1-Y_k)(1-A_{k+1}) \widehat W^{\text{sd}}_k}{P_n(A_{k+1} = 0 \vert \overline{L}_k, \overline{D}_k, Y_k = A_k = 0)}\right]\nonumber\\
  &&=\sum_{\overline{l}_k, \overline{d}_k} E_n\left[\dfrac{(1-Y_k)(1-A_{k+1}) \widehat W^{\text{sd}}_k}{P_n(A_{k+1} = 0 \vert \overline{L}_k = \overline{l}_k, \overline{D}_k = \overline{d}_k, Y_k = A_k = 0)}\middle\vert \overline{L}_k = \overline{l}_k, \overline{D}_k = \overline{d}_k, Y_k = A_k = 0 \right]\nonumber\\[-5pt]
  &&\qquad\qquad \times P_n(\overline{L}_k = \overline{l}_k, \overline{D}_k = \overline{d}_k, Y_k = A_k = 0)\nonumber\\
  &&=\sum_{\overline{l}_k, \overline{d}_k} E_n\left[(1-Y_k)(1-A_k) \widehat W^{\text{sd}}_k\middle\vert \overline{L}_k = \overline{l}_k, \overline{D}_k = \overline{d}_k, Y_k = A_k = 0 \right]
  P_n(\overline{L}_k = \overline{l}_k, \overline{D}_k = \overline{d}_k, \overline{Y}_k = \overline{A}_k = 0)\nonumber\\
  &&= E_n\left[(1-Y_k)(1-A_k) \widehat W^{\text{sd}}_k\right].\nonumber
\end{eqnarray}
The third equality holds because $Y_k = Y_k(1-D_k)$ by the deterministic relation between $Y_k$ and $D_k$,
and the fourth equality holds because, under non-parametric estimation of $W^{\text{sd}}_{k}$, we have
\begin{eqnarray*}
  (1-D_k) \widehat W^{\text{sd}}_{k} &=& \dfrac{1-D_k}{\prod_{s = 1}^{k} P_n(A_s = 0 \vert \overline L_{s-1}, \overline D_{s-1}, A_{s-1} = Y_{s-1} = 0)}\\
  &=& \dfrac{1-D_k}{\prod_{s = 1}^{k} P_n(A_s = 0 \vert \overline L_{s-1}, A_{s-1} = D_{s-1} = Y_{s-1} = 0)} = (1-D_k)\widehat W_k.
\end{eqnarray*}

However, if $\overline L_k$ is high-dimensional, or contains non-categorical covariates, non-parametric estimation of the weights becomes practically infeasible due to the curse of dimensionality and parametric (or semi-parametric) weight estimation is therefore warranted. However, algebraic equivalence of $\widehat{\varphi}^{\text{\tiny KM}}_K$ and $\widehat{\varphi}_K$ no longer holds in these more general settings.

Note that, interpretation of $\widehat{h}^{\text{sd},w_{ik}}_k$ in terms of discrete-time counterfactual subdistribution hazards is warranted under conditions~\eqref{app-seqexch}-\eqref{app-consistency}. In contrast, as indicated in Appendix~\ref{sec:app-ident3} \citep[and in][]{Young2020}, interpretation of $\widehat{\tilde h}^{(1),w_{ik}}_k$ and $\widehat{\tilde h}^{(2),w_{ik}}_k$ in terms of discrete-time counterfactual event-specific hazards is only warranted under conditions~\eqref{app-seqexch}-\eqref{app-consistency} and the additional exchangeability assumption~\eqref{app-seqexch2}.
Conditions~\eqref{app-seqexch}-\eqref{app-consistency} hence permit to reconstruct the hypothetical population (in which no one were exposed) from the observed data in such a way that each factor of the Kaplan-Meier like estimator~\eqref{app-kaplanmeierlike_cf} (see Appendix~\ref{sec:app-estcfdata}) that involves unobserved counterfactual data $\overline Y_k^{\overline a = 0}$ (for $k \le K$) can be estimated from the observed factual data $(\overline L_k, \overline A_k, \overline D_k, \overline Y_k)$ by an IPC weighted equivalent.

\subsection{Connection with estimators based on marginal structural models}

Define $V \subseteq L_0$ a subset of the set of baseline covariates $L_0$, and stratum-specific counterfactual discrete-time event-specific hazards
\begin{eqnarray*}
  h^{(1),\overline a = 0}_k (V) &\equiv& \Pr(Y^{\overline a = 0}_k = 1 \vert D^{\overline a = 0}_k = Y^{\overline a = 0}_{k-1} = 0, V)\\
  h^{(2),\overline a = 0}_k (V) &\equiv& \Pr(D^{\overline a = 0}_k = 1 \vert Y^{\overline a = 0}_{k-1} = D^{\overline a = 0}_{k-1} = 0, V),
\end{eqnarray*}
and the stratum-specific counterfactual discrete-time subdistribution hazard
\begin{eqnarray*}
  h^{\text{sd},\overline a = 0}_k (V) &\equiv& \Pr(Y^{\overline a = 0}_k = 1 \vert Y^{\overline a = 0}_{k-1} = 0, V).
\end{eqnarray*}
Given that
\begin{eqnarray*}
  \varphi_K &=& \sum_{v} \Pr(Y^{\overline a = 0}_K =1 \vert V=v) f(V=v)\\
  &=& \sum_{v} \sum_{k=1}^K h^{(1),\overline a = 0}_k(v) \left(1 - h^{(2),\overline a = 0}_k(v) \right) \prod_{s=1}^{k-1} \left(1 - h^{(1),\overline a = 0}_s(v) \right) \left(1 - h^{(2),\overline a = 0}_s(v) \right) f(V=v)\\
  &=& \sum_{v} \sum_{k=1}^K h^{\text{sd},\overline a = 0}_k(v) \prod_{s=1}^{k-1} \left(1 - h^{\text{sd},\overline a = 0}_s(v) \right) f(V=v)
\end{eqnarray*}
alternative strategies for estimating the counterfactual risk $\varphi_K$ revolve around
\begin{enumerate}
  \item[(1)] modelling either the stratum-specific counterfactual event-specific hazards $h^{(1),\overline a = 0}_k(V)$ and $h^{(2),\overline a = 0}_k(V)$, or the stratum-specific counterfactual subdistribution hazard $h^{\text{sd},\overline a = 0}_k(V)$,
  \item[(2)] obtaining a prediction $\widehat{\Pr}(Y^{\overline a = 0}_{iK} =1 \vert V_{i})$ for each individual $i$, and then
  \item[(3)] averaging these predictions over individuals (i.e. over the empirical distribution of $V$).
\end{enumerate}

\paragraph{Estimators based on modelling event-specific hazards}

Following Appendix~\ref{sec:app-ident3}, it can easily be shown that, under assumptions~\eqref{app-seqexch}-\eqref{app-consistency} and~\eqref{app-seqexch2}, for each stratum $V = v: \Pr(V=v) > 0$, and for each $k = 1,...,K$, $h^{(1),\overline a = 0}_k (v)$ and $h^{(2),\overline a = 0}_k (v)$ are non-parametrically identified as
\begin{eqnarray*}
  h^{(1),\overline a = 0}_k (v) &=& h^{(1),w_{ik}}_k\left(\overline 0, v\right)\\
  h^{(2),\overline a = 0}_k (v) &=& h^{(2),w_{ik}}_k\left(\overline 0, v\right),
\end{eqnarray*}
with
\begin{eqnarray*}
  h^{(1),w_{ik}}_k\left(\overline a_k, V\right) &\equiv& \dfrac{E[Y_k(1-D_k)(1-Y_{k-1})I(\overline A_k = \overline a_k) W^A_k \vert V]}{E[(1-D_k)(1-Y_{k-1})I(\overline A_k = \overline a_k) W^A_k \vert V]}\\
  h^{(2),w_{ik}}_k\left(\overline a_k, V\right) &\equiv& \dfrac{E[D_k(1-Y_{k-1})(1-D_{k-1})I(\overline A_k = \overline a_k) W^A_k \vert V]}{E[(1-Y_{k-1})(1-D_{k-1})I(\overline A_k = \overline a_k) W^A_k \vert V]}
\end{eqnarray*}
and weights
\begin{eqnarray*}
  W^A_{ik} &\equiv& \prod_{s=1}^k \dfrac{1}{\Pr\left(A_{is} \middle\vert \overline L_{i,s-1}, \overline A_{i,s-1}, D_{i,s-1} = Y_{i,s-1} = 0\right)}.
\end{eqnarray*}
As before, given that we are only considering monotone (dichotomous) exposure patterns (i.e. for each $k = 1,...,K$, we have $A_k \ge A_{k-1}$), when $\overline L_k$ is high-dimensional and/or includes non-categorical covariates, the weights $W^A_{k}$ can be estimated by
\begin{eqnarray*}
  W^A_{ik}(\widehat \theta) &\equiv& \prod_{s=1}^k \dfrac{1}{\left\{1-p(s, \overline L_{i,s-1}; \widehat \theta)\right\}^{1-A_{is}} p(s, \overline L_{i,s-1}; \widehat \theta) \vphantom{\left\{p(s, \overline L_{i,s-1}; \widehat \theta)\right\}}^{(1-A_{i,s-1})A_{is}}},
\end{eqnarray*}
where $p(k, \overline L_{k-1};\theta)$ is a pre-specified model (e.g. a pooled logistic regression model or a Cox proportional hazards model), indexed by (vector-valued) parameter $\theta$, for the discrete-time event-specific hazard of exposure in interval $k$ conditional on the confounder history $\Pr(A_{k}=1 | \overline L_{k-1}, A_{k-1} = D_{k-1} = Y_{k-1} = 0)$.
However, given that $V$ may also be high-dimensional and/or include non-categorical covariates, non-parametric estimation of $h^{(1),w_{ik}}_k\left(\overline 0, V\right)$ and $h^{(2),w_{ik}}_k\left(\overline 0, V\right)$ may not be feasible either. Instead, non-saturated marginal structural hazard models (MSMs) may be postulated for these counterfactual event-specific hazards. In practice, the indexing parameters of these MSMs are often estimated by fitting either weighted Cox models \citep[see e.g.][]{Pouwels2017a} or weighted pooled logistic regression models, with $W^A_{k}(\widehat \theta)$ as IPC weights. Under correct specification of $p(k, \overline L_{k-1};\theta)$, the estimates of the nuisance parameters $\beta$ and $\gamma$ of the respective working models for $h^{(1),w_{ik}}_k\left(\overline a_k, V\right)$ and $h^{(2),w_{ik}}_k\left(\overline a_k, V\right)$ are consistent estimates of the structural parameters of the corresponding postulated MSMs for the counterfactual event-specific hazards $h^{(1),\overline a_k}_k (V)$ and $h^{(2),\overline a_k}_k (V)$, leading to the following model-based estimator for $\varphi_K$:
\begin{eqnarray*}
  E_n \left[ \sum_{k=1}^K \widehat{h}^{(1),w_{ik}}_k\left(\overline 0, V; \widehat \beta, \widehat \theta \right) \left\{1 - \widehat{h}^{(2),w_{ik}}_k\left(\overline 0, V; \widehat \gamma, \widehat \theta \right) \right\} \prod_{s=1}^{k-1} \left\{1 - \widehat{h}^{(1),w_{is}}_s\left(\overline 0, V; \widehat \beta, \widehat \theta \right) \right\} \left\{1 - \widehat{h}^{(2),w_{is}}_s\left(\overline 0, V; \widehat \gamma, \widehat \theta \right) \right\} \right].
\end{eqnarray*}
Even under a correctly specified (semi)parametric model $p(k, \overline L_{k-1};\theta)$, the weights $W^A_{k}(\widehat \theta)$ may tend to be highly unstable when $\overline L_k$ is high-dimensional and/or includes non-categorical covariates, resulting in an inefficient estimator. Alternatively, weights may be stabilized by multiplying $W^A_{k}(\widehat \theta)$ with an arbitrary factor that only depends on $k$, $\overline A_k$ and $V$, e.g.
\begin{eqnarray*}
  SW^A_{ik}(\widehat \eta, \widehat \theta) &\equiv& W^A_{ik}(\widehat \theta) \prod_{s=1}^k \left\{1-q(s, V_i; \widehat \eta)\right\}^{1-A_{is}} q(s, V_i; \widehat \eta) \vphantom{\left\{q(s, V_i; \widehat \eta)\right\}}^{(1-A_{i,s-1})A_{is}}\\
  &=& \prod_{s=1}^k \dfrac{\left\{1-q(s, V_i; \widehat \eta)\right\}^{1-A_{is}} q(s, V_i; \widehat \eta) \vphantom{\left\{q(s, V_i; \widehat \eta)\right\}}^{(1-A_{i,s-1})A_{is}}}{\left\{1-p(s, \overline L_{i,s-1}; \widehat \theta)\right\}^{1-A_{is}} p(s, \overline L_{i,s-1}; \widehat \theta) \vphantom{\left\{p(s, \overline L_{i,s-1}; \widehat \theta)\right\}}^{(1-A_{i,s-1})A_{is}}},
\end{eqnarray*}
where $q(k, V;\eta)$ is a pre-specified model, indexed by (vector-valued) parameter $\eta$, for $\Pr(A_{k}=1 | A_{k-1} = D_{k-1} = Y_{k-1} = 0, V)$, the discrete-time event-specific hazard of exposure in interval $k$ conditional on the set of baseline confounders $V$. Applying stabilized weights $SW^A_{k}(\widehat \eta, \widehat \theta)$ instead of $W^A_{k}(\widehat \theta)$ in the above procedure leads to the following model-based estimator for $\varphi_K$:
\begin{eqnarray*}
  E_n \left[ \sum_{k=1}^K \widehat{h}^{(1),w_{ik}}_k\left(\overline 0, V; \widehat \beta, \widehat \eta, \widehat \theta \right) \left\{1 - \widehat{h}^{(2),w_{ik}}_k\left(\overline 0, V; \widehat \gamma, \widehat \eta, \widehat \theta \right) \right\} \prod_{s=1}^{k-1} \left\{1 - \widehat{h}^{(1),w_{is}}_s\left(\overline 0, V; \widehat \beta, \widehat \eta, \widehat \theta \right) \right\} \left\{1 - \widehat{h}^{(2),w_{is}}_s\left(\overline 0, V; \widehat \gamma, \widehat \eta, \widehat \theta \right) \right\} \right].
\end{eqnarray*}

\paragraph{Estimators based on modelling the subdistribution hazard}

Following Appendix~\ref{sec:app-ident2}, it can easily be shown that, under assumptions~\eqref{app-seqexch}-\eqref{app-consistency}, for each stratum $V = v: \Pr(V=v) > 0$, and for each $k = 1,...,K$, $h^{\text{sd},\overline a = 0}_k (v)$ is non-parametrically identified as
\begin{eqnarray*}
  h^{\text{sd},\overline a = 0}_k (v) &=& h^{\text{sd},w_{ik}}_k\left(\overline 0, v\right)
\end{eqnarray*}
with
\begin{eqnarray*}
  h^{\text{sd},w_{ik}}_k\left(\overline a_k, V\right) &\equiv& \dfrac{E[Y_k(1-Y_{k-1})I(\overline A_k = \overline a_k) W^{\text{sd},A}_k \vert V]}{E[(1-Y_{k-1})I(\overline A_k = \overline a_k) W^{\text{sd},A}_k \vert V]}
\end{eqnarray*}
and weights
\begin{eqnarray*}
  W^{\text{sd},A}_{ik} &\equiv& \prod_{s=1}^k \dfrac{1}{\Pr\left(A_{is} \middle\vert \overline L_{i,s-1}, \overline A_{i,s-1}, \overline D_{i,s-1}, Y_{i,s-1} = 0\right)}.
\end{eqnarray*}
As before, given that we are only considering monotone (dichotomous) exposure patterns (i.e. for each $k = 1,...,K$, we have $A_k \ge A_{k-1}$), when $\overline L_k$ is high-dimensional and/or includes non-categorical covariates, the weights $W^{\text{sd},A}_{k}$ can be estimated by
\begin{eqnarray*}
  W^{\text{sd},A}_{ik}(\widehat \theta) &\equiv& \prod_{s=1}^k \dfrac{1}{\left\{1-p(s, \overline L_{i,s-1}; \widehat \theta)\right\}^{(1-D_{i,s-1})(1-A_{is})} p(s, \overline L_{i,s-1}; \widehat \theta) \vphantom{\left\{p(s, \overline L_{i,s-1}; \widehat \theta)\right\}}^{(1-A_{i,s-1})(1-D_{i,s-1})A_{is}}}.
\end{eqnarray*}
However, given that $V$ may also be high-dimensional and/or include non-categorical covariates, non-parametric estimation of $h^{\text{sd},w_{k}}_k\left(\overline 0, V\right)$ may not be feasible either. Instead, a non-saturated MSM may be postulated for these counterfactual subdistribution hazards. In practice, its indexing parameters are often estimated by fitting either a weighted Cox model or a weighted pooled logistic regression model \citep[see e.g.][]{Bekaert2010}, with $W^{\text{sd},A}_{k}(\widehat \theta)$ as IPC weights. Under correct specification of $p(k, \overline L_{k-1};\theta)$, the estimates of the nuisance parameters $\zeta$ of this working model for $h^{\text{sd},w_{ik}}_k\left(\overline a_k, V\right)$ are consistent estimates of the structural parameters of the corresponding postulated MSM for the counterfactual subdistribution hazard $h^{\text{sd},\overline a_k}_k (V)$, leading to the following alternative model-based estimator for $\varphi_K$:
\begin{eqnarray*}
  E_n \left[ \sum_{k=1}^K \widehat{h}^{\text{sd},w_{ik}}_k\left(\overline 0, V; \widehat \zeta, \widehat \theta \right) \prod_{s=1}^{k-1} \left\{1 - \widehat{h}^{\text{sd},w_{is}}_s\left(\overline 0, V; \widehat \zeta, \widehat \theta \right) \right\} \right].
\end{eqnarray*}
Again, even under a correctly specified (semi)parametric model $p(k, \overline L_{k-1};\theta)$, the weights $W^{\text{sd},A}_{k}(\widehat \theta)$ may tend to be highly unstable when $\overline L_k$ is high-dimensional and/or includes non-categorical covariates, resulting in an inefficient estimator. Alternatively, weights may be stabilized by multiplying $W^{\text{sd},A}_{k}(\widehat \theta)$ with an arbitrary factor that only depends on $k$, $\overline A_k$ and $V$, e.g.
\begin{eqnarray*}
  SW^{\text{sd},A}_{ik}(\widehat \xi, \widehat \theta) &\equiv& W^{\text{sd},A}_{ik}(\widehat \theta) \prod_{s=1}^k \left\{1-r(s, V_i; \widehat \xi)\right\}^{1-A_{is}} r(s, V_i; \widehat \xi) \vphantom{\left\{r(s, V_i; \widehat \theta)\right\}}^{(1-A_{i,s-1})A_{is}}\\
  &=& \prod_{s=1}^k \dfrac{\left\{1-r(s, V_i; \widehat \xi)\right\}^{1-A_{is}} r(s, V_i; \widehat \xi) \vphantom{\left\{r(s, V_i; \widehat \xi)\right\}}^{(1-A_{i,s-1})A_{is}}}{\left\{1-p(s, \overline L_{i,s-1}; \widehat \theta)\right\}^{(1-D_{i,s-1})(1-A_{is})} p(s, \overline L_{i,s-1}; \widehat \theta) \vphantom{\left\{p(s, \overline L_{i,s-1}; \widehat \theta)\right\}}^{(1-A_{i,s-1})(1-D_{i,s-1})A_{is}}},
\end{eqnarray*}
where $r(k, V;\xi)$ is a pre-specified model, indexed by (vector-valued) parameter $\xi$, for $\Pr(A_{k}=1 | A_{k-1} = Y_{k-1} = 0, V)$.
Applying stabilized weights $SW^{\text{sd},A}_{k}(\widehat \xi, \widehat \theta)$ instead of $W^{\text{sd},A}_{k}(\widehat \theta)$ in the above procedure leads to the following model-based estimator for $\varphi_K$:
\begin{eqnarray*}
  E_n \left[ \sum_{k=1}^K \widehat{h}^{\text{sd},w_{ik}}_k\left(\overline 0, V; \widehat \zeta, \widehat \xi, \widehat \theta \right) \prod_{s=1}^{k-1} \left\{1 - \widehat{h}^{\text{sd},w_{is}}_s\left(\overline 0, V; \widehat \zeta, \widehat \xi, \widehat \theta \right) \right\} \right].
\end{eqnarray*}

\clearpage

\section{Plots with 95\% bootstrap confidence intervals}\label{sec:app-confint}

\begin{figure}[!htbp]
\centerline{\includegraphics[width=440pt, angle = 90]{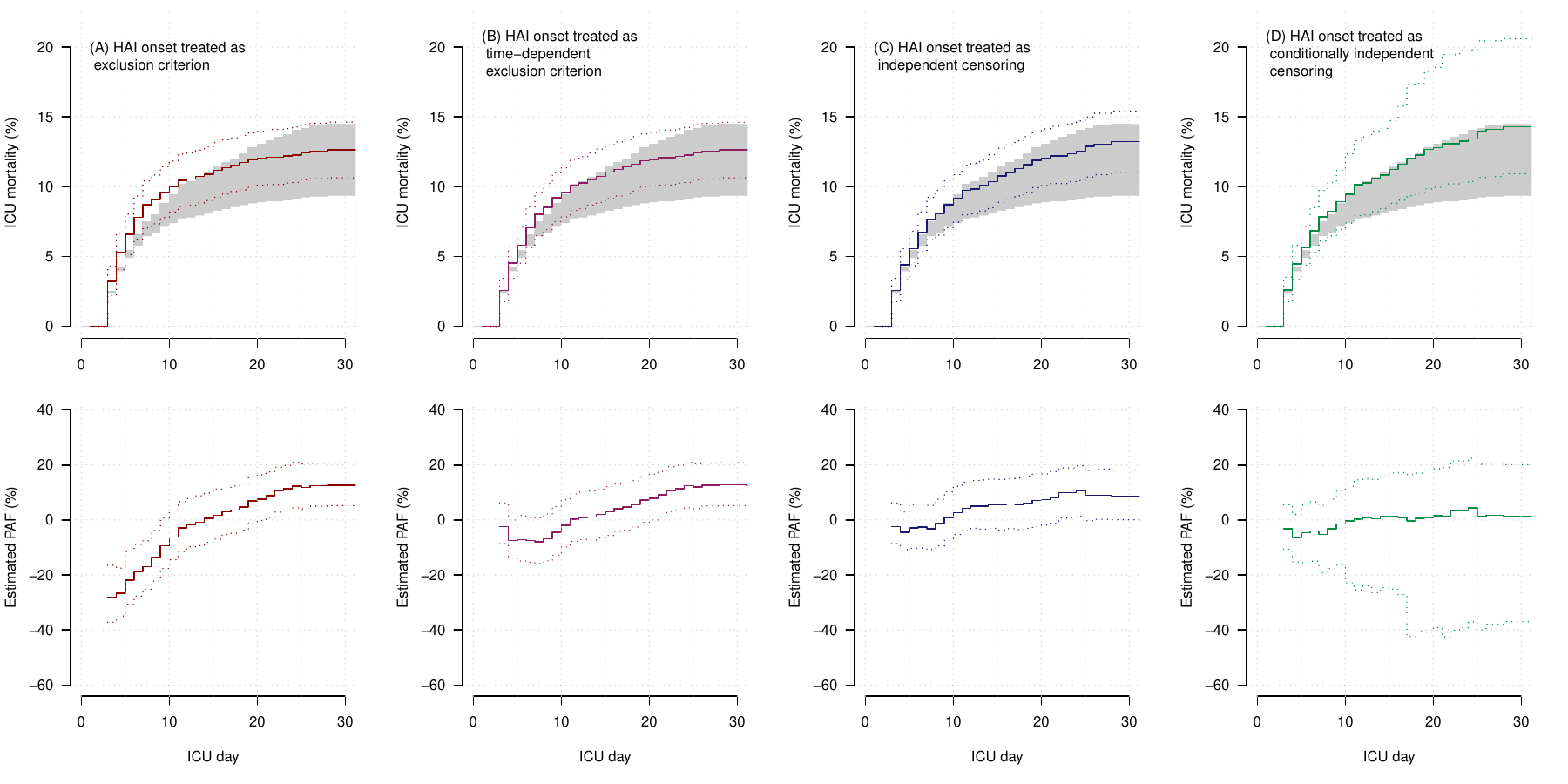}}
\caption{Point estimates of the \emph{counterfactual} risk of ICU death within the first 30 ICU days (see motivating example), as obtained by the Aalen-Johansen estimators $\widehat\varphi_K^{\star \text{\tiny AJ}}$, $\widehat\varphi_K^{\dagger \text{\tiny AJ}}$, $\widehat\varphi_K^{\circ \text{\tiny AJ}}$ and $\widehat\varphi_K^{\text{\tiny AJ}}$ (solid colored lines in the upper/left panels A--D, respectively), and their corresponding 95\% bootstrap confidence intervals (dotted colored lines). As a visual reference, the upper bound of the gray shaded area depicts the \emph{factual} risk of ICU death as estimated by $\widehat\phi_K^{\text{\tiny AJ}}$, while the lower bound depicts the factual risk of HAI-free ICU death as estimated by $\widehat{\tilde\phi}_K^{\text{\tiny AJ}}$. The lower/right panels display the corresponding estimated population-attributable fractions (PAFs).\label{fig:emp_example_app}}
\end{figure}

\end{document}